\documentclass{aa}  

\usepackage{txfonts}
\usepackage[breaklinks=false]{hyperref}
\usepackage{color}

\newcommand{\imin}{i_{\rm min}}
\newcommand{\jmin}{j_{\rm min}}
\newcommand{\imax}{i_{\rm max}}
\newcommand{\jmax}{j_{\rm max}}
\newcommand{\MEANZ}{2.40}

\newcommand{\lamrf}{\lambda_{\rm RF}}
\newcommand{\lamlya}{\lambda_{\rm Ly\alpha}}
\newcommand{\fqlam}{f_q(\lambda)}
\newcommand{\dqlam}{\delta_q(\lambda)}

\newcommand{\dhatqlam}{\hat\delta_q(\lambda)}

\newcommand{\LyaValue}{121.567}
\newcommand{\LybValue}{102.572}
\newcommand{\LyaForestMinValue}{104}
\newcommand{\LyaForestMaxValue}{120}

\newcommand{\lamRF}{\lambda_{\rm RF}}

\newcommand{\LyaForest}{Ly$\alpha$~forest}

\newcommand{\Lya}{$\mathrm{Ly}\alpha$}
\newcommand{\Lyb}{$\mathrm{Ly}\beta$}

\newcommand{\bqso}{b_{\rm QSO}}
\newcommand{\betaqso}{\beta_{\rm QSO}}
\newcommand{\sigmavqso}{\sigma_{v,{\rm QSO}}}
\newcommand{\drparqso}{\Delta r_{\parallel,{\rm QSO}}}
\newcommand{\bhcd}{b_{\rm HCD}}
\newcommand{\betahcd}{\beta_{\rm HCD}}
\newcommand{\Lhcd}{L_{\rm HCD}}
\newcommand{\blya}{b_{\rm Ly\alpha}}
\newcommand{\betalya}{\beta_{\rm Ly\alpha}}
\newcommand{\bbetalya}{b(1+\beta)_{\rm Ly\alpha}}
\newcommand{\lambdauv}{\lambda_{\rm UV}}
\newcommand{\auv}{a_{\rm UV}}
\newcommand{\tuv}{t_{\rm UV}}

\newcommand{\LyaMath}{\mathrm{Ly}\alpha}

\newcommand{\LCDM}{$\Lambda$CDM}
\newcommand{\lcdm}{$\Lambda$CDM}
\newcommand{\hMpc}{h^{-1}~\mathrm{Mpc}}
\newcommand{\hGpc}{h^{-1}~\mathrm{Gpc}}
\newcommand{\apar}{$\alpha_{\parallel}$}
\newcommand{\aperp}{$\alpha_{\perp}$}
\newcommand{\aparMath}{\alpha_{\parallel}}
\newcommand{\aperpMath}{\alpha_{\perp}}
\newcommand{\chiSquareMin}{\chi^{2}_{\rm min}}
\newcommand{\DHh}{D_{H}}
\newcommand{\DMm}{D_{M}}

\newcommand{\om}{\Omega_{m}}
\newcommand{\oc}{\Omega_{c}}
\newcommand{\ob}{\Omega_{b}}
\newcommand{\on}{\Omega_{\nu}}
\newcommand{\ol}{\Omega_{\Lambda}}

\newcommand{\rpar}{r_{\parallel}}
\newcommand{\rperp}{r_{\perp}}
\newcommand{\RparalRperp}{\left(r_{\parallel},r_{\perp}\right)}
\newcommand{\xiOneD}{\xi^{ff,1D}}
\newcommand{\xiff}{\xi^{ff}}
\newcommand{\xihat}{\hat{\xi}}
\newcommand{\xiqf}{\xi^{\,qf}}
\newcommand{\xiqq}{\xi^{\,qq}}
\newcommand{\xiqfth}{\xi^{\,qf,th}}
\newcommand{\xibb}{\xi^{\rm BB}}
\newcommand{\xiqath}{\xi^{\,qa}}
\newcommand{\xismooth}{\xi_{\rm smooth}}
\newcommand{\xipeak}{\xi_{\rm peak}}
\newcommand{\xitp}{\xi^{\rm TP}}
\newcommand{\Psmooth}{P_{\rm sm}}
\newcommand{\Ppeak}{P_{\rm peak}}
\newcommand{\PQL}{P_{\rm QL}}
\newcommand{\PL}{P_{\rm L}}
\newcommand{\Apeak}{A_{\rm peak}}
\newcommand{\zeff}{z_{\rm eff}}
\newcommand{\FNL}{F_{\rm NL}}
\newcommand{\VNL}{V_{\rm NL}}

\DeclareMathOperator*{\sinc}{sinc}

\makeatletter
\renewcommand*\aa@pageof{, page \thepage{} of \pageref*{LastPage}}
\makeatother

\begin{document}

\title{Baryon acoustic oscillations from the complete SDSS-III
  Ly$\alpha$-quasar cross-correlation function at $z=2.4$}

\author{
H\'elion du Mas des Bourboux\inst{1},
Jean-Marc~Le Goff\inst{1},
Michael~Blomqvist\inst{2},
Nicol\'as~G.~Busca\inst{3}, 
Julien~Guy\inst{4},
James~Rich\inst{1},
Christophe~Y\`eche\inst{1,5},
Julian~E.~Bautista\inst{6},
\'{E}tienne~Burtin\inst{1},
Kyle S. Dawson\inst{6},
Daniel~J.~Eisenstein\inst{7},
Andreu~Font-Ribera\inst{5,8},
David~Kirkby\inst{9},
Jordi~Miralda-Escud\'{e}\inst{10,11},
Pasquier~Noterdaeme\inst{12},
Isabelle~P\^aris\inst{2},
Patrick~Petitjean\inst{12},
Ignasi~P\'erez-R\`afols\inst{11},
Matthew~M.~Pieri\inst{2},
Nicholas~P.~Ross\inst{13},
David~J.~Schlegel\inst{5},
Donald~P.~Schneider\inst{14,15},
An\v{z}e~Slosar\inst{16},
David~H.~Weinberg\inst{17},
%\and
Pauline~Zarrouk\inst{1}
}

\institute{
IRFU, CEA, Universit\'e Paris-Saclay,  F-91191 Gif-sur-Yvette, France
\and
Aix Marseille Universit\'e, CNRS, LAM (Laboratoire d’Astrophysique de Marseille) UMR 7326, F-13388, Marseille, France
\and
APC, Universit\'{e} Paris Diderot-Paris 7, CNRS/IN2P3, CEA, Observatoire de Paris, 10, rue A. Domon \& L. Duquet,  Paris, France
\and
LPNHE,  CNRS/IN2P3,  Universit\'e  Pierre  et  Marie  Curie Paris 6, Universit\'e Denis Diderot Paris 7, 4 place Jussieu, 75252 Paris CEDEX, France
\and
Lawrence Berkeley National Laboratory, 1 Cyclotron Road, Berkeley, CA 94720, USA
\and
Department of Physics and Astronomy, University of Utah, 115 S 1400 E, Salt Lake City, UT 84112, USA
\and
Harvard-Smithsonian Center for Astrophysics, 60 Garden St., Cambridge, MA 02138, USA
\and
Department of Physics and Astronomy, University College London, Gower Street, London, United Kingdom 
\and
Department of Physics and Astronomy, University of California, Irvine, CA 92697, USA
\and
Instituci\'{o} Catalana de Recerca i Estudis  Avan\c{c}ats, Barcelona, Catalonia
\and
Instituci\'{o} de Ciències del Cosmos, Universitat de Barcelona (UB-IEEC), Catalonia
\and
Universit\'e Paris 6 et CNRS, Institut d'Astrophysique de Paris, 98bis blvd. Arago, 75014 Paris, France
\and
SUPA, Institute for Astronomy, University of Edinburgh, Royal Observatory, Edinburgh, EH9 3HJ, United Kingdom
\and
Department of Astronomy and Astrophysics, The Pennsylvania State University, University Park, PA 16802
\and
Institute for Gravitation and the Cosmos, The Pennsylvania State University, University Park, PA 16802
\and
Brookhaven National Laboratory, 2 Center Road,  Upton, NY 11973, USA
\and
Department of Astronomy, Ohio State University, 140 West 18th Avenue, Columbus, OH 43210, USA
}

\abstract{
        We present a measurement of baryon acoustic oscillations (BAO)
        in the cross-correlation of quasars with the Ly$\alpha$-forest
        flux transmission at
        a mean redshift of $z=2.40$.
        The measurement uses the complete Sloan Digital Sky Survey (SDSS-III)
        data sample:
        168,889 forests and 234,367 quasars
        from the SDSS data release  DR12.
        In addition to the statistical improvement on our previous study using
        DR11, we have implemented numerous improvements
        at the analysis level enabling a more accurate measurement of
        this cross-correlation.
        We have also developed the first simulations of the cross-correlation
        that allow us to test different aspects of our data analysis
        and to search
        for potential systematic errors in the determination of the
        BAO peak position.
        We measure the
        two ratios $D_{H}(z=2.40)/r_{d} = 9.01 \pm 0.36$
        and $D_{M}(z=2.40)/r_{d} = 35.7 \pm 1.7$,
        where the errors
        include marginalization over the non-linear velocity of quasars and
        the 
        cross-correlation of metals and quasars, among other effects.
        These results are within $1.8\sigma$ of the prediction
        of the flat-$\Lambda$CDM model describing the  observed
        cosmic microwave background (CMB)
        anisotropies.
        We combine  this study with the
    Ly$\alpha$-forest
    auto-correlation function,
    yielding
        $D_{H}(z=2.40)/r_{d} = 8.94 \pm 0.22$
        and $D_{M}(z=2.40)/r_{d} = 36.6 \pm 1.2$,
        within $2.3\sigma$ of the
        same flat-$\Lambda$CDM
        model.
}


\authorrunning{H. du Mas des Bourboux et al.}
\titlerunning{BAO at $z=2.40$}
\maketitle

\section{Introduction}

Baryon acoustic oscillations (BAO) in the pre-recombination universe
\citep{1970ApJ...162..815P,1970Ap&SS...7....3S}
left their imprint on the anisotropy spectrum
of the cosmic microwave background
(CMB)
and on late-time correlations of the matter density.
These two effects provide a well-understood tool for
studying cosmological models.
The CMB anisotropy spectrum \citep{2016A&A...594A..13P}
provides percent-level
measurements of the matter and baryon densities relative
to the known photon density and thereby
precisely fixes the parameters of the flat-\LCDM~cosmological model.
The position of the BAO peak in the late-time correlation function
determines the angular and Hubble distances at
the observed redshift, both relative to the sound horizon, $r_d$.
Such measurements allow one to constrain more complicated cosmological
models that include non-zero curvature and/or evolving dark energy
\citep{2016A&A...594A..13P,2015PhRvD..92l3516A}.

The original studies of the BAO peak
\citep{2005ApJ...633..560E,2005MNRAS.362..505C},
and most of those that followed,
have used galaxies as mass tracers.
The most precise measurements were 
in the redshift range  $0.35<z<0.65$  from
the Baryon Oscillation Spectroscopy Survey (BOSS) of the
Sloan Digital Sky Survey (SDSS-III)
\citep{2012MNRAS.427.3435A,2014MNRAS.439...83A,2014MNRAS.441...24A,2017MNRAS.470.2617A}.
Other measurements using galaxies
\citep{2007MNRAS.381.1053P,2010MNRAS.401.2148P,2011MNRAS.416.3017B,2011MNRAS.415.2892B,2012MNRAS.427.2132P,2012MNRAS.427.2168M,2012MNRAS.426..226C,2013MNRAS.431.2834X,2015MNRAS.449..835R}
map  distances and expansion rates for $z<0.8$.
The first observations of the BAO peak in the range $0.8<z<2.2$
using the eBOSS quasars as tracers have
recently been reported \citep{2017arXiv170506373A}.
There is an impressive agreement between the results of these studies 
and the expectations
of flat-\LCDM~models based on CMB data, as emphasized
by \citet{2016A&A...594A..13P}.

BAO correlations can be studied at redshift near $z\sim2.4$ by
using the flux transmission in \Lya{} forests
as a mass tracer \citep{2007PhRvD..76f3009M}.
The BAO peak has been detected in
the transmission auto-correlation of SDSS \Lya{} forests
\citep{2013A&A...552A..96B,2013JCAP...04..026S,2013JCAP...03..024K,
2015A&A...574A..59D,2017A&A...603A..12B}.
Complementary to the auto-correlation, BAO can also be studied using
the cross-correlation of quasars
and the flux in \Lya{} forests.
Such correlations were first detected
in SDSS DR9 \citep{2013JCAP...05..018F}, and the first BAO detection
was presented in \citet{2014JCAP...05..027F} using SDSS DR11.

\begin{figure*}[ht]
        \centering
        \includegraphics[width=.90\textwidth]{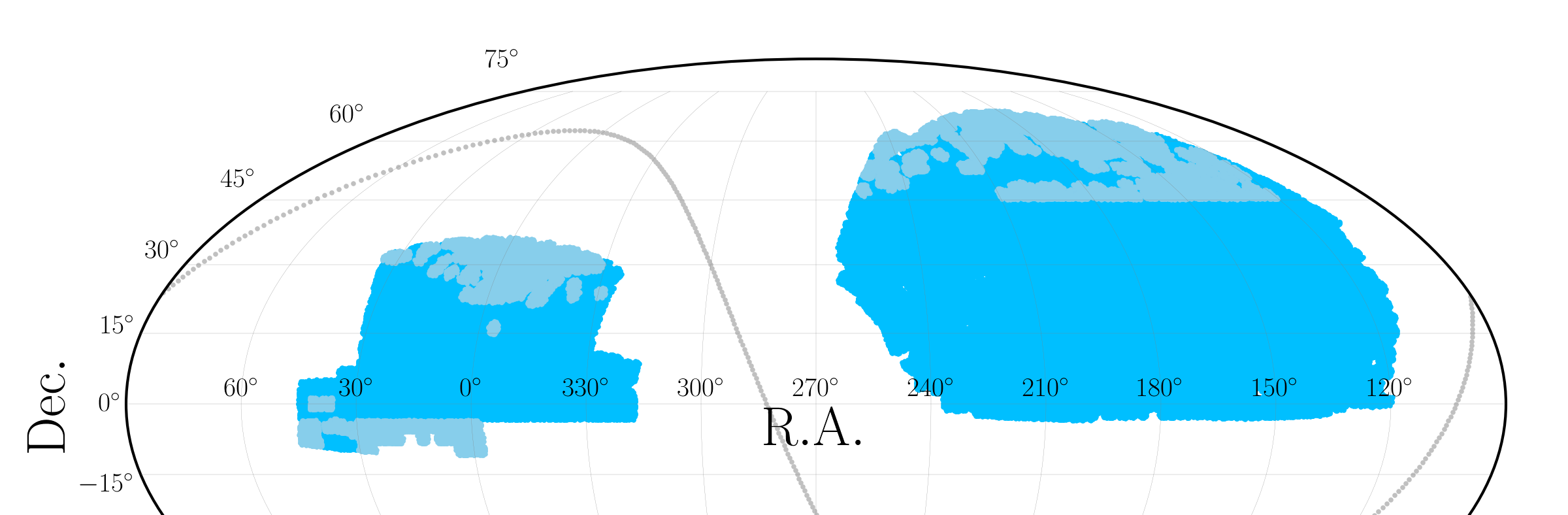}
        \caption{Mollweide projection of the BOSS DR12 footprint
        in equatorial coordinates
        used in this study. The light gray points represent the position of the Galactic plane.
        The blue points are
        the positions of the forests from DR12 used here $z_{\mathrm{forest}} \in [2,6]$.
        The light blue points are the positions of the new forests
        not included in the DR11 study of \citet{2014JCAP...05..027F}.
        }
        \label{figure::footprint}
        \centering
\end{figure*}

This paper presents the DR12 study of the quasar-forest cross-correlation
and derives joint cosmological constraints using the 
\Lya{}-forest auto-correlation of DR12 \citep{2017A&A...603A..12B}.
In addition to the use of an increased number of quasar-forest pairs,
the analysis presented here includes 
three important improvements on the analysis of \citet{2014JCAP...05..027F},
the first two of which were also used
in the auto-correlation analysis of \citet{2017A&A...603A..12B}:
\begin{itemize}
\item
  We use a new data reduction pipeline
  whose most important
  features are described in Sect. \ref{section::data}.
  The new pipeline has
  improved linearity for the small fluxes characteristic
of the \Lya{} forest resulting in 
 a better understanding
of the effects of imperfect modeling of the calibration stars.
\item
We model the distortion of the correlation function due to
the continuum fitting in the forest.
This procedure, described in Sect.
\ref{subsection::The_Lya_forest_quasar_cross_correlation::The_distortion_matrix},
 allows us to fit the 
 observed correlation function without the addition of arbitrary
 power-law ``broadband'' terms.
\item
  We test the analysis procedure with the  mock data sets
described in Sect. \ref{section::mocks}
that contain correlated quasars and forests.
  The mock data sets previously
  used to test the auto-correlation analysis contained
  correlated forests but  no physical correlation with the associated
  quasars.
  The lack of mock data sets  was
  the major limitation of the analysis of \citet{2014JCAP...05..027F}.
\end{itemize}

This paper is organized as follows.
Section \ref{section::data} describes the DR12 data set used in this study.
Section \ref{section::Measurement_of_the_transmission_field} summarizes
the measurement of the flux-transmission
field and
Section \ref{section::The_Lya_forest_quasar_cross_correlation}
describes its correlation with quasars.
Section \ref{section::fits} describes our theoretical model
of the cross-correlation and the fits to the observed correlation function.
The mock data sets used to validate the analysis procedure
are
presented in Section \ref{section::mocks}.
Section \ref{section::cosmo} summarizes the cosmological
implications of these and other BAO measurements.
Section \ref{section::conclusion} presents our conclusions.

\begin{figure}
        \centering
        \includegraphics[width=\columnwidth]{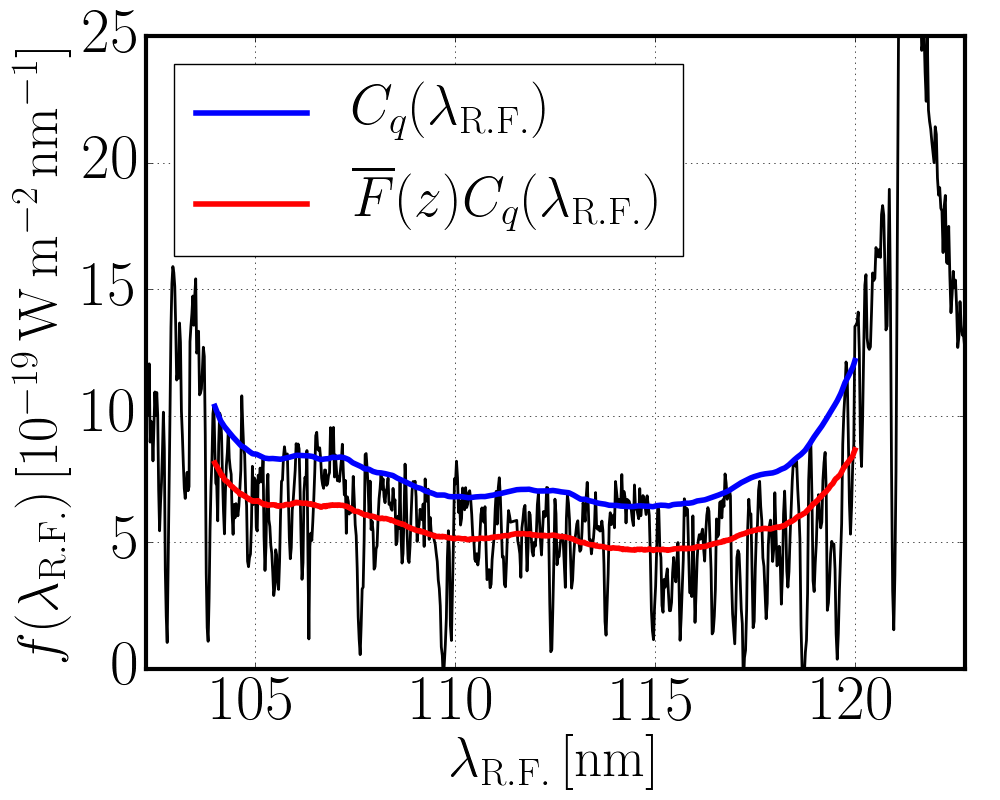}
        \caption{
                Example of a BOSS quasar spectrum of 
                at
                $z=2.91$.
                The spectrograph resolution at $\lambda\sim400~{\rm nm}$
                is $\sim0.2~{\rm nm}$.
                The red and blue lines cover the
                forest region used here, $\lamRF \in [\LyaForestMinValue,\LyaForestMaxValue]~\mathrm{nm}$.
                This region is sandwiched between the quasar's \Lyb{} and \Lya{}
                emission lines at 
                \LybValue{}~nm and \LyaValue{}~nm respectively.
                The blue line is
                the model of the continuum for this particular quasar, $C_q(\lamRF)$,
                and the red line is the
                product of the continuum and the
                mean absorption, $\overline{F}(z) C_q(\lamRF)$,
                as calculated by the method described in 
                Sect.~\ref{section::Measurement_of_the_transmission_field}.
%               $\mathrm{(plate, MJD, fiberId, thingId)} = (6972,56426,0912,535481821)$.
        }
        \label{figure::spec_6972-56426-0912}
\end{figure}

\begin{figure*}[ht]
        
        % Sources/histo_pairs_lya_qso.txt
        % Codes/codes.py :: histogram_QSO_and_pixel()
        %
        % Codes/codes.py :: histogram_pairs()
        
        \centering
        \includegraphics[width=\columnwidth]{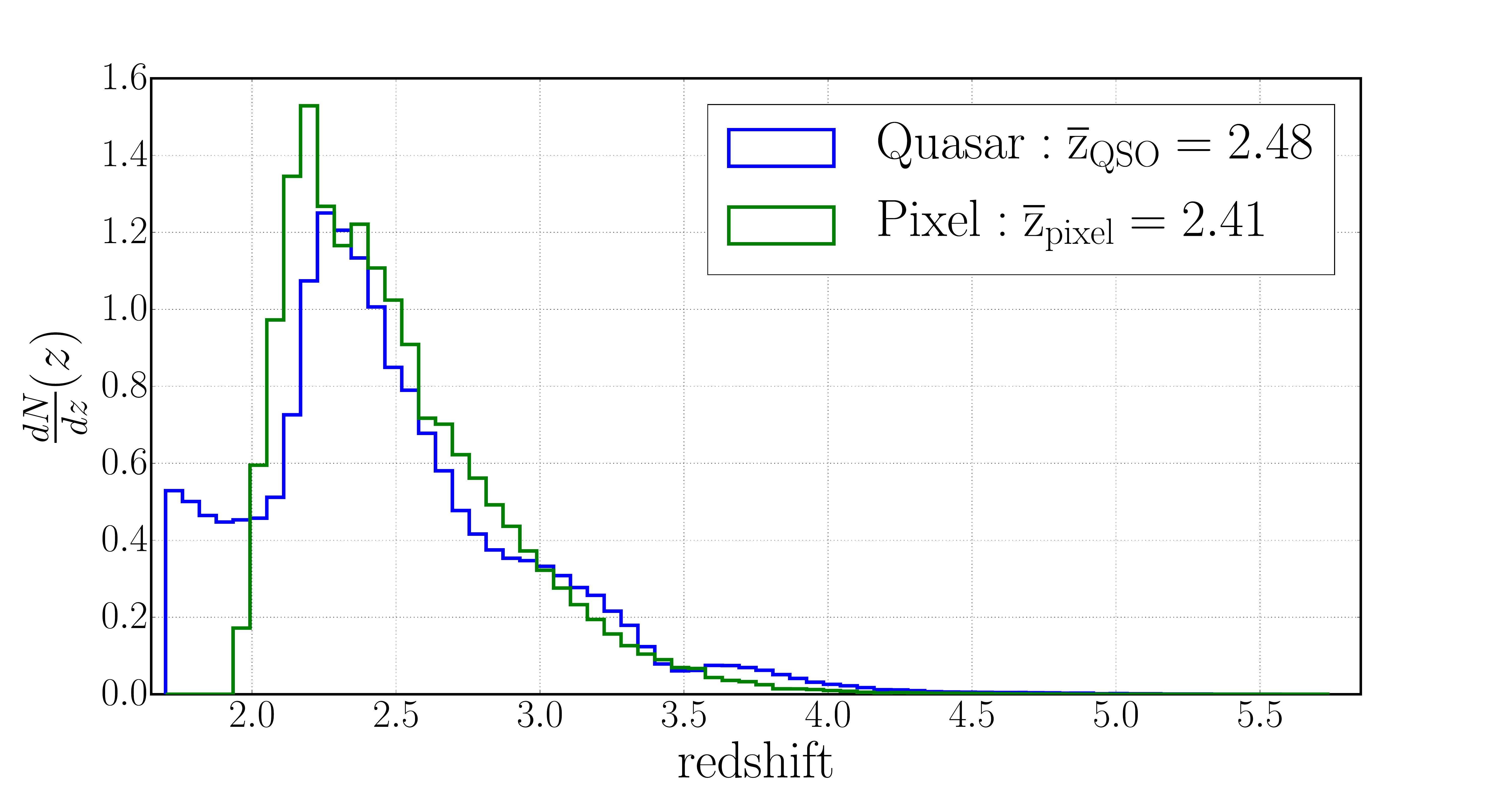}
        \includegraphics[width=\columnwidth]{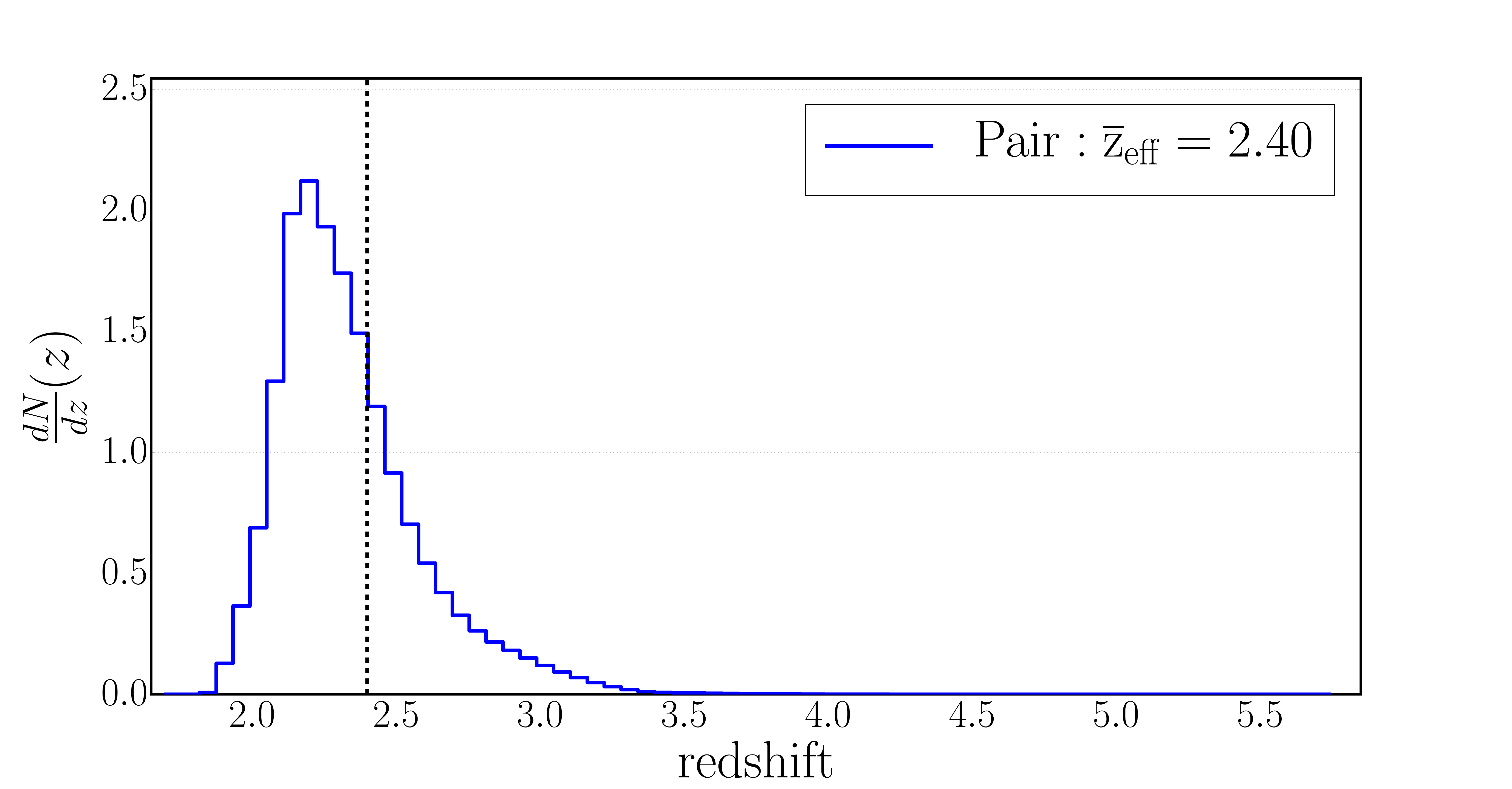}
        \caption{Left panel presents the
          distribution of the redshift of quasars (blue) and
          forest pixels (green)
          with the redshift for the latter calculated assuming
          \Lya{} absorption.
          The pixels are weighted as described in Sect.
          \ref{subsection::The_Lya_forest_quasar_cross_correlation::The_correlation_function}.
                The right panel displays the
                weighted distribution of the redshift of the $1.8\times10^9$
                pixel-quasar pairs in the BAO region:
                $r_{pair} \in [80,120]~\hMpc$.
                The redshift of a pair is defined by: $z_{pair} = (z_{pixel}+z_{\rm QSO})/2$.
                The weighted mean redshift of the pairs (dashed black line)
                defines the mean redshift, $\zeff=2.40$, of the measurement of the BAO peak position.
        }
        \label{figure::histogram_redshift_pairs}
\end{figure*}

% ----------------------------------------------------------------------
\section{Quasar and forest samples and data reduction}
\label{section::data}

The results presented here are based on data collected
by the Sloan Digital Sky Survey (SDSS) \citep{2000AJ....120.1579Y}.
Most of the quasars and the entirety of
the \Lya{} forests were
gathered over a five-year period
by the SDSS-III Collaboration
\citep{2011AJ....142...72E, % Eisenstein++ 2011, SDSS-III
%2012AJ....144..144B, % Bolton++ 2012, pipeline
1998AJ....116.3040G, % Gunn++ 1998, SDSS camera
2006AJ....131.2332G, % Gunn++ 2006, SDSS telescope
2013AJ....146...32S% Smee++ 2012, SDSS/BOSS spectrographs
%2000AJ....120.1579Y} % York++ 2000, SDSS-I/II technical summary
}.
This data is publicly available in the twelfth data release (DR12) of SDSS
as presented in 
\citet{2015ApJS..219...12A}.
The DR12 celestial footprint covering
$\sim\pi~\rm{sr}\sim10^4~{\rm deg^2}$
is displayed in 
Fig.~\ref{figure::footprint}.

The DR12 quasar catalog is described in \citet{2017A&A...597A..79P}.
Most of the quasar spectra were obtained by
the Baryon Oscillation Spectroscopic Survey, BOSS \citep{2013AJ....145...10D}.
However,
DR12 also includes six months of data from
SEQUELS
\citep{2015ApJS..221...27M,2015ApJS..219...12A},
the pilot survey for the eBOSS survey.
We have also used quasars, but not forests,
from the SDSS DR7 quasar catalog \citep{2010AJ....139.2360S}.
Figure \ref{figure::spec_6972-56426-0912} displays a
typical quasar spectrum in the forest wavelength range
where the BOSS spectrograph resolution is $\sim 0.2~{\rm nm}$.

The quasar target selection used in BOSS, summarized in
\citet{2012ApJS..199....3R},  combines different targeting methods described
in \citet{2010A&A...523A..14Y}, \citet{2011ApJ...743..125K}, and
\citet{2011ApJ...729..141B}.
The selection algorithms use SDSS photometry and, when available,
data from
the GALEX
survey \citep{2005ApJ...619L...1M} in the UV,
the UKIDSS
survey \citep{2007MNRAS.379.1599L} in the NIR,
and the FIRST
survey \citep{1995ApJ...450..559B} in the radio.

As described in \citet{2017A&A...603A..12B},
the DR12 data were  processed using a new software package that differs
from the standard DR12 SDSS-III pipeline 
\citep{2012AJ....144..144B} 
and which has become the standard
pipeline for SDSS DR13 \citep{2016arXiv160802013S}.
For this study, the most important
difference with respect to the DR12 pipeline
is that pixels on the CCD
image are combined to give a flux
with pixel-weights determined only by the CCD readout noise.
While this method is sub-optimal because it ignores photo-electron Poisson
noise, compared to the DR12 method
it yields an unbiased flux estimate 
since the weights do not depend
on the observed CCD counts, which are needed to estimate Poisson noise.
A more detailed description of the changes to the extraction pipeline 
is given in Appendix~A of \citet{2017A&A...603A..12B}.

For each object,
the pipeline provides a flux calibrated spectrum, $f(\lambda)$, errors,
and an object classification (galaxy, quasar, star).
A model spectrum is fit to $f(\lambda)$, providing  
a redshift estimate.
For this study,
we use the ``coadded'' spectra constructed from
typically four exposures of 15 minutes resampled at
wavelength pixels of width 
$\Delta\log_{10}\lambda=10^{-4}$ 
($c\Delta\lambda/\lambda\sim{\rm 69~km\,s^{-1}}$).
Unlike the auto-correlation measurement of \citet{2017A&A...603A..12B},
we use these pixels directly, not combining them into wider analysis pixels.
Approximately 10\% of the   quasars 
have repeated observations widely separated in time, in which case
we use the observation with the best signal-to-noise ratio.

The mean ratio, $R(\lambda)$, of model and observed fluxes
as a function of observed wavelength  have
small ($\sim1\%$)
deviations from unity caused by imperfect modeling
of the spectra of calibration stars.
As described in \citet{2017A&A...603A..12B},
the fluxes were given a global correction for these imperfections
by dividing them by $R(\lambda)$ estimated in the weakly absorbed range
$(141<\lamRF<153~{\rm nm})$.

The spectra of all quasar targets were visually inspected
\citep{2012A&A...548A..66P,2014A&A...563A..54P,2017A&A...597A..79P} 
to correct for misidentifications, to flag broad absorption lines (BALs),
and to determine the quasar redshift used in the analysis.
Damped \Lya{} troughs (DLAs)\citep{1986ApJS...61..249W}
were visually flagged, but also identified and characterized
automatically \citep{2012A&A...547L...1N}.
The visual inspection of DR12 confirmed
216,579 quasars in the redshift range $1.7<z_q<5.8$ to which we add
17,788 DR7 quasars that are not included in DR12, producing
a total of 234,367 quasars.
Their redshift distribution is shown in Fig. \ref{figure::histogram_redshift_pairs}.
The forest sample is taken from 198,357 DR12 quasars
in the range $2.0<z_q<6.0$ (Fig. \ref{figure::histogram_redshift_pairs}).
Elimination of spectra with identified BALs leaves 174,726 forests.
Requiring 50 or more pixels in the \Lya{} forest regions then leaves
171,579 forests.
Finally, 2690 forests failed the continuum fitting procedure, producing
a total of 168,889 forests for this study.

Because of the very low number of observed quasars at $z>3.5$ and
the requirement that a sufficient range of forest be within the
spectral range of SDSS,
the quasar-forest pixel pairs actually used for the calculation of the
cross-correlation function involved mostly quasars in the range $1.8<z_q<3.5$
and forests with quasars in the range $2.0<z_q<3.5$.
In these ranges, our sample includes 217,780 quasars and 157,845 forests,
to be compared with 164,017 and 130,825 for the study
of \citet{2014JCAP...05..027F}.

For the measurement of the flux transmission, we adopt the
rest-frame wavelength interval
\begin{equation}
        104\,<\lamrf<120~{\rm nm} \; .
        \label{rflambdarange}
\end{equation}
As illustrated in Fig.~\ref{figure::spec_6972-56426-0912},
this range is 
bracketed by the  emission lines $\lambda_{\rm Ly\beta}=102.572$~nm and 
$\lamlya=121.567$~nm.  This region
was chosen as the maximum range that avoids the large pixel
variances on the wings of the two lines due to  quasar-to-quasar
diversity of line-emission strengths and profiles.
The observed wavelength range is
\begin{equation}
360.0 < \lambda < 723.5~{\rm nm},
\end{equation}
corresponding to the redshift range
$1.96<z<4.96$ for \Lya{} absorption. The lower limit
is set by the requirement that
the system throughput be greater than 10\% of its peak value. 
The upper limit on $\lambda$ is, in fact, of no importance because
there are few quasar-pixel pairs beyond $z=3.5$  ($\lambda=547$~nm).
The distribution of the redshift of \Lya-absorber-quasar pairs
contributing to the BAO peak
is shown in the righthand panel of Fig. \ref{figure::histogram_redshift_pairs}.
The pixels are weighted as described in Sect.
\ref{subsection::The_Lya_forest_quasar_cross_correlation::The_correlation_function}.
The distribution has a weighted mean of $\zeff=2.40$, which defines
the effective redshift of our measurement of the BAO peak position.

\section{Measurement of the transmission field}
\label{section::Measurement_of_the_transmission_field}

Fluctuations in the flux transmission fraction are defined  by
\begin{equation}
        \dqlam=\frac{\fqlam}{ C_q(\lambda)\overline{F}(z)}-1 \;.
        \label{delta:def}
\end{equation}
Here, $\fqlam$ is the observed flux density for quasar $q$
at observed wavelength $\lambda$,
$C_q(\lambda)$ is the unabsorbed flux density 
(the so-called ``continuum''), and
 $\overline{F}(z)$ is the mean transmitted fraction
 at the absorber redshift, $z(\lambda)=\lambda/\lamlya-1$. 
Measurement of the flux-transmission field $\dqlam$ requires estimates
of the product $C_q(\lambda)\overline{F}(z)$ for each quasar.
We closely follow  the procedure used for the auto-correlation
measurement \citep{2017A&A...603A..12B}.
We assume the quasar continuum, $C_q(\lambda)$, 
is the product of  a universal
function of the rest-frame wavelength, $\lamrf=\lambda/(1+z_q)$ and
a quasar-dependent linear function of $\lamrf$,
included to account for 
quasar spectral diversity:
\begin{equation}
C_q(\lambda) 
= C(\lamrf)
[a_q + b_q(\lamrf-\overline{\lamrf})],
\label{templatewarpeq}
\end{equation}
where $\overline{\lamrf}$ is the weighted mean for each forest and where
$C(\lamrf)$ is normalized so that its integral over the forest
is equal to unity.
The $(a_q,b_q)$ and $C(\lamrf)$ are determined by maximizing
the likelihood function given by
\begin{equation}
L = \prod_{q,\lambda} P(\, \fqlam\;|\;C_q(\lambda)\,) \;.
\end{equation} 
Here $P(\fqlam\, | \,C_q(\lambda))$ is the 
probability to observe a flux $\fqlam$
for a given continuum found by convolving the intrinsic probability,
$D(F=\fqlam/C_q(\lambda),z)$,
with the observational resolution assumed to be Gaussian: 
\begin{equation}
 P(\, \fqlam\;|\;C_q(\lambda)\,)
\propto
\int_0^1 dF D(F,z)\exp\left[
\frac{-(C_qF-\fqlam)^2}{2\sigma_q^2(\lambda)}
\right]\;,
\end{equation}
where $\sigma_q^2(\lambda)$ is the variance due to readout noise and
photon statistics.
The function
$D(F,z)$ is taken to be 
the log-normal model of absorption used to generate the mock data
of \citet{2015JCAP...05..060B}.

As emphasized in \citet{2017A&A...603A..12B}, the use of forest data
to determine the quasar continuum
necessarily produces biased estimates of $\dqlam$
because of two effects.
The most important is that fitting an amplitude and slope
$(a_q,b_q)$ for each forest
biases the mean $\dqlam$ and its first moment toward
vanishing values within a given forest.
Since this bias is only approximate, we find it
convenient to make it exact  by explicitly subtracting
from each $\dqlam$
(defined by Eq. \ref{delta:def})
the mean and first moments:
\begin{equation}
        \dhatqlam = \dqlam
        - \overline{ \delta_q} 
        - \Lambda 
        \frac{\overline{\Lambda\delta}}
        {\overline{\Lambda^2}} \; ,
        \hspace*{5mm}
        \Lambda\equiv\lamrf-\overline{\lamrf} \; ,
        \label{dhatdef}
\end{equation}
where the over-bars refer to weighted averages over individual forests.
The resulting values of $\hat\delta_q(\lambda)$ are thus linear combinations
of the originals:  $\hat\delta_i = \sum_iP_{ij}\delta_j$ with
the projection matrix given by
\begin{equation}
        P_{ij} =
        \delta^{K}_{ij}
        -\frac{w_{j}}{\sum\limits_{k} w_{k}}
        -\frac{
          w_{j}
          \Lambda_i\Lambda_j
        }{
          \sum\limits_{k} w_{k}
          \Lambda_k^2
        } \; ,
        \hspace*{5mm}\Lambda_k\equiv\lambda_{{\rm RF}\,k}-\overline{\lamrf,}
        \label{equation::analyse_donne::projection_42}
\end{equation}
where $\delta^K_{ij}$ is the Kroeneker delta and
the $w_j$ are weights used in the calculation of the
correlation function (Eq. \ref{equation::xi_cross_estimator}).

The second effect is that fitting $\overline{F}(z)$ biases
toward zero the mean $\delta$
at each observed wavelength, $\overline{\delta(\lambda)}\rightarrow0$,
where the over-bar means the average at fixed $\lambda$.
As the last step, we therefore explicitly transform
the $\dhatqlam$ of Eq. \ref{dhatdef}:
$\dhatqlam \rightarrow \dhatqlam-\overline{\delta(\lambda)}$.
Because of the large number of forests, this transformation has
much less effect than the intra-forest subtraction (\ref{dhatdef}).

% ----------------------------------------------------------------------
%            The \LyaForest - quasar cross-correlation
% ----------------------------------------------------------------------

\section{The \Lya-forest-quasar cross-correlation}
\label{section::The_Lya_forest_quasar_cross_correlation}

The flux-transmission field is sampled at points in a space defined by observed
wavelength and position on the sky.
It is therefore natural to measure the cross-correlation with quasars
as a function of angular and redshift separation,
$\xi(\Delta\theta,\Delta z),$ where $\Delta z$ is the difference
between the quasar redshift and the forest-pixel redshift calculated assuming
\Lya{} absorption.
In the approximation that \Lya{} absorption dominates in the forest,
the BAO peak in these coordinates would be at
$\Delta z=r_d/\DHh(z)$
in the radial direction and at
$\Delta\theta=r_d/\DMm(z)$ in the transverse direction,
where $\DHh(z)=c/H(z)$ and $\DMm(z)$ are the Hubble  and  comoving-angular
distances.
While this formulation has the advantage of remaining close
to the directly observed quantities, it has the disadvantage
that both $\DHh$ and $\DMm$ vary significantly over the redshift
range of BOSS.  This would lead to significant broadening of the peak
unless several wavelength bins were used.

To avoid this complication we transform $(\Delta\theta,\Delta z)$
to Cartesian coordinates, $(\rperp,\rpar)$ using the
distances, $D_q=\DMm(z_q)$ and $D_{\rm Ly\alpha}=\DMm(z_{\rm Ly\alpha})$,
calculated according to 
a flat ``fiducial'' cosmological model:
\begin{equation}
  \rperp=(D_{\rm Ly\alpha}+D_q)
  \sin\left(\frac{\Delta\theta}{2}\right),
\hspace*{5mm}   \rpar=
  (D_{\rm Ly\alpha}-D_q)
  \cos\left(\frac{\Delta\theta}{2}\right) \;.
        \label{coordinatedef}
\end{equation}
To the extent that \Lya{}-absorption dominates the absorption
field, and if the fiducial cosmology is the true cosmology,
the function $\xi(\rperp,\rpar)$ will be the expected
biased version of the mass correlation function
and the BAO peak will be at the predicted position.
Absorption by metals and the high column-density systems
(HCDs) complicates this simple picture,
and therefore the fits of Sect. \ref{section::fits}
must take these and other effects into account.

\begin{table}
        \centering
        \caption{
                Parameters of the flat-\LCDM{} cosmological model used
                for the production and analysis of the mock spectra
                and for the analysis of the data.
                The \citet{2016A&A...594A..13P} cosmological
                parameters are used for the data.
                The parameters defining the models are given in the first section:
                the density of cold dark matter, baryon, and massive
                neutrinos, the reduced Hubble constant, and the number of light
                neutrino species.
                The second section of the table gives derived
                parameters and quantities calculated at the relevant
                redshift $\zeff$.
                The sound horizon at the drag epoch, $r_{d}$, is calculated
                using
                CAMB
                \citep{2000ApJ...538..473L}.
                The linear growth
                rate of structure is calculated using the approximation
                $f \;\sim\Omega_{m}(z)^{0.55}$ \citep{2007APh....28..481L}.
        }
        \label{table::cosmology_parameters}
        \begin{tabular}{l c c }
                & Mocks & Planck    \\
                &       &  $\left( \mathrm{TT+lowP} \right)$    \\

                \noalign{\smallskip}
                \hline \hline
                \noalign{\smallskip}

                $\om h^2$               &       0.1323  &       0.1426  \\
                $=\oc h^2$              &       0.1090  &       0.1197  \\
                $\;+\ob h^2$    &       0.0227  &       0.02222 \\
                $\;+\on h^2$    &       0.0006  &       0.0006  \\
                $h$                             &       0.7             &       0.6731  \\
                $N_{\nu}$               &       3               &       3               \\
                $\sigma_{8}$    &       0.795   &       0.830   \\
                $n_{s}$                 &       0.97    &       0.9655  \\

                \hline \noalign{\smallskip}
                
                $\om$                                   &       0.27            &       0.3147  \\
                $r_d~[\hMpc]$                   &       104.80          &       99.17   \\
                $r_d~[\mathrm{Mpc}]$            &       149.7           &       147.33  \\
                $\zeff$                         &       2.25            &       \MEANZ  \\
                $\DHh(\zeff)/r_{d}$     &       8.495           &       8.369   \\
                $\DMm(\zeff)/r_{d}$     &       39.24           &       39.77   \\
                $f(\zeff)$                      &       0.95916         &       0.97076 \\

        \end{tabular}
\end{table}

The fiducial cosmology used for the analysis of the data is the
best-fit flat-\LCDM{} model of \citet{2016A&A...594A..13P}.
The parameters of this model are given in the second column
of Table \ref{table::cosmology_parameters}.
The mock spectra were produced using a different cosmology
(Column 1 of the table) and we use this cosmology to
analyze the mock data.

% ----------------------------------------------------------------------
\subsection{The correlation function}
\label{subsection::The_Lya_forest_quasar_cross_correlation::The_correlation_function}

The correlation between the transmission field in the \LyaForest{} and the quasar distribution
is estimated using a simple weighted mean of $\hat{\delta}$
at a given distance of a quasar:
\begin{equation}
        \xihat_{A} = \frac{\sum\limits_{(i,k) \in A} w_{i} \, \hat\delta_{i} }{
        \sum\limits_{(i,k) \in A} w_{i} } \; ,
        \label{equation::xi_cross_estimator}
\end{equation}
where $w_{i}$ is the weight given to a measurement $\hat\delta_{i}$ (see below).
The sum runs over all possible pixel-quasar pairs $(i,k)$ falling inside the bin $A$.
This bin is defined in separation space $A = \RparalRperp_{A}$,
but in this paper
we will also refer to $(r,\mu)$,
with $r^2=\rperp^2+\rpar^2$ and
$\mu=\rpar/r$, the cosine of the angle formed by the line of sight
and the vector $\vec{r}$.
Following Eq. \ref{coordinatedef},
positive values of $\rpar$ correspond to an absorber distance greater
than the quasar distance.
The bins are squares in $(\rperp,\rpar)$-space of size $4~\hMpc$.
We calculate the correlation 
for separations $\rpar \in [-200,200]~\hMpc$ and for $\rperp \in [0,200]~\hMpc$.
We thus have $100$ bins in the $\rpar$ direction
and $50$ in the $\rperp$ direction, with a  total number
of bins, $N_{\mathrm{bin}} = 100 \times 50 = 5000$.

Because of the continuum fit
and  the projection of pixels
(Eq. \ref{dhatdef})
the pixel-quasar correlation vanishes
on all scales for pixels of a quasar's own forest.
For this reason, we do not use such pairs.
As described in  \citet{2015A&A...574A..59D},
the weights, $w_i$, are chosen so as to account for both 
Poisson noise  in the flux measurement and for
the intrinsic fluctuations in $\delta_i$ due to cosmological
large-scale structure.
The weights are set to zero for 
pixels flagged by the pipeline as having problems due, for example, 
to sky emission lines or cosmic rays.

% ----------------------------------------------------------------------
\subsection{The distortion matrix}
\label{subsection::The_Lya_forest_quasar_cross_correlation::The_distortion_matrix}

The transformation (\ref{dhatdef}) mixes
pixels so that the correlation between a quasar and
a pixel is equal to the original quasar-pixel correlation
plus a linear combination of the correlations between the
quasar and the other pixels of the forest.
This statement means that the measured correlation function
is a ``distorted'' version of the true correlation function.
Since the transformation (\ref{dhatdef}) is linear,
the relation between measured, $\xihat$, and true, $\xi$, correlation functions
is given by a distortion matrix $D_{AA^\prime}$:
\begin{equation}
  \xihat_A  = \sum_{A^\prime}D_{AA^\prime}\xi_{A^\prime}\;,
\label{DABfirstuseeq}
\end{equation}
where 
\begin{equation}
        D_{AA^\prime} = \frac{\sum\limits_{(i,k) \in A}
        w_{i}
        ~
        \sum\limits_{(j,k) \in A^\prime}
        P_{ij}
        ~
        }{
        \sum\limits_{(i,k) \in A} w_{i} } \; ,
        \label{equation::distortion_matrix_measure}
\end{equation}
where $i$ and $j$ refer to pixels from the same forest,  $k$ 
refers to a quasar, and $P_{ij}$ is the projection matrix
(Eq. \ref{equation::analyse_donne::projection_42}).

The matrix $D_{AA^\prime}$ depends only on the geometry and weights
of the survey. Its effect is illustrated on the mocks by Figure
\ref{figure::xi_wedge_000_rescale2_compare_mocks_raw_cooked_no_met_with_fits}.
The diagonal elements of the matrix are close
to one, $D_{AA} \approx 0.97$, and the non-diagonal elements are small,
$|D_{AA^\prime}| \lesssim 0.01$. Since the continuum fitting only mixes
pixels from the same forest, all matrix elements $D_{A A^\prime}$ with
$|\rperp^{A}-\rperp^{A^\prime}|>20~\hMpc$ are negligible.

\begin{figure*}
        % Codes/codes.py plot_compare_correlation_new_version()
        \centering
        \includegraphics[width=\columnwidth]{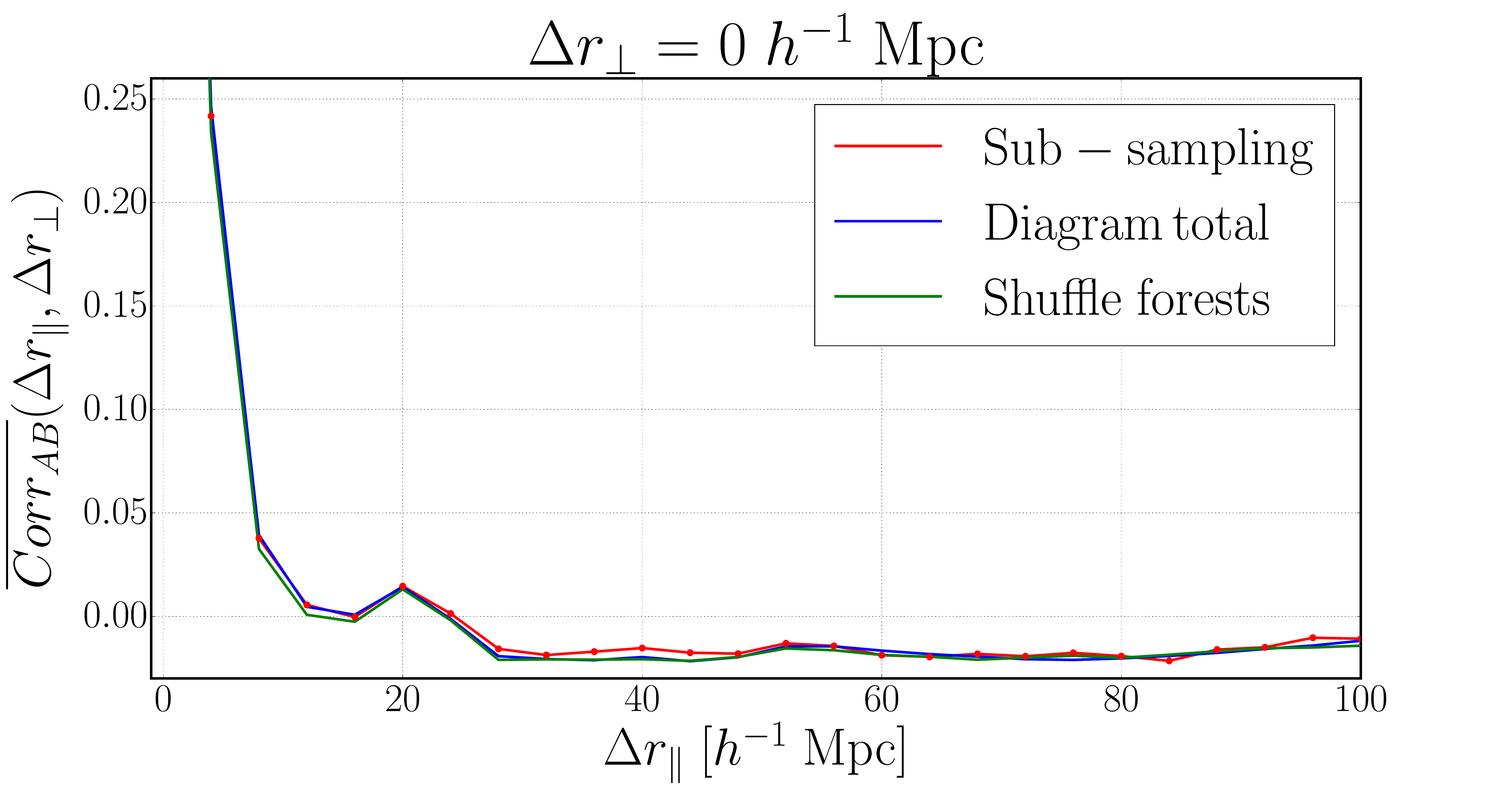}
        \includegraphics[width=\columnwidth]{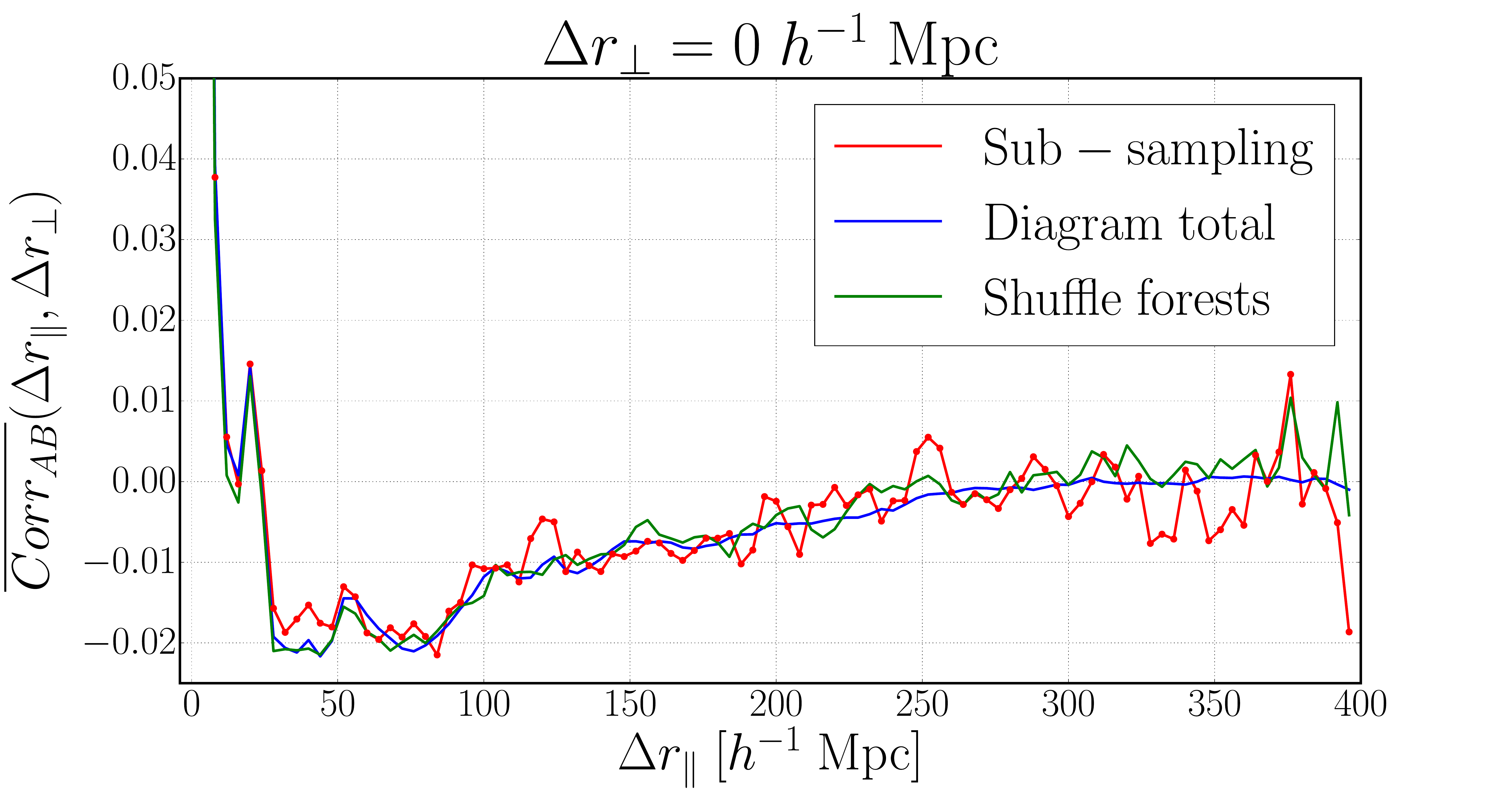}
        \includegraphics[width=\columnwidth]{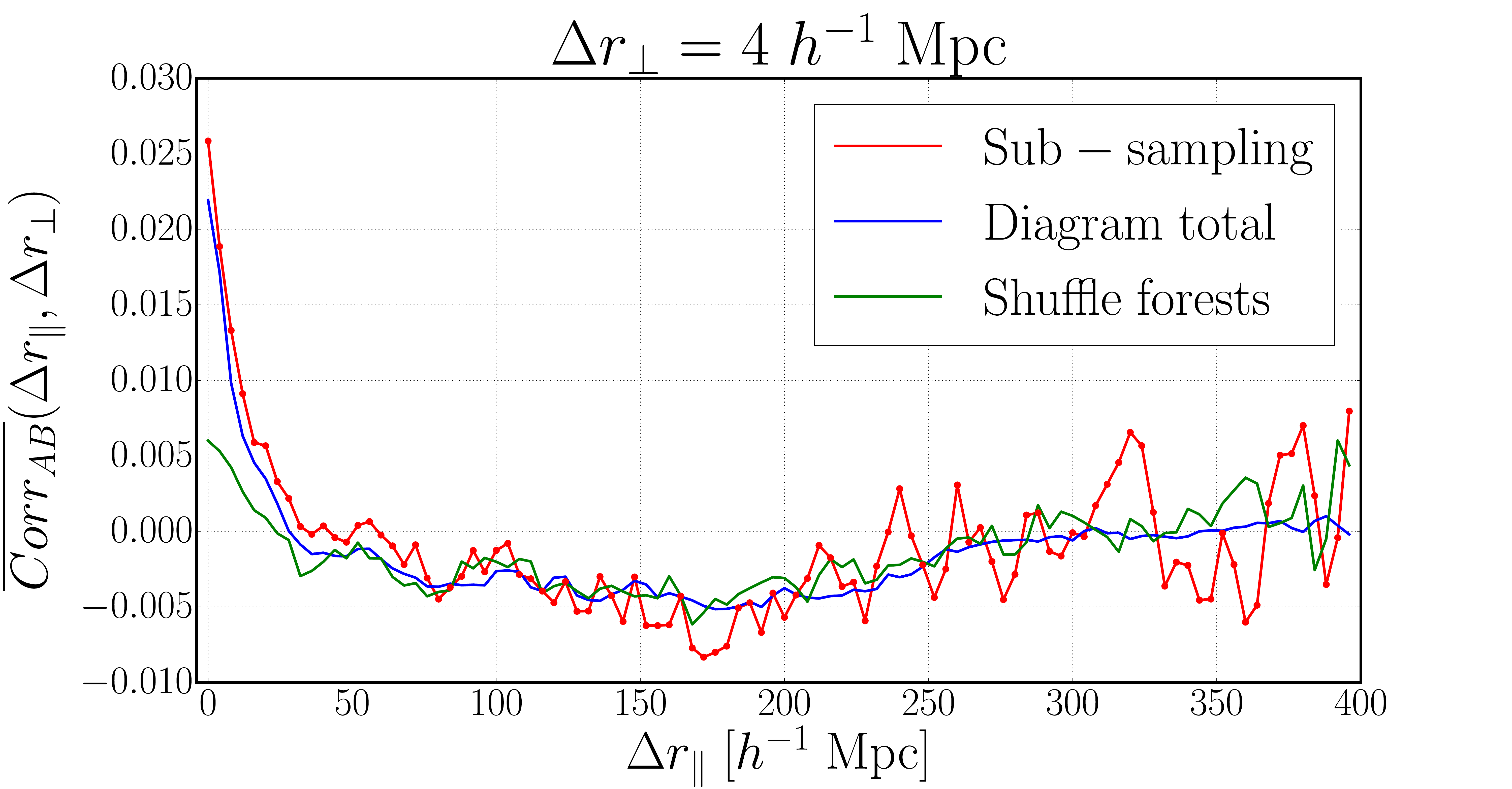}
        \includegraphics[width=\columnwidth]{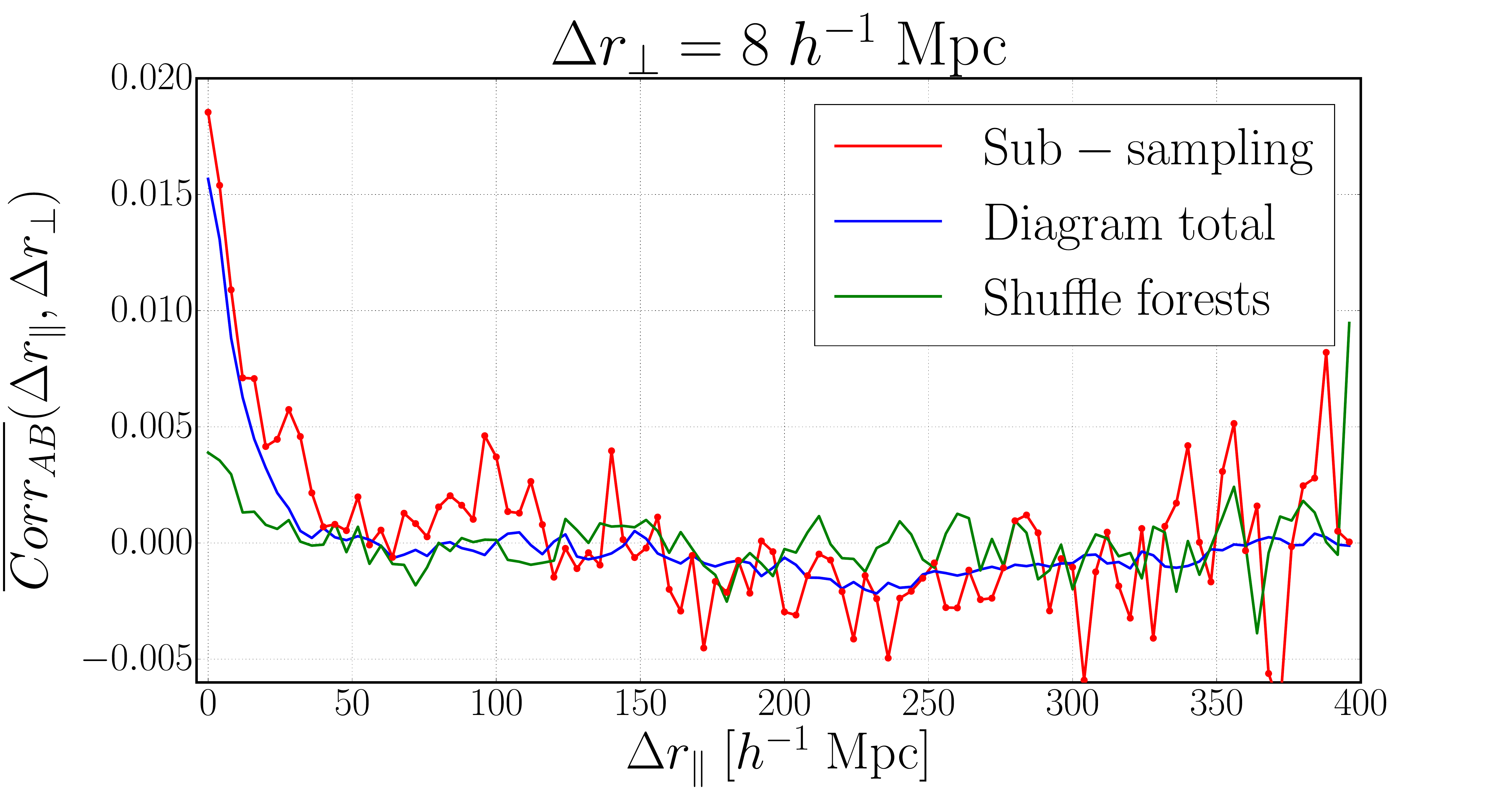}
        \caption{
          Mean normalized covariance matrix,
          $Corr_{AB}\equiv C_{AB}/\sqrt{C_{AA}C_{BB}}$, 
          as a function of
          $\Delta \rpar = |\rpar^{A}-\rpar^{B}|$
          for the three lowest values of  
          $\Delta \rperp = |\rperp^{A}-\rperp^{B}|$.
          The top figures are for $\Delta \rperp=0$, with the
          righthand panel showing only points with
                $Corr_{AB}<0.1$.
                The bottom two figures are for $\Delta \rperp=4~\hMpc$ (left)
                and $\Delta \rperp=8~\hMpc$ (right).
                Shown are the correlations given by the sub-sampling,
                by the sum of all the diagram expansion, and by the
                shuffle of forests.
                The shuffle technique fails for
                $(\Delta\rperp>0, \Delta\rpar<30~\hMpc)$ where inter-forest
                correlations dominate.
        }
        \label{figure::correlation_matrix}
\end{figure*}

\subsection{The covariance matrix}
\label{subsection::The_Lya_forest_quasar_cross_correlation::The_covariance_matrix}

The covariance associated with the measured
correlation function $\xihat$ for two bins
$A$ and $B$ is given by:
\begin{equation}
        C_{A B}
        =
        \langle \xihat_{A} \xihat_{B} \rangle
        -
        \langle \xihat_{A} \rangle
        \langle \xihat_{B} \rangle \; .
        \label{equation::covar_xi_estimator}
\end{equation}
We estimated the  covariance matrix of the data using  two independent
approaches.

The first technique involves writing the covariance matrix as
a function of the known flux auto-correlation function.
Combining Equations \ref{equation::xi_cross_estimator} and 
\ref{equation::covar_xi_estimator},
we have:
\begin{equation}
        C_{A B}
        =
        \frac{
        \sum\limits_{(i,k) \in A}
        \sum\limits_{(j,l) \in B}
        w_{i} w_{j}
        \langle\hat\delta_{i} \hat\delta_{j}\rangle
        }
        {W_{A}W_{B}} 
        -
        \langle \xihat_{A} \rangle
        \langle \xihat_{B} \rangle
        \label{equation::appendix::xi_cross_covar_estimator_2},
\end{equation}
where $(i,k)$ is a pixel-quasar pair falling in the bin $A$ and $(j,l)$
a pixel-quasar pair falling in the bin $B$.
The sums of weights, $W_{A}$ and $W_{B}$, are for the bins
$A$ and $B$ respectively.
This expression of the covariance matrix depends on the correlation between
two pixels, $\langle\hat\delta_{i} \hat\delta_{j}\rangle$.
Intra-forest correlations, $\xiOneD$, are generally larger than inter-forest
correlations so
the largest off-diagonal elements of the covariance matrix are due to
the terms where  $k$ and $l$
are the same quasar and $i$ and $j$ are in the same forest.
This behavior implies that the largest elements  have
$r_{\perp A}=r_{\perp B}$.
The other terms involving inter-forest correlations
can be described by ``diagrams'' of increasing
complexity, as discussed in  Appendix \ref{appendix::covariance_matrix}.

The second technique uses sub-samples of the data.
We divide the BOSS footprint of Fig.
\ref{figure::footprint}
into sub-samples and measure
$\xihat^{s}_{A}$ and $\xihat^{s}_{B}$ in each sub-sample $s$.
Neglecting the small correlations between sub-samples,
the covariance (\ref{equation::covar_xi_estimator})
is given by:
\begin{equation}
        C_{AB} = \frac{1}{W_{A} W_{B}} \sum\limits_{s} W_{A}^{s} W_{B}^{s} 
        \left[ \xihat^{s}_{A} \xihat^{s}_{B}
        - \xihat_{A} \xihat_{B} \right] \,,
        \label{equation::covar_xi_estimator_subsampling}
\end{equation}
where $W_{A}^{s}$ is the sum of weights in the sub-sample $s$ for the bin $A$.
We used 80 sub-samples of similar statistical sizes and shapes.
We tested with 1000 sub-samples
and observed no significant changes  of
$\chi^2$ of the fit and
the value and  precision of the BAO-peak parameters.

Although more accurate, the calculation of the diagram
expansion is time consuming, and therefore
not practical for the analysis of the mock data sets.
We thus fit the data and the mocks using the covariance from
the 80 sub-samplings.
To limit the noise of this estimate,
we use the normalized covariance matrix
(hereafter ``correlation matrix''),
\begin{equation}
        Corr_{AB} = \frac{C_{AB}}{ \sqrt{C_{AA}C_{BB}} } \; .
        \label{equation::correlation_matrix}
\end{equation}
To  good approximation, $Corr_{AB}$ is
a function only of $(\Delta\rperp,\Delta\rpar)$
where
$\Delta\rperp=|r_{\perp A} -r_{\perp B}|$ and
$\Delta\rpar =|r_{\parallel A} -r_{\parallel B}|$.
We therefore average the correlation matrix to determine
$Corr_{AB}(\Delta\rperp,\Delta\rpar),$ which is then
used to calculate $C_{AB}(\Delta\rperp,\Delta\rpar)$.
This procedure is validated with a fit of the data and of a subset
of the mocks using the
covariance matrix from the diagram expansion
(Eq. \ref{equation::appendix::xi_cross_covar_estimator_2}).

As a partial check of the first two methods, we used
a third technique based on a shuffle of the positions on the sky 
of the forests.
We keep the values of pixels but change the position of each forest to
the position of another forest of the survey.
We then produce a large number of realizations of shuffles, $r$, and
measure for each
of them the cross-correlation $\xihat^{r}$.
We then measure the
covariance matrix of these nearly independent cross-correlations with Eq.
\ref{equation::covar_xi_estimator_subsampling} (replacing $s$ with $r$).
The shuffling procedure removes inter-forest and quasar-forest correlations
but retains
the intra-forest correlations.
As such, we expect that the shuffle technique will correctly
calculate the important $\Delta\rperp=0$ elements of the
covariance matrix.

A fourth technique can be applied only to mock data sets
where the covariance is given directly by the mock-to-mock
variations of the correlation function.
The results of this technique,
presented in Subsection
\ref{subsection::mockfits}, agree with the other techniques and
confirm their validity.

The $N_{\mathrm{bin}}^{2} = 5000\times5000$ elements of matrix $C_{AB}$
have a relatively simple structure.
By far the most important elements are on the diagonal.
They are, to a good approximation, inversely proportional to
the number of pixel-quasar pairs, $N^{A}_{\mathrm{pair}}$, used in the calculation of the correlation
function in the bin $A$:
\begin{equation}
  C_{AA}  \approx \frac{3.0 \, \langle\delta^2\rangle}{N^{A}_{\mathrm{pair}}}
  \;\sim 1.7\times10^{-7}\frac{100~\hMpc}{\rperp}\; ,
        \label{equation::covariance}
\end{equation}
where $\langle\delta^2\rangle \approx 0.2$ is the variance of BOSS pixels
in the \LyaForest{} and where the second form uses the fact that
$N^{A}_{\mathrm{pair}}$ is approximately proportional to $\rperp$.
The variance, $C_{AA}$, is about three times what one would calculate assuming
all pixels are independent.
This decrease in the effective number of pixels
is due to the correlations between neighboring
pixels in a given forest.

To display the off-diagonal elements it is convenient to
use the correlation matrix (\ref{equation::correlation_matrix}).
The off-diagonal elements of the correlation matrix have
a simple structure.
The correlation is primarily due to pairs of pixel-quasar
pairs
sharing the same quasar and
the same forest  (T2 in Fig.
\ref{figure::appendix::correlation_matrix_in_listDiagrams}).
As a result, the largest elements have $\Delta\rperp=0$.
The elements of the correlation matrix as a
function of $\Delta \rpar = |\rpar^{\,A}-\rpar^{\,B}|$
for the smallest values of $\Delta \rperp = |\rperp^{\,A}-\rperp^{\,B}|$
are presented in Fig. \ref{figure::correlation_matrix}.
Its four panels show the good agreement between the correlation matrix
from the sub-sampling and
the diagram expansion.
As expected, the shuffle technique works well for $\Delta\rperp=0$ but
not for $\Delta\rperp>0$.
The top panels
present $Corr_{AB}$ for $\Delta \rperp=0$.
These two panels
are the reflection of the $\xiOneD$ shown in Fig.
\ref{figure::xi1D_compare_data_mock_with_lines}.
The \Lya\  metal peaks listed in
Table \ref{table::metals_contribution} are visible.
The bottom left and right panels give the correlation matrix for
$\Delta \rperp= 4~\hMpc$ and for $\Delta \rperp= 8~\hMpc$ where the correlation
is very small.

\section{Fits for the peak position}
\label{section::fits}

\begin{figure}
        % Work/Data/Cross_alone/Fits_pyLya/Fit_rmin10_rmax160_metals_laurentzian_HCD_UV_QSORadiation/
        \centering
        \includegraphics[width=0.98\columnwidth]{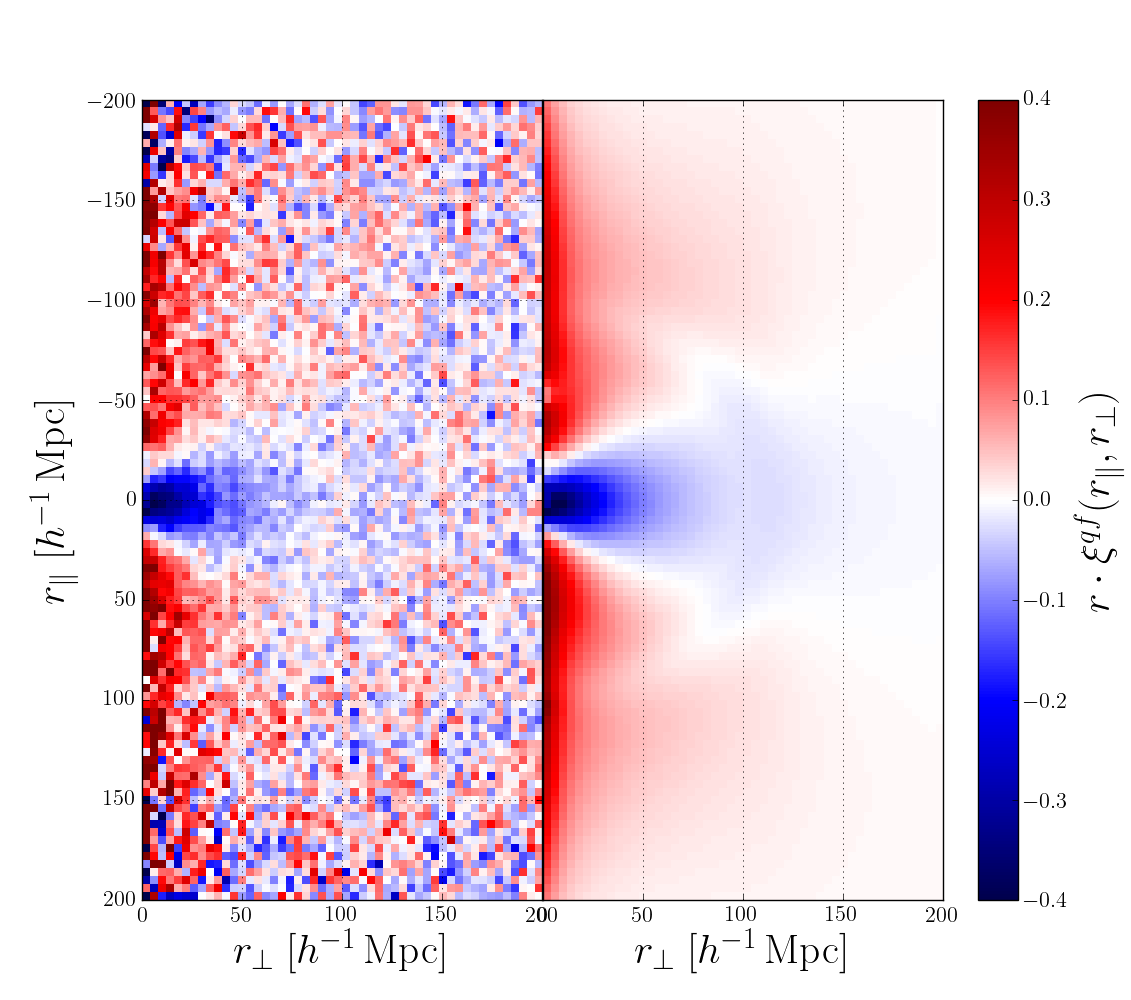}
        \caption{
        Measured (left) and the best fit model (right) of the
        \Lya{}-forest-quasar cross-correlation.
        The distortion matrix
        (\ref{equation::distortion_matrix_measure})        
        has been applied to the model.
        The correlation is multiplied
        by a factor $r$. The BAO scale appears here as a half ring of radius
        $r \approx 100~\hMpc$.
        The color code is saturated for clarity.
        }
        \label{figure::appendix::data_bestFit_data_rescale_0_2D}
\end{figure}

To determine the position of the BAO peak,
we fit the measured forest-quasar cross-correlations, shown in the 
left panel of Fig. \ref{figure::appendix::data_bestFit_data_rescale_0_2D},
to a model that describes the underlying
physical correlations and possible systematics.
We use the model of  \Lya~correlations
introduced by \citet{2017A&A...603A..12B}, and
generalized here to include quasars.
Its parameters are described in Table \ref{table::fit_parameters}
and the best fit model is
shown in the right panel of
Fig. \ref{figure::appendix::data_bestFit_data_rescale_0_2D}.
The best fit parameters are listed in Table
\ref{table::bestfit_best_model_fit_parameters_cross_auto_combined}.
We use the fitting package ``picca''
,\footnote{https://github.com/igmhub/picca.}
which evolved from the baofit package
\citep{2013JCAP...03..024K,2015JCAP...11..034B}.

\subsection{Model of the cross-correlation}
\label{subsection::Fits_for_the_peak_position::Model_of_the_cross_correlation}

\begin{table*}
        \centering
        \caption{List of the parameters of the fits of the \Lya-forest-quasar
        cross-correlation. The first section of the table lists the parameters
        for the standard fitting procedure. The second section\
        gives the parameters
        that are fixed in the standard fits.
        All biases, $b$, refer to the bias at $z=2.4$.
        }
        \label{table::fit_parameters}
        \begin{tabular}{l l}
        Parameters &
        Description     \\
        \noalign{\smallskip}
        \hline \hline
        \noalign{\smallskip}
        \apar{}, \aperp{} &     BAO peak position parameters (Eq. \ref{equation::alpha})\\
        $b_{\rm Ly\alpha}, \beta_{\rm Ly\alpha}$ &      Bias parameters for \Lya~absorption (Eq. \ref{equation::bias_beta_definition}) \\
        $b_m$, ($\beta_m=0.5$) &        Biases parameters (Eq. \ref{equation::bias_beta_definition}) of four metal species (Table \ref{table::metals_contribution}) \\
        $\sigmavqso $ &
        Quasar radial velocity smearing (non-linear velocity and redshift measurement errors
        (Eq. \ref{equation::pk_non_linear_quasar_velocity}) \\
        $(b,\beta,L)_{\rm HCD}$ &
        Parameters of the unidentified high column density systems (HCD) (Eq. \ref{equation::bias_beta_definition_for_HCD})     \\
        $b_{\Gamma},b^\prime,\lambdauv$ &
        Parameters of the UV fluctuations.   $(b^\prime,\lambdauv)$ fixed to $(-2/3,300~\hMpc)$\\
        $\xitp_0,\,\lambdauv$ &
        Transverse proximity effect (Eq. \ref{equation::qso_lya_radiation_contribution}). $\lambdauv$ fixed to $300~\hMpc$  \\
        $\drparqso$ &
        Shift of the cross-correlation due to systematic errors in the quasar redshift measurement
                (Eq. \ref{equation::definition_delta_rParal_QSO})\\
        \noalign{\smallskip}
        \hline
        \noalign{\smallskip}
        $\bqso$ $(=3.87)$ &Quasar bias, fit in combined cross+auto fit.  \\
        $\auv,\tuv^{-1}$ $(=0,0)$ &
               Quasar radiation anisotropy and lifetime (Eq. \ref{equation::qso_lya_radiation_contribution})\\
        $\vec{\Sigma}=(\Sigma_{\parallel},\Sigma_{\perp})$      &
        Non-linear broadening of the BAO peak (Eq. \ref{equation::pk_p_nl}) \\
        $\vec{R} = (R_{\parallel},R_{\perp})$ &
        Smoothing parameter for the binning of the correlation (Eq. \ref{equation::pk_binning_term}) \\
        $\Apeak$ $(=1)$&
        BAO peak amplitude (Eq. \ref{equation::pk_p_nl}) \\
        $a_{i,j}$ $(=0)$ &
        Power-law broadband parameters (Eqs. \ref{equation::broadband_1} and \ref{equation::broadband_2}) \\
        $\alpha_{\LyaMath}$ $(=2.9$) &
        Redshift evolution parameter for $b_{a}$, the bias of absorbers (Eq. \ref{equation::bias_absorber_evolution_definition}) \\
        \end{tabular}
\end{table*}

\begin{table}
        \centering
        \caption{Major metal transitions seen
          in the intergalactic medium (IGM)
          and present in the forest-quasar cross-correlation
        for $\rpar \in [-200,200]~\hMpc$.
        The second column lists the rest-frame wavelength of the transition.
        The third column is the ratio between the metal transition
        and the \Lya{} transition, where $\lambda_{1}$ is the greater of
        the two wavelengths.
        The last column gives the apparent
        radial distance difference between the 
        \Lya{} and metal absorption corresponding to vanishing
        physical separation, at $\lambda_{\mathrm{Obs.}} = 410~\mathrm{nm}$.
        }
        \label{table::metals_contribution}
        \begin{tabular}{l l l c }
        Transition      &
        $\lambda_{m}$   &
        $\lambda_{1}/\lambda_{2}$       &
        $\rpar^{\LyaMath , m} $ \\
        &       $[\mathrm{nm}]$ &       &       $[\hMpc]$ \\
        
        \noalign{\smallskip}
        \hline \hline
        \noalign{\smallskip}

        SiII(126.0)             &       126.04  &       1.037   &       +103    \\
        SiIII(120.7)    &       120.65  &       1.008   &       -21             \\
        SiII(119.3)             &       119.33  &       1.019   &       -52             \\
        SiII(119.0)             &       119.04  &       1.021   &       -59             \\

        \end{tabular}
\end{table}

The expected value
of the measured cross-correlation, $\xihat_A$,
in the $(\rperp,\rpar)$ bin $A$
is related to the theoretical
cross-correlation, $\xiqfth$, by:
\begin{equation}
        \xihat_A = \sum\limits_{A^\prime} D_{AA^\prime}
        \left[ \xiqfth_{A^\prime} + \xibb_{A^\prime} \right],
        \label{equation::xi_complete}
\end{equation}
where $D_{AA^\prime}$ is the distortion matrix (Sect.
\ref{subsection::The_Lya_forest_quasar_cross_correlation::The_distortion_matrix}).
The broadband
term, $\xibb_{A}$, is an optional function used to test
for imperfections in the model and for systematic errors.

The cosmological cross-correlation 
is the sum of several contributions:
\begin{equation}
  \xiqfth =
  \xi^{\,q \LyaMath{}} +
  \sum\limits_{a} \xiqath
  + \xitp\,.
        \label{equation::theoretical_cosmo}
\end{equation}
The first term represents the correlation between quasars, $q$, and
\Lya{} absorption in the IGM.
The  second term is the sum over all other absorbers:
the metals of Table \ref{table::metals_contribution} and unidentifed
high column density systems (HCDs).
All absorbers trace the underlying matter fluctuations, but we
separate out the \Lya{} absorbers because \Lya~absorption is
assumed in the calculation of
the quasar-pixel separation,
$(\rperp,\rpar)$, therefore requiring
a special treatment  for metals.
The third term, $\xitp$,  is
the correlation between a quasar and a neighboring forest
due to the effect of the quasar's radiation
on the ionized fraction of the IGM.
This effect of a quasar on its own forest is
generally referred to as the ``proximity effect''
\citep{1986ApJ...309...19M,1988ApJ...327..570B}.
In the general case studied here, we use
the term ``transverse proximity effect'', $\xitp$.

The physical component of the model is dominated by the cross-correlation
due to \Lya{} absorption in the IGM.  It is assumed to be a biased version of
the  total matter auto-correlation of the appropriate flat-\LCDM{} model
modified to free the position of the BAO
peak:
\begin{equation}
        \xi^{\,q \LyaMath{}}(\rperp,\rpar,\aperpMath,\aparMath)=
        \xismooth(\rperp,\rpar) + \xipeak(\aperpMath\rperp,\aparMath\rpar) \;.
        \label{xismoothpeak}
\end{equation}
The BAO peak position parameters to be fit are
\begin{equation}
\aparMath = \frac { \left[\DHh(\zeff)/r_d\right] }
{\left[\DHh(\zeff )/r_d\right]_{\rm fid}}
\hspace*{3mm}{\rm and}\hspace*{5mm}
\aperpMath = \frac { \left[\DMm(\zeff)/r_d\right] }
{\left[\DMm(\zeff)/r_d\right]_{\rm fid}} ~,
\label{equation::alpha}
\end{equation}
where the subscript ``fid'' refers to the fiducial cosmological
model from Table \ref{table::cosmology_parameters} used to transform angle differences
and redshift differences to $(\rperp,\rpar)$.

The  nominal correlation function,
$\xi^{\,q \LyaMath{}}(\rperp,\rpar,\aperpMath=\aparMath=1)$, is derived from
its Fourier transform:
\begin{multline}
                P^{\,q \LyaMath{}}(\vec{k},z) = \PQL(\vec{k},z)
                d_{q}(\mu_{k},z)
                d_{\rm Ly\alpha}(\mu_{k},z)     \\
                \times
                \sqrt{\VNL(k_{\parallel})}
                \sqrt{\FNL(\vec{k})}
                G(\vec{k})
                \label{equation::pk_1}\;,
\end{multline}
where $\vec{k} = (k_{\parallel},k_{\perp})$ is the wavenumber of modulus $k$
and $\mu_{k} = k_{\parallel}/k$,
with $k_{\parallel}$ being the component along the line of sight
and $k_{\perp}$ across.
As described in more detail below, $\PQL$ is the (quasi) linear matter
spectrum, $d_q$ and $d_{\rm Ly\alpha}$ are 
the standard factors \citep{1987MNRAS.227....1K}
describing redshift-space distortion,
$\VNL$ and $\FNL$ describe non-linear corrections, and
$G(\vec{k})$ gives the effects of $(\rperp,\rpar)$ binning on the measurement.
  Calculation of $\xi^{\,q \LyaMath{}}$
  for a given $(\rperp,\rpar)$ bin uses the weighted mean
  $(\rperp,\rpar)$ of pixel pairs in the bin and, 
  for  $P^{\,q \LyaMath{}}(\vec{k},z)$, the weighted mean redshift of the bin.
  From bin to bin, this redshift varies in the range 2.38 to 2.43 about
  the mean redshift of the survey, $\zeff=\MEANZ$.

The first term in (\ref{equation::pk_1}) provides for the aforementioned
separation of the peak and smooth contributions
to the correlation function (Eq. \ref{xismoothpeak}):
\begin{equation}
        \PQL(\vec{k},z) =
        \Psmooth(k,z)
        + \Apeak
        e^{ -\left[ ( k_{\parallel}\Sigma_{\parallel} )^{2} + ( k_{\perp}\Sigma_{\perp} )^{2} \right]/2 }
        \Ppeak(k,z)\;,
        \label{equation::pk_p_nl}
\end{equation}
where  the smooth component, $\Psmooth$,
is derived from the linear power spectrum, $\PL(k,z)$,
via the side-band technique \citep{2013JCAP...03..024K} 
and $\Ppeak=\PL-\Psmooth$.
The redshift-dependent linear power spectrum is obtained from
CAMB \citep{2000ApJ...538..473L}
with the appropriate cosmology for data or mocks
(Table \ref{table::cosmology_parameters}).
The peak amplitude parameter, $\Apeak$, is normally set to unity
but can be fit in non-standard analyses.
The correction for non-linear broadening of the BAO peak is parameterized by
$\vec{\Sigma}=(\Sigma_{\parallel},\Sigma_{\perp})$,
set equal to 
$(6.41,3.26)~\hMpc$ in the standard fit \citep{2007ApJ...664..660E}.

The second and third terms in (\ref{equation::pk_1}) are the quasar
and \Lya~Kaiser factors describing redshift-space distortions:
\begin{equation}
        d_{t}(\mu_{k},z) = b_{t}(z) \left( 1 + \beta_{t} \mu_{k}^{2} \right),
        \label{equation::bias_beta_definition}
\end{equation}
where $b_{t}(z)$ is the bias and $\beta_{t}$ is the redshift space
distortion (RSD) parameter for the tracer $t$ $({\rm =Ly\alpha,quasar})$.
The fit of the cross-correlation is clearly only sensitive to the
product of the quasar and \Lya~biases, so by convention we set
$\bqso=3.87$ as measured by \citet{2016JCAP...11..060L} and
assume a redshift dependence
given by
Equation 15 of \citet{2005MNRAS.356..415C}:
\begin{equation}
        \bqso(z) = 0.53 + 0.289(1+z)^{2}\;.
        \label{equation::bias_quasar_evolution_definition}
\end{equation}
For \Lya{} absorption we assume
\begin{equation}
        \blya(z) = \blya(2.4)[(1+z)/(1+2.4)]^{\alpha_{\LyaMath}},
        \label{equation::bias_absorber_evolution_definition}
\end{equation}
where $\alpha_{\LyaMath}=2.9$
as observed in measurements of
the flux-correlation, $\xiOneD$,
within individual forests \citep{2006ApJS..163...80M}.

Fluctuations of ionizing UV radiation 
\citep{2014PhRvD..89h3010P,2014MNRAS.442..187G}
lead to
a scale-dependence of $\blya$
given by Eq. 12 of \citet{2014MNRAS.442..187G}.
The effect of the fluctuations is to increase $\blya$ from
its nominal value at small scale to a different value at
large scale.  The transition scale is determined by the
UV photon mean free path, which we set to a comoving value of $\lambdauv=300~\hMpc$
\citep{2013ApJ...769..146R}.
We then fit for one parameter, $b_\Gamma$ corresponding
to the $b_\Gamma(b_s-b_a)$ of \citet{2014MNRAS.442..187G}; it
determines the change
in $\blya$ between large and small scales.
A second bias, $b_a^\prime$, that determines the precise dependence
of the bias on scale, is set to the nominal value of $-2/3$
used by \citet{2014MNRAS.442..187G}.

The \Lya{} RSD parameter, $\betalya$,
is expected to have a redshift dependence that
is somewhat weaker than that for the bias $\blya$,
varying between $z=2.25$ and $z=3.0$ by a factor $\sim1.2$
in the simulations of  \citet{2015JCAP...12..017A}
compared to a factor $\sim1.8$ for $\blya$ 
(Eq. \ref{equation::bias_absorber_evolution_definition}).
Because of the narrow range of mean redshifts of $(\rperp,\rpar)$ bins, we
neglect the  variation of $\betalya$, fitting only its value at $\zeff$.
For quasars, the RSD parameter, $\beta_{q}$, is directly linked
to the bias $\bqso$ and to $f$,
the linear growth rate of structure:
\begin{equation}
        \bqso \betaqso = f \;\sim\Omega_{m}(z)^{0.55 } \; ,
        \label{equation::linear_growth_rate}
\end{equation} 
where $f = f(\zeff)$ is given in Table \ref{table::cosmology_parameters}.

The term $\VNL(k_{\parallel})$ is the effect on the power spectrum of
non-linear quasar velocities and the precision of
quasar redshift measurements.
Following Eq. 18 of \citet{2009MNRAS.393..297P}, we use
a Lorentz-damping form:
\begin{equation}
        \VNL(k_{\parallel}) = \frac{1}{1+ \left( k_{\parallel} \sigmavqso \right)^{2} },
        \label{equation::pk_non_linear_quasar_velocity}
\end{equation}
where $\sigmavqso~[\hMpc]$ is a free parameter.
Alternative fits use a Gaussian form.

The term $F_{\rm NL}(\vec{k})$ is a correction for non-linear effects
in \Lya{} absorption at large $k$
due to the isotropic enhancement of power due to non-linear growth,
the isotropic suppression of power due to gas pressure, and
the suppression of power due to line-of-sight non-linear
peculiar velocity and thermal broadening.
It can be chosen to be of one of the two forms
given by Equation 21 of \citep{2003ApJ...585...34M} or as presented in
\citet{2015JCAP...12..017A}.
Our standard fit uses the former.

The last term in (\ref{equation::pk_1}),
$G(\vec{k})$, accounts for smoothing due to the
binning of the measurement of $\xiqf$.
Following \citet{2017A&A...603A..12B},
we use
\begin{equation}
        G(\vec{k}) =
        \left[
        \sinc{\left( \frac{R_{\parallel}k_{\parallel}}{2}\right)}
        \sinc{\left( \frac{R_{\perp}k_{\perp}}{2}\right)}
        \right]^{2},
        \label{equation::pk_binning_term}
\end{equation}
where $R_{\parallel}$ and $R_{\perp}$ are the 
scales of the smoothing.
For standard fits, we  fix both to the bin width,
$R_{\parallel}=R_{\perp}=4~\hMpc$.

The second contribution to $\xiqfth$
in Eq. (\ref{equation::theoretical_cosmo})
is the sum over non-\Lya{} absorbers.  
Because there is little
absorption by metals,
the treatment of metal components is simplified without
the separation into peak and smooth components.
The fiducial correlation function is directly used
to calculate the metal-quasar 
correlation, although with individual $(b,\beta)$ for each
species.

Absorption by metals is complicated by the fact that
the $(\rperp,\rpar)$ bins $A$ corresponding to
an observed $(\Delta\theta,\Delta\lambda)$ are
calculated
assuming absorption due to the \Lya{} transition (Eq. \ref{coordinatedef}).
This $(\rperp,\rpar)$ does not correspond to the physical quasar-absorber
separation
if the absorption is not due to \Lya.
The model correlation function $\xi^{qm}_A$ must be evaluated
at a different $(\rperp,\rpar)$ found by replacing $D_{\rm Ly\alpha}$
in (\ref{coordinatedef}) with the distance calculated
from the redshift $z_m=\lambda/\lambda_m -1$
and taking the weighted average for pixel-quasar pairs in the bin $A$.

The contribution of each absorber to the cross-correlation is maximized
in the $(\rperp,\rpar)$ bin that corresponds to vanishing physical
separation. For the \Lya{} contribution, this bin corresponds
to $(\rperp,\rpar)=(0,0)$.
For the other species, it corresponds to $\rperp=0$ and to
$\rpar\sim(1+z)\DHh(z)(\lambda_m-\lamlya)/\lamlya$ as
given in Table \ref{table::metals_contribution}. 
Because amplitudes for SiII and SiIII are mostly determined
by the excess correlation at $(\rperp\sim0, \rpar\neq0)$,
the $\beta$ for each metal is poorly determined.
We therefore fix their value to $\beta=0.5$
corresponding to host halos with bias of
two, the value found for DLAs \citep{2012JCAP...07..028F},
which is also typical
of star-forming galaxies.
The redshift dependence of the biases is assumed to be the same
as that for $\blya$ as given by Eq. \ref{equation::bias_absorber_evolution_definition}.
Because all $(\rperp,\rpar)$ bins have nearly the same mean redshift,
this assumption has very little impact on the fits.

The standard fit also takes into account the correlation between the
quasar distribution and absorption
by unidentified HCD systems.
This new absorber is modeled with a modified Kaiser factor
\citep{2017A&A...603A..12B}
defined as:
\begin{equation}
        d_{\rm HCD}(\mu_{k}) =
        \bhcd
        \left( 1 + \betahcd \mu_{k}^{2} \right)
        \sinc{\left( \Lhcd k_{\parallel}\right)},
        \label{equation::bias_beta_definition_for_HCD}
\end{equation}
where $\bhcd,\,\betahcd$ are the traditional bias and beta parameters
of the absorption, and $\Lhcd$ is the associated smoothing scale.
Because of degeneracies, we add a Gaussian prior for $\betahcd$,
of mean $0.5$ and standard deviation $0.2$.

The final term in   (\ref{equation::theoretical_cosmo}),
$\xitp$, represents the contribution
to the \Lya-quasar cross-correlation from radiation effects.
In the vicinity of a quasar, the radiation emitted from
the quasar dominates over the UV background, 
increasing the ionization fraction of the surrounding gas. This increase
makes it more transparent to the quasar
\Lya{} photons. Therefore, this effect 
introduces an extra term in the correlation 
between the quasars and the \Lya{} forest
\citep{2013JCAP...05..018F}.
We use the form
\begin{equation}
  \xitp=
  \frac{\xitp_0}{r^2}
  \exp\left[\frac{-r}{\lambdauv}\right]
[1-\auv(1-\mu^2)]
  \exp\left[\frac{-r(1+\mu)}{\tuv}\right]
  ,
    \label{equation::qso_lya_radiation_contribution}
\end{equation}
where $r$ is the comoving separation in units of $\hMpc$,
$\lambdauv=300~\hMpc$
\citep{2013ApJ...769..146R},
and $\xitp_0$ is an amplitude to be fitted.
The parameters $(\auv,\tuv^{-1})$ describe anisotropic
and time-dependent emission.  They are set to zero in the
standard fit. Leaving them free in the fit gives a slight
preference for anisotropy: $\auv=1.27\pm 0.56$ 
(Table \ref{table::fit_data_model}).

\begin{figure*}[tb]
        % Work/Data/Cross_alone/Fits_pyLya/Fit_rmin10_rmax160_metals_laurentzian_HCD_UV_QSORadiation/
        \centering
        \includegraphics[width=\columnwidth]{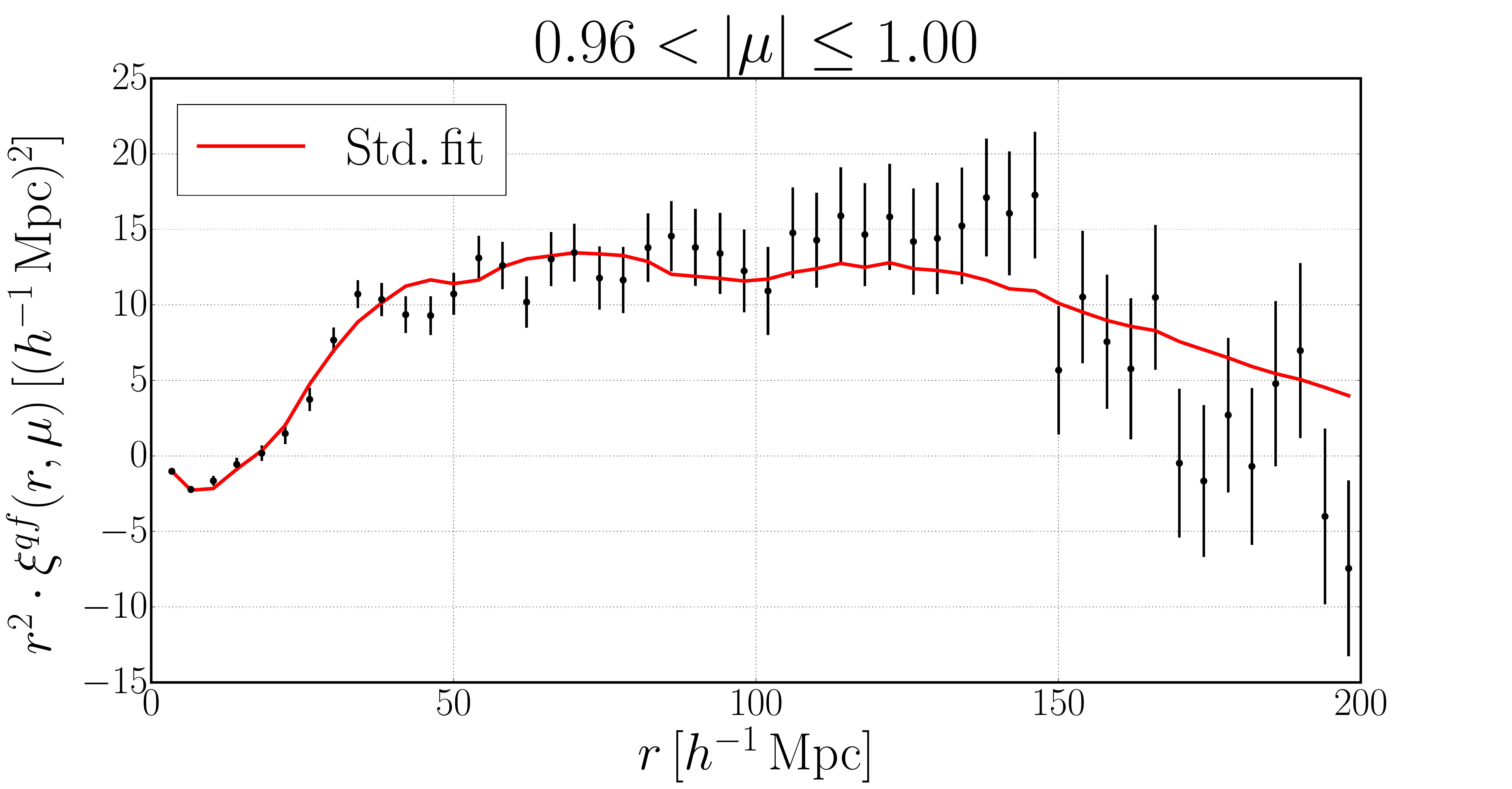}
        \includegraphics[width=\columnwidth]{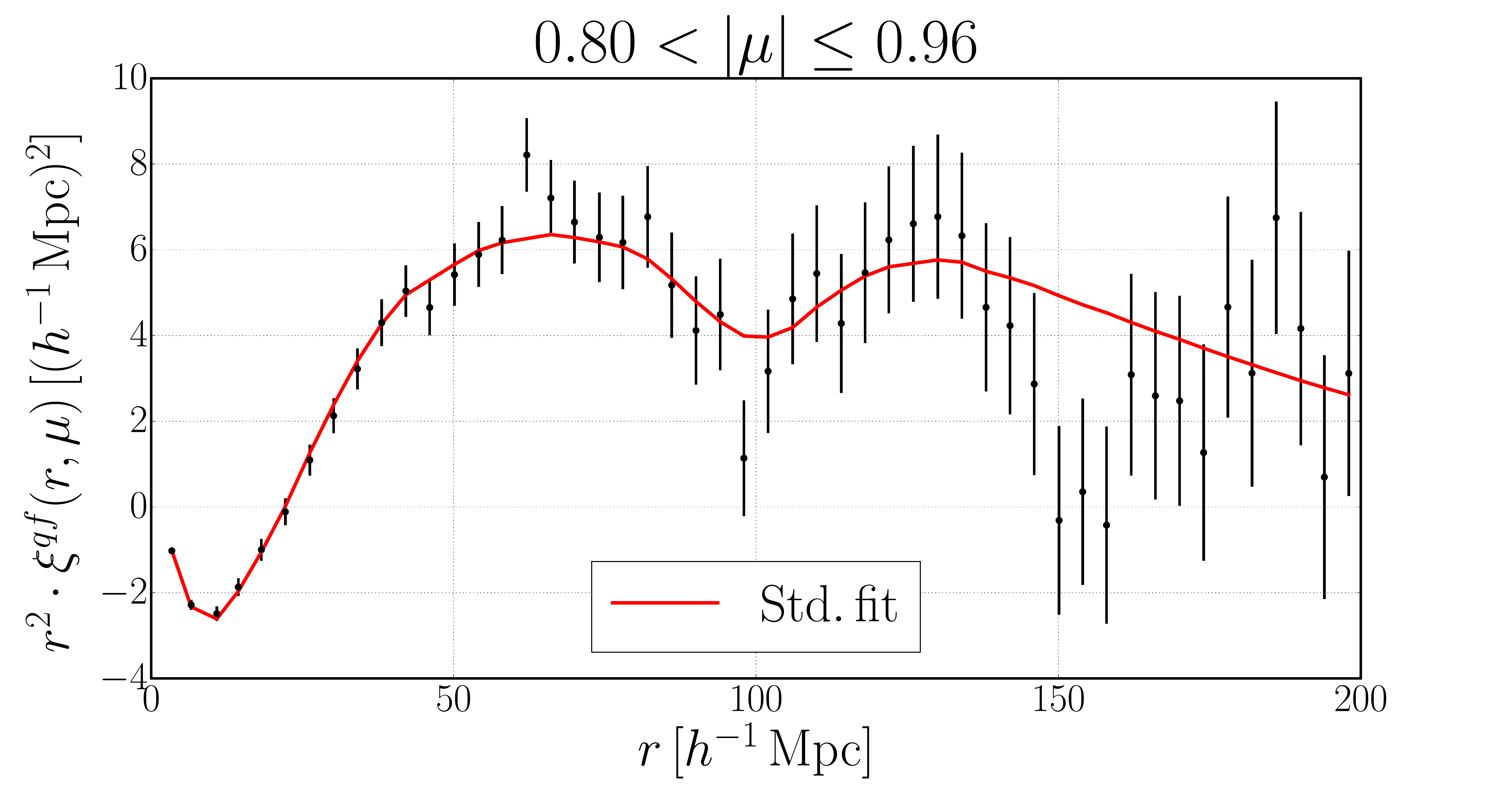}
        \includegraphics[width=\columnwidth]{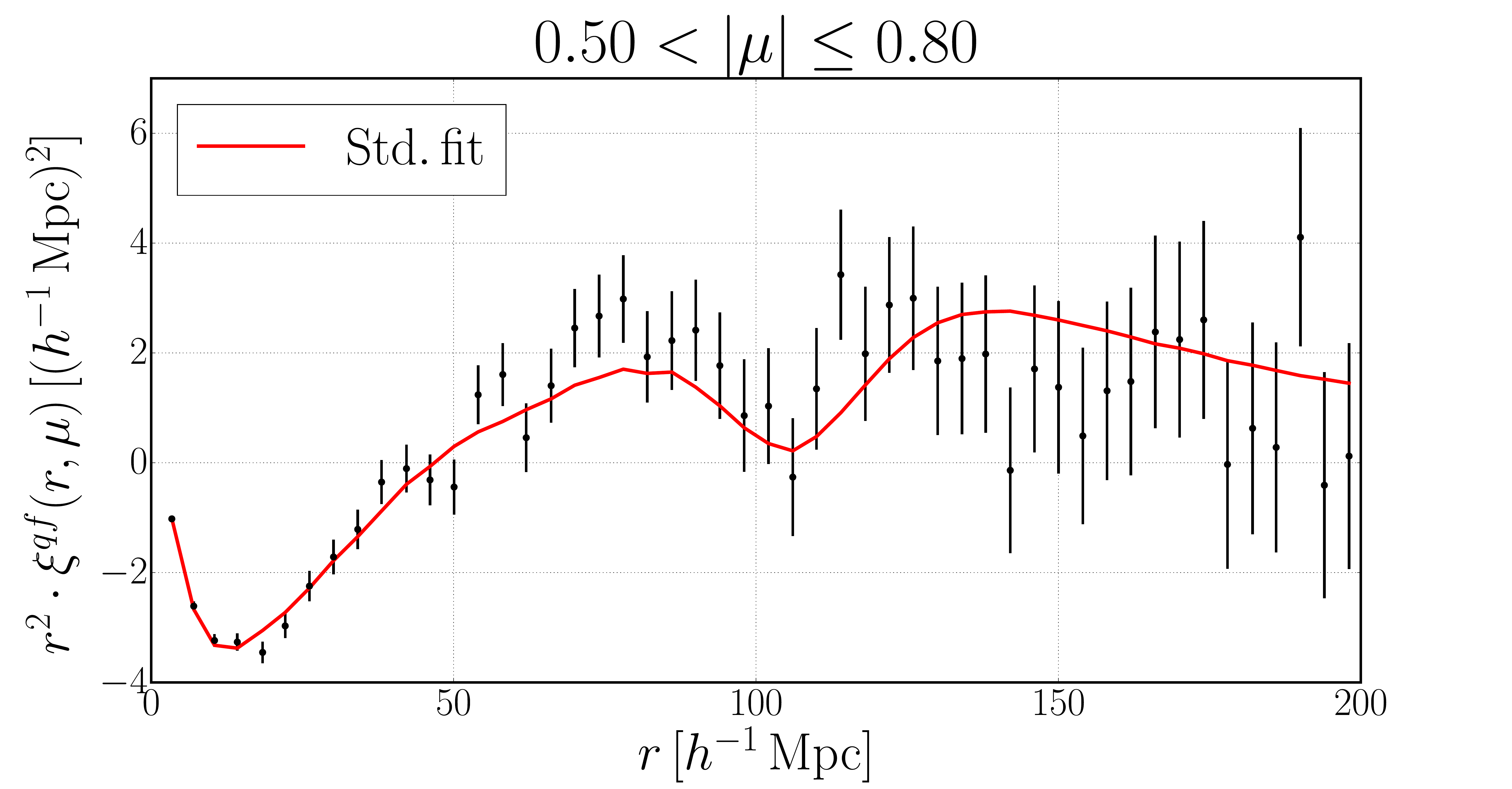}
        \includegraphics[width=\columnwidth]{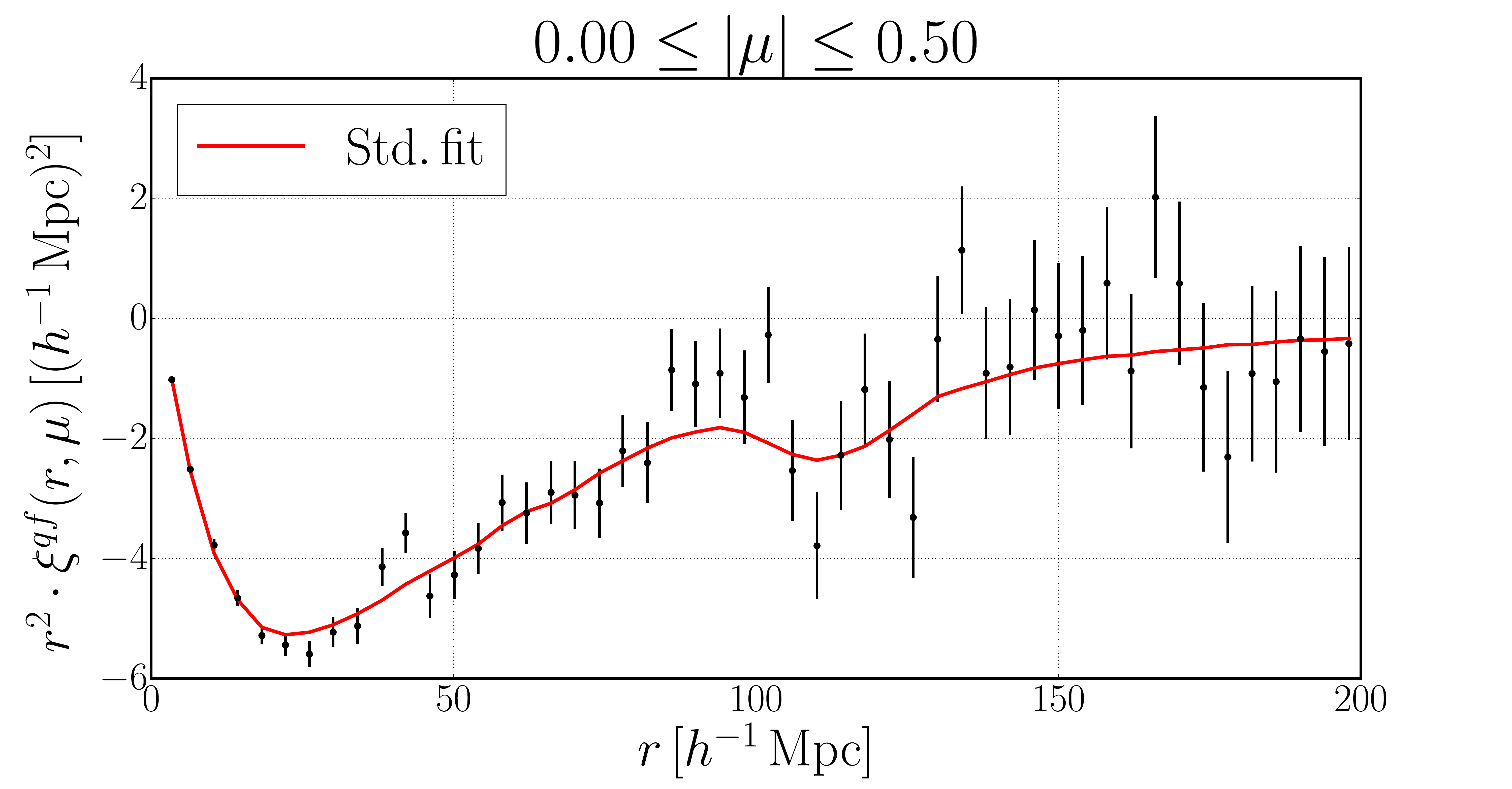}
        \caption{
          Cross-correlation function in four ranges
          of $\mu=r_{\parallel}/r$.  The data are the  black points and
                the red curves give the standard fit
                ($10<r<160~\hMpc$)
                used to measure the BAO parameters.
                The cross-correlation is multiplied by
                $r^{2}$ to show the BAO scale.
        }
        \label{figure::xi_data_best_fit_4_weges_mu}
\end{figure*}
\begin{figure*}
        % Work/Data/Cross_alone/Fits_pyLya/Fit_rmin10_rmax160_metals_laurentzian_HCD_UV_QSORadiation/
        \centering
        \includegraphics[width=\columnwidth]{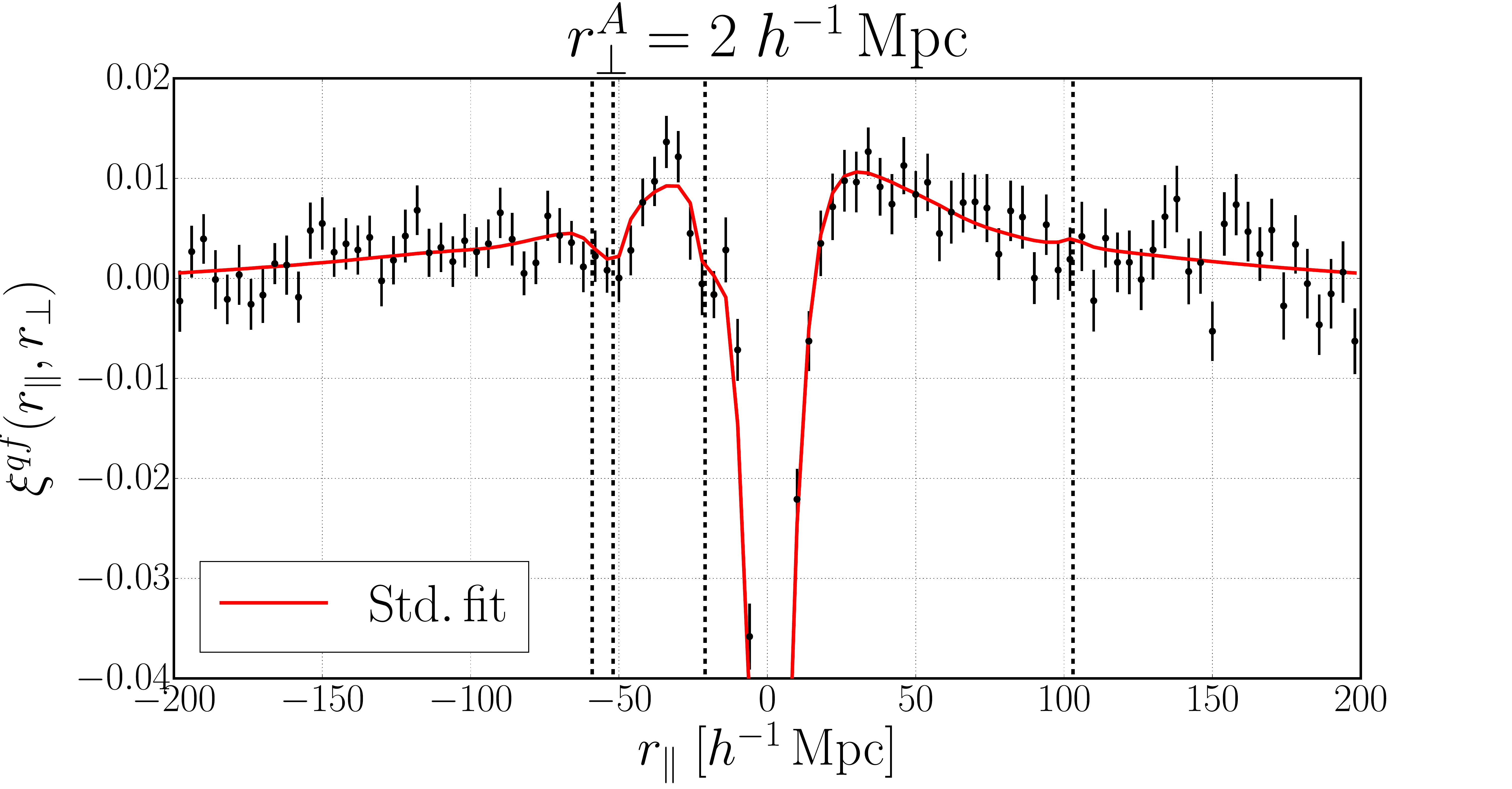}
        \includegraphics[width=\columnwidth]{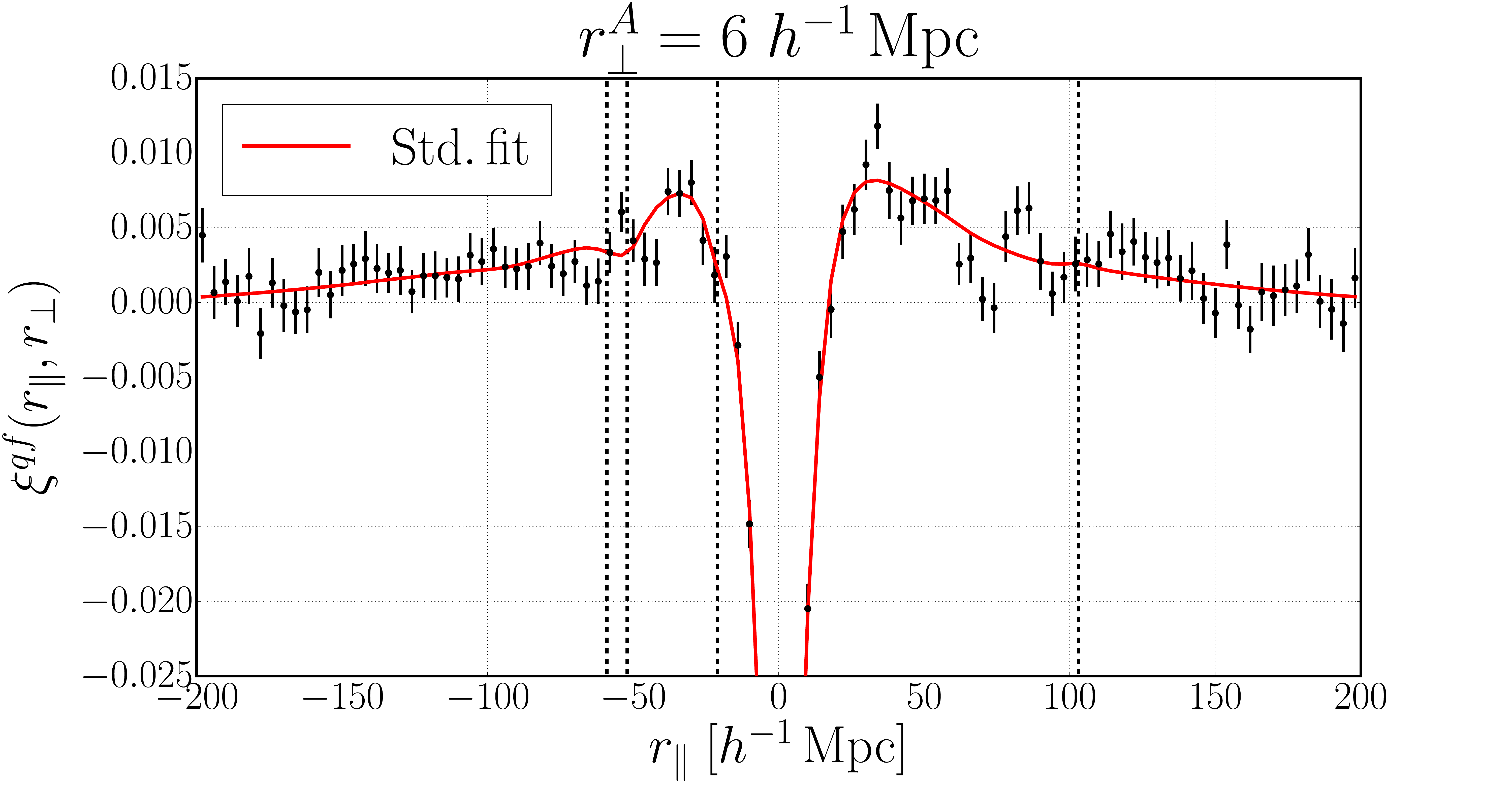}
        \caption{
          Correlation function for two ranges of $\rperp^{A}$.
          The data are the black points and
                the red curves give the standard fit
                (over the range $10<r<160$~$\hMpc$)
                used to measure the BAO parameters.
                These slices of constant $\rperp^{A}$ demonstrate
                the impact of metal
                transitions on the data.
                The four vertical dashed lines give the
                position of the four peaks of
                the metal-quasar correlations of
                Table \ref{table::metals_contribution}:
                $\rpar \approx -60~\hMpc$ 
                (SiII(119.3) and SiII(119.0));  
                $\rpar = -21~\hMpc$ 
                (SiIII(120.7)); and $\rpar\approx +103~\hMpc$
                (SiII(126.0)).
        }
        \label{figure::xi_data_best_fit_2_slice_rperp}
\end{figure*}

The optional $\xibb_{A}$ term of (\ref{equation::xi_complete})
is a  ``broadband function'' that is
a slowly varying function of $\RparalRperp$:
\begin{equation}
%\begin{multline}
        \xibb_1(r,\mu) =
        \sum^{\imax}\limits_{i=\imin}
        \sum^{\jmax}\limits_{j=\jmin}
        a_{ij} \frac{L_{j}(\mu)}{r^{i}}\, ,
        \label{equation::broadband_1}
%\end{multline}
\end{equation}
where $L_{j}$  is the Legendre polynomial of degree $j$.
The standard form, $(\imin,\imax,\jmin,\jmax)=(0,2,0,6)$, corresponds
  to parabolas in $r^2\xi(r,\mu)$ for seven independent $\mu$ ranges.
  We have also used functions of the form
  (see Appendix \ref{append::nonstandardfits}):
\begin{equation}
        \xibb_{2}\RparalRperp =
        \sum^{i_{\max}}\limits_{i=i_{\min}}
        \sum^{j_{\max}}\limits_{j=j_{\min}}
        a_{ij} r^{i}_{\parallel} r^{j}_{\perp} \, .
        \label{equation::broadband_2}
\end{equation}
In previous studies,
broadband functions were of central importance since we did not attempt to model
the distortion from the continuum fitting.
In this study this effect is modeled
with the distortion matrix $D_{AA^\prime}$. The purpose of the
broadband functions is now only
to search for systematic errors due to hypothetical
correlations
between the peak position and the sidebands.
Its function is also to account
for unknown physical, instrumental, or analytical effects missing in the model.
The standard fit, used to measure the BAO parameters,
has no broadband functions: $a_{ij} = 0$.

When estimating the model on the grid of separation coordinate, we allow
for a mean shift of the absorber-quasar separation along
the $\rpar$ direction:
\begin{equation}\drparqso=
          r_{\parallel,{\rm true}}-r_{\parallel,{\rm measured}} \,.
        \label{equation::definition_delta_rParal_QSO}
\end{equation}
The shift, described by the fit parameter $\drparqso$,
is mostly due to systematic errors in the measurement of the redshift of the
quasar. Indeed, the different emission lines of the quasars have different
relative velocities \citep{1982ApJ...263...79G,2016ApJ...831....7S}.

  The model of the correlation function is not $\rpar$-symmetric because
  of the contribution of metal absorption and the variation of
  the mean redshifts with $\rpar$.
  Further asymmetry is introduced by the continuum-fitting distortion.
  The mean of any residual $\rpar$-asymmetry is absorbed by
  the quasar-redshift parameter (\ref{equation::definition_delta_rParal_QSO}).
  The fits do not reveal any significant additional asymmetries but a complete
  study of such effects \citep{2014PhRvD..89h3535B,2016JCAP...02..051I} is not
  included here.

\begin{table*}[tb]
        % Codes/code_print_result_fit_all_param.py
        % Work/Data/Cross_alone/Fits_pyLya/Fit_rmin10_rmax160_metals_laurentzian_HCD_UV_QSORadiation/
        % Work/Data/Auto_alone/Fits_pyLya/Fit_rmin10_rmax160_metals_HCD_UV/
        % Work/Data/Combined_auto_cross/Fits_pyLya/Fit_rmin10_rmax160_metals_laurentzian_HCD_UV_QSORadiation/
        %\small
        \centering
        \begin{tabular}{ l|r|r|r }
        Parameter &
        cross alone &
        auto alone &
        auto + cross \\
        \noalign{\smallskip}
        \hline \hline
        \noalign{\smallskip}
        $\alpha_{\parallel}                                   $     & $1.077 \pm 0.042$      & $1.053 \pm 0.039$      & $1.069 \pm 0.029$      \\ 
        $\alpha_{\perp}                                       $     & $0.898 \pm 0.042$      & $0.970 \pm 0.060$      & $0.920 \pm 0.034$      \\
        $b_{\mathrm{Ly}\alpha} (1+\beta_{\mathrm{Ly}\alpha})  $     & $-0.350 \pm 0.019$      & $-0.3559 \pm 0.0042$      & $-0.3544 \pm 0.0038$      \\ 
        $\beta_{\mathrm{Ly}\alpha}                            $     & $1.90 \pm 0.34$      & $1.628 \pm 0.085$      & $1.650 \pm 0.081$      \\ 
        $\bqso                                                $     & $3.87$      &  & $3.70 \pm 0.12$      \\ 
        $\drparqso \, [\hMpc{}]                  $     & $-0.93 \pm 0.14$      &  & $-0.79 \pm 0.13$      \\ 
        $\sigmavqso  \, [\hMpc{}]                           $     & $6.43 \pm 0.90$      &  & $4.67 \pm 0.32$      \\ 
        $\xitp_0                                              $     & $0.37 \pm 0.22$      &  & $0.65 \pm 0.15$      \\ 
$10^{3} \, b_{\mathrm{CIV(154.8)}}                             $     &                                        & $-23.0 \pm 11.0$      & $-22.0 \pm 11.0$      \\
$10^{3} \, b_{\mathrm{SiII(126.0)}}                             $     & $0.7 \pm 1.7$                & $-1.3 \pm 1.1$          & $-1.05 \pm 0.91$      \\
$10^{3} \, b_{\mathrm{SiIII(120.7)}}                            $     & $-7.0 \pm 1.7$              & $-2.9 \pm 1.1$          & $-3.6 \pm 1.1$      \\
$10^{3} \, b_{\mathrm{SiII(119.3)}}                             $     & $-4.0 \pm 1.6$               & $-3.23 \pm 0.84$      & $-2.96 \pm 0.74$      \\
$10^{3} \, b_{\mathrm{SiII(119.0)}}                             $     & $-1.4 \pm 1.7$              & $-3.99 \pm 0.83$      & $-3.4 \pm 0.73$      \\
        $\bhcd                                              $     & $0.029 \pm 0.010$      & $-0.0318 \pm 0.0047$      & $-0.0275 \pm 0.0040$      \\ 
        $\betahcd                                          $     & $0.55 \pm 0.20$      & $0.69 \pm 0.17$      & $0.79 \pm 0.17$      \\ 
        $\Lhcd \, [\hMpc{}]    $     & $62.7 \pm 7.3$      & $24.5 \pm 1.1$      & $23.9 \pm 1.2$      \\ 
        $b_{\Gamma}                                        $     & $-0.18 \pm 0.12$      & $0.150 \pm 0.058$      & $0.108 \pm 0.049$      \\

        \noalign{\smallskip}
        \hline \hline
        \noalign{\smallskip}

        $N_{\rm bin}$           & $2504$      & $1252$    & $3756$      \\ 
        $N_{\rm param}$         & $15$        & $13$      & $17$      \\ 
        $\rho(\aparMath,\aperpMath)$                              &   -0.377                &       -0.369           &     -0.362              \\
        $\chi^{2}_{\min}$       & $2576.31$   & $1232.56$ & $3833.16$      \\ 
        probability                     & $0.11$      & $0.55$    & $0.14$      \\
        $\chi^2(\aparMath=\aperpMath=1)$   & 2582.58    &   1234.82       &   3841.97         \\

        \end{tabular}
        \caption{
        Fit results for
        the cross-correlation (this study), the auto-correlation \citep{2017A&A...603A..12B} extrapolated to $z=2.40$,
        and the combined fit.
Errors correspond to CL=68.27\%.
        Parameters without errors are fixed.
        The fit is over the range  $10<r<160~\hMpc$.
        }
        \label{table::bestfit_best_model_fit_parameters_cross_auto_combined}
\end{table*}

\subsection{Fits of the cross-correlation}
\label{subsection::Fits_for_the_peak_position::Results}

Our ``standard'' fit of the cross-correlation function uses the
15 parameters in the first group of Table \ref{table::fit_parameters}.
The best-fit values are shown in the column ``cross alone'' of 
Table \ref{table::bestfit_best_model_fit_parameters_cross_auto_combined}.
Instead of fitting the bias of the \Lya{} absorber, $b_{\LyaMath}$, we fit the
combination $b_{\LyaMath}(1+\beta_{\LyaMath})$, which is less correlated
with $\beta_{\LyaMath}$ and better constrained.
We limit the fit to separations
$r \in [10,160]~\hMpc$
and fit all directions $\mu \in [-1,1]$.
As we will see below, these choices 
have no significant impact on the values and precision of
the two BAO-peak parameters.

\begin{figure}
        % Work/Data/contours.py
        \centering
\includegraphics[width=\columnwidth]{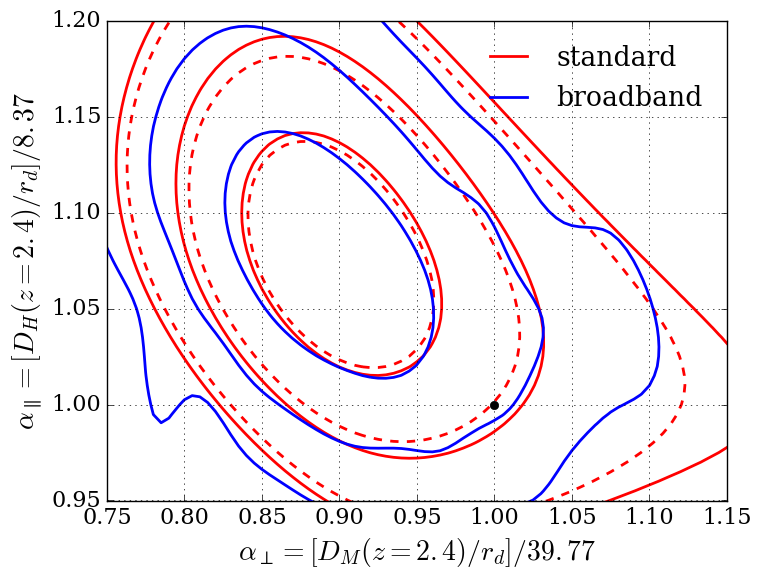}
\caption{Constraints on $(\aparMath,\aperpMath)$ from the standard
  fit (red) and fit with a
  broadband term (\ref{equation::broadband_1}) (blue). The dashed red
  lines correspond to 
          $\Delta\chi^2=\chi^{2} - \chiSquareMin=(2.3,6.18,11.83),$
  while the solid lines correspond to 
  $\Delta\chi^2=(2.62,7.25,12.93)$, that is, to 
          confidence levels of $(68.27,95.45\%,99.7\%)$.
        The black point $(\aparMath,\aperpMath)=(1,1)$ indicates
        the value for the Planck~2016 flat-\lcdm~cosmology.
        }
        \label{figure::chi2_scan_data_alpha_paral_alpha_perp}
\end{figure}

The best fit is shown in
Fig. \ref{figure::xi_data_best_fit_4_weges_mu}
for four ranges of $\mu$
and in Fig. \ref{figure::xi_data_best_fit_2_slice_rperp}
for the two lowest $\rperp$ bins.
The best-fit values of the BAO peak position are
$(\aperpMath,\aparMath)=(0.898,\,1.077)$ with constant $\chi^2$
contours indicated in red in 
Fig. \ref{figure::chi2_scan_data_alpha_paral_alpha_perp}.
The dashed contours for
$\chi^2(\aperpMath,\aparMath)-\chiSquareMin=(2.29,\,6.18,\,11.83)$
correspond to the nominal $(68.27,95.45,99.73\%)$ limits on
$(\aperpMath,\aparMath)$.
This correspondence is, however, not expected to be exact because
even if $\xi$ has Gaussian errors,
the model is not a linear function of $(\aperpMath,\aparMath)$.
In the analysis of the 100 mock data sets (Sect. \ref{section::mocks}),
the number of sets yielding
$\Delta\chi^2\equiv\chi^2(\aperpMath=\aparMath=1)-\chiSquareMin>6.18$ was
greater than
the expected 4.5\%
(Table \ref{table::fit_mock_sets}, last column).
This result suggests that
the confidence level corresponding to $\Delta\chi^2=6.18$ is overestimated.
To make a more precise estimate of the relation between
$\Delta\chi^2$ and confidence level, we
generated a large number of simulated correlation functions using the fiducial
cosmological model and the best fit values of non-BAO parameters, 
randomized using the covariance matrix measured with the data.
Based on 
these studies, described in detail in Appendix \ref{append::fastMC}
and
summarized in Table \ref{table::fastmc},
we adopt $\Delta\chi^2=(2.62,7.25)$ as 
confidence levels of $(68.27,95.45\%)$.
These levels are  the solid red lines
in Fig. \ref{figure::chi2_scan_data_alpha_paral_alpha_perp}.
The best-fit values of (\aperp,\apar) are $1.8\sigma$ from the
CMB-inspired flat-\LCDM~model
\citep{2016A&A...594A..13P}, which has
a $\chi^2$ that is 6.27 greater than the best fit.

The best-fit values and confidence level (68.27,95.45\%) ranges
for the BAO parameters are:
\begin{align}
        % Work/Data/Cross_alone/Fits_pyLya/Fit_rmin10_rmax160_metals_laurentzian_HCD_UV_QSORadiation/
        % Work/Data/Cross_alone/Fits_pyLya/Fit_rmin10_rmax160_metals_laurentzian_HCD_UV_QSORadiation_sigmaErrors1/
        % Work/Data/Cross_alone/Fits_pyLya/Fit_rmin10_rmax160_metals_laurentzian_HCD_UV_QSORadiation_sigmaErrors2/
        % Work/Data/Cross_alone/Fits_pyLya/Fit_rmin10_rmax160_metals_laurentzian_HCD_UV_QSORadiation_sigmaErrors3/
%\aperpMath~=~ & 0.898~(1\sigma)_{0.038}^{0.040}~(2\sigma)_{0.076}^{0.088}
\aperpMath~=~ & 0.898~_{- 0.041 }^{+ 0.043 }\;_{- 0.084 }^{+ 0.098 }
        \label{equation::measure_alpha_perp},\\[2pt]
%\aparMath~=~ & 1.077~(1\sigma)_{0.038}^{0.039}~(2\sigma)_{0.076}^{0.082}
\aparMath~=~ & 1.077~_{ -0.041 }^{ +0.043 }\;_{ -0.084 }^{ +0.090 }
        \label{equation::measure_alpha_parallel},
\end{align}
corresponding to 
\begin{align}
        % Work/Data/Cross_alone/Fits_pyLya/Fit_rmin10_rmax160_metals_laurentzian_HCD_UV_QSORadiation/
        % Work/Data/Cross_alone/Fits_pyLya/Fit_rmin10_rmax160_metals_laurentzian_HCD_UV_QSORadiation_sigmaErrors1/
        % Work/Data/Cross_alone/Fits_pyLya/Fit_rmin10_rmax160_metals_laurentzian_HCD_UV_QSORadiation_sigmaErrors2/
        % Work/Data/Cross_alone/Fits_pyLya/Fit_rmin10_rmax160_metals_laurentzian_HCD_UV_QSORadiation_sigmaErrors3/
%\frac{\DMm(z=\MEANZ)}{r_{d}}~=~ &  35.7~(1\sigma)_{1.5}^{1.6}~(2\sigma)_{3.0}^{3.5}
\frac{\DMm(z=\MEANZ)}{r_{d}}~=~ &  35.7~_{- 1.6 }^{+ 1.7 }\;_{ -3.3 }^{ +3.9 }
        \label{equation::measure_DA_over_rd},\\[2pt]
%\frac{\DHh(z=\MEANZ)}{r_{d}}~=~ & 9.01~(1\sigma)_{0.32}^{0.33}~(2\sigma)_{0.64}^{0.68}
\frac{\DHh(z=\MEANZ)}{r_{d}}~=~ & 9.01~_{- 0.35 }^{+ 0.36 }\;_{- 0.71 }^{+ 0.75 }
        \label{equation::measure_Dh_over_rd}.
\end{align}
The two BAO parameters are $-38\%$ correlated with one another.
  Following the results of Table \ref{table::fastmc}, the $(1\sigma,2\sigma)$ errors
  correspond to $\Delta\chi^2=(1.17,4.94)$ for $\DMm/r_d$
  and to $\Delta\chi^2=(1.19,4.87)$ for $\DHh/r_d$.

\citet{2014JCAP...05..027F}  measured 
$\DHh(z=2.36)/r_d=9.0\pm0.3$ and
$\DMm(z=2.36)/r_d=36.3\pm1.4$.
Scaling $\DMm(2.36)$ and $\DHh(2.36)$ to $z=2.4$ (using
the fiducial cosmology) results in
$\DHh(z=2.4)/r_d=8.85\pm0.3$ and
$\DMm(z=2.4)/r_d=35.7\pm1.4$.
The prior $\DMm$ measurement agrees well with the present result,
while $\DHh$ has shifted by $0.5\sigma$. As discussed in
\citet{2017A&A...603A..12B}, this shift is typical of what can be
expected due to the statistical
difference between the DR11 and  DR12 samples.

\begin{figure}
        % Work/Data/contours.py
        \centering
        \includegraphics[width=\columnwidth]{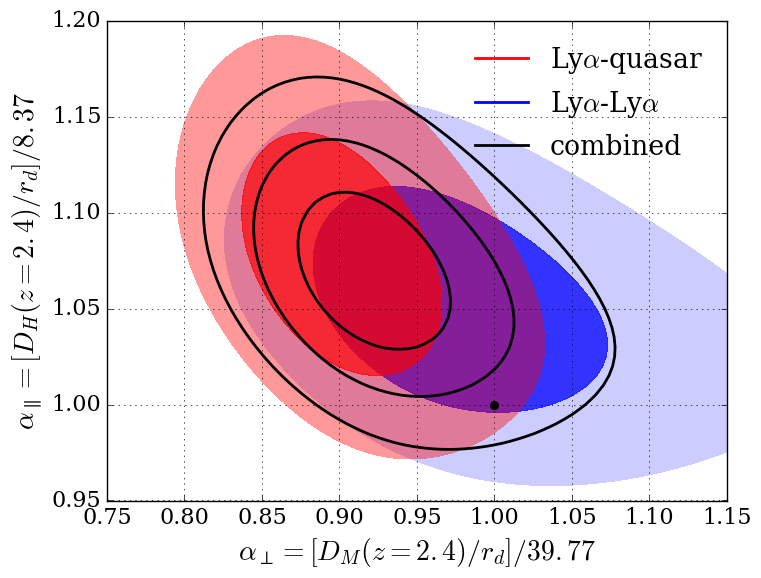}
        \caption{
Constraints on $(\aparMath,\aperpMath)$ corresponding to 
$CL=68.27$ and $95.45\%$
for the cross-correlation (red) and
the auto-correlation \citep{2017A&A...603A..12B}
with a unconstrained broadband (blue).
The black lines show the $CL=68.27$, $95.45\%$ and $99.7\%$ limits
for the combined fit.
The black point $(\aparMath,\aperpMath)=(1,1)$ indicates
the  value for the Planck~2016 flat-\lcdm~cosmology.
}
        \label{figure::chi2_scan_data_alpha_paral_alpha_perp_combined}
\end{figure}

The best fit values for the bias and the RSD parameters of the \Lya{} field
are
$\blya(1+\betalya) = -0.350 \pm 0.019$
and $ \betalya = 1.90 \pm 0.34 $.
They are  compatible
with the values of \citet{2017A&A...603A..12B}
found using the auto-correlation functionn, reported here in the 
column ``auto alone'' of
Table~\ref{table::bestfit_best_model_fit_parameters_cross_auto_combined}.

The effect of metals is visible in the lowest $\rperp$ bins
(Fig.~\ref{figure::xi_data_best_fit_2_slice_rperp}).
The measured bias of SiIII(120.7),
$b_{\mathrm{SiIII(120.7)}}  = -0.0070 \pm 0.0017$, is 
incompatible with zero at more than $4$~sigma. We thus have  evidence
of a large-scale cross-correlation between  metals and
quasars. The three other metals 
are detected with less significance or not at all.

While
the metal parameters found in the cross-correlation are broadly
consistent with those found in the auto-correlation, this is not
the case for the HCD parameters.
The best-fit value of $\bhcd$ even has the opposite sign of that
found in the auto-correlation.
This suggests that the HCD parameters model non-HCD effects,
as noted by \citet{2017A&A...603A..12B}.
Fortunately, the BAO parameters are insensitive to the HCD modeling.
Fixing the HCD parameters to those found in the auto-correlation
results in no significant change in (\aperp,\apar)  (Line ``HCD fixed''
of Table \ref{table::fit_data_model}).

The uncertainties in Equations (\ref{equation::measure_alpha_perp})
through (\ref{equation::measure_Dh_over_rd}) are purely statistical.
In the \Lya~auto-correlation measurement of \citet{2017A&A...603A..12B},
possible systematic uncertainties in the correlation function related to
correlated flux-calibration errors were studied in detail.
For the quasar-flux cross-correlation, these errors are not relevant.
The primary identified systematic error here is in the measurement
of quasar redshifts, but this issue leads to an asymmetry in $\xi(\rpar),$ which
is parameterized by $\drparqso$ and included in the fit.
As such, the error is included in the statistical error.

To search for unexpected systematic errors in the determination of the BAO-peak
position,  we performed fits with modified models.  These fits are described
in Appendix \ref{append::nonstandardfits} and summarized
in Table \ref{table::fit_data_model}.  No obvious discrepancies with the
standard fit were found.
Of special interest are fits that included a broadband component of the
form (\ref{equation::broadband_1}).
This fit provides constraints on (\aperp,\apar) that are very similar to
the standard fit, as seen in Fig. \ref{figure::chi2_scan_data_alpha_paral_alpha_perp}.
This insensitivity to the addition of a broadband term differs
from the result  for the auto-correlation \citep{2017A&A...603A..12B}
where,
because of the very weak BAO signal in the transverse direction,
such terms significantly degraded constraints on \aperp. We also performed fits on subsamples of the data as described in
Appendix \ref{append::nonstandardfits} and summarized in
Table \ref{table::fit_data_split}.  No obvious discrepancies
were found.

\subsection{Combination with the auto-correlation}

This analysis of the \Lya-quasar cross-correlation in DR12 quasars
can be combined with the results of the \Lya{} auto-correlation in DR12
\citep{2017A&A...603A..12B}.
This study can be done by simply
combining the (\aperp,\apar) likelihood contours for the two
correlation functions, or by performing a joint fit of the two correlation
functions.
In the first case we need to estimate the covariance between the
two values of (\aperp,\apar).
The second case
requires the full covariance matrix between  the two correlation functions.
This problem was studied in \citet{2015A&A...574A..59D}, who showed that the
covariance was sufficiently small to be ignored.
The studies with the mock data sets discussed in
Sect. \ref{subsection::Validation_of_the_cross_auto_covariance_matrix}
confirm this conclusion and demonstrate that, as expected,  the
(\aperp,\apar) derived from the auto- and cross-correlations are largely
uncorrelated.

We first combine the two measurements by performing a joint fit of the
two correlation functions.
This fit has 17 free parameters: the 15 from the cross-correlation
model and the biases of the CIV
forest, $b_{\rm CIV}$, and of quasars, $\bqso$. 
The best-fit results
are presented in
Table \ref{table::bestfit_best_model_fit_parameters_cross_auto_combined}.
Figure \ref{figure::chi2_scan_data_alpha_paral_alpha_perp_combined}
gives (in black) the
68.27\%, 95.45\%, and 99.7\% CL
contours (using $\Delta\chi^2=(2.5,6.5,13.0)$ from Table \ref{table::fastmc}).
The results differ from the prediction of
the Planck~2016 flat~\lcdm~cosmology
by $2.3\sigma$.
The figure also displays the contours for the
auto-correlation in blue \citep{2017A&A...603A..12B} and the
cross-correlation in red (this study).

The best-fit values  for the BAO parameters are:
\begin{align}
        % Work/Data/Combined_auto_cross/Fits_pyLya/Fit_rmin10_rmax160_metals_laurentzian_HCD_UV_QSORadiation/
        % Work/Data/Combined_auto_cross/Fits_pyLya/Fit_rmin10_rmax160_metals_laurentzian_HCD_UV_QSORadiation_sigmaErrors1/
        % Work/Data/Combined_auto_cross/Fits_pyLya/Fit_rmin10_rmax160_metals_laurentzian_HCD_UV_QSORadiation_sigmaErrors2/
        % Work/Data/Combined_auto_cross/Fits_pyLya/Fit_rmin10_rmax160_metals_laurentzian_HCD_UV_QSORadiation_sigmaErrors3/
%\aperpMath~=~ & 0.920~(1\sigma)_{0.029}^{0.032}~(2\sigma)_{0.059}^{0.069}
\aperpMath~=~ & 0.920~_{- 0.030 }^{+ 0.033 }\;_{- 0.062 }^{+ 0.072 }
        \label{equation::measure_alpha_perp_combined},\\[2pt]
%\aparMath~=~ & 1.069~(1\sigma)_{0.025}^{0.026}~(2\sigma)_{0.051}^{0.054}
\aparMath~=~ & 1.069~_{- 0.026 }^{+ 0.027 }\;_{- 0.052 }^{+ 0.055 }
        \label{equation::measure_alpha_parallel_combined},
\end{align}
corresponding to:
\begin{align}
        % Work/Data/Combined_auto_cross/Fits_pyLya/Fit_rmin10_rmax160_metals_laurentzian_HCD_UV_QSORadiation/
        % Work/Data/Combined_auto_cross/Fits_pyLya/Fit_rmin10_rmax160_metals_laurentzian_HCD_UV_QSORadiation_sigmaErrors1/
        % Work/Data/Combined_auto_cross/Fits_pyLya/Fit_rmin10_rmax160_metals_laurentzian_HCD_UV_QSORadiation_sigmaErrors2/
        % Work/Data/Combined_auto_cross/Fits_pyLya/Fit_rmin10_rmax160_metals_laurentzian_HCD_UV_QSORadiation_sigmaErrors3/
%\frac{D_{M}(z=\MEANZ)}{r_{d}}~=~ & 36.6~(1\sigma)_{1.2}^{1.3}~(2\sigma)_{2.3}^{2.7}
\frac{\DMm(z=\MEANZ)}{r_{d}}~=~ &  36.6~_{ -1.3 }^{ +1.4 }\;_{ -2.4 }^{ +2.8 }
        \label{equation::measure_DA_over_rd_combined},\\[2pt]
%\frac{D_{H}(z=\MEANZ)}{r_{d}}~=~ & 8.94~(1\sigma)_{0.21}^{0.22}~(2\sigma)_{0.42}^{0.45}
\frac{\DHh(z=\MEANZ)}{r_{d}}~=~ & 8.94~_{- 0.22 }^{+ 0.23 }\;_{- 0.43 }^{+ 0.46 }
        \label{equation::measure_Dh_over_rd_combined}.
\end{align}

The combined fit of the auto-
and cross-correlations breaks the degeneracy between
$\blya$ and $\bqso$
and we find:
\begin{equation}
        \bqso(z=2.40) = 3.70 \pm 0.12
        \label{equation::measure_bias_QSO_combined},
\end{equation}
where the error is statistical.
This result is 
in agreement with the results of \citet{2005MNRAS.356..415C} and of
\citet{2016JCAP...11..060L}, but a study of possible systematic errors
will not be presented here.

The second method of performing the joint fit consists of
simply summing the $\chi^2(\aperpMath,\aparMath)$ of the cross-correlation
measurement (Fig. \ref{figure::chi2_scan_data_alpha_paral_alpha_perp})
and the auto-correlation measurement of \citet{2017A&A...603A..12B}.
The measurement of the auto-correlation depends on 
whether or not one includes a broadband term in the fitting, as
seen in Table 6 and Fig. 15. of \citet{2017A&A...603A..12B}.
The broadband does not improve the quality of the fit so
we adopt the broadband-free result as our primary result.
The summed $\chi^2$ broadband-free fit
gives a result hardly different from
(\ref{equation::measure_alpha_perp_combined}) and
(\ref{equation::measure_alpha_parallel_combined}):
$\aperpMath=0.925\pm0.035\pm0.075$ and 
$\aparMath=1.066\pm0.028\pm0.059$,
corresponding to a shift of $\sim0.15\sigma$.
Use of a broadband
(``no additional priors'' of Table 6 of \citet{2017A&A...603A..12B})
results in 
$\aperpMath=0.935\pm0.038\pm0.082$ and 
$\aparMath=1.063\pm0.028\pm0.058$,
corresponding to a shift of $\sim0.4\sigma$.

\begin{figure*}
        % Codes/codes.py :: plot_compare_xi1D_mock_vs_data()
        \centering
        \includegraphics[width=\columnwidth]{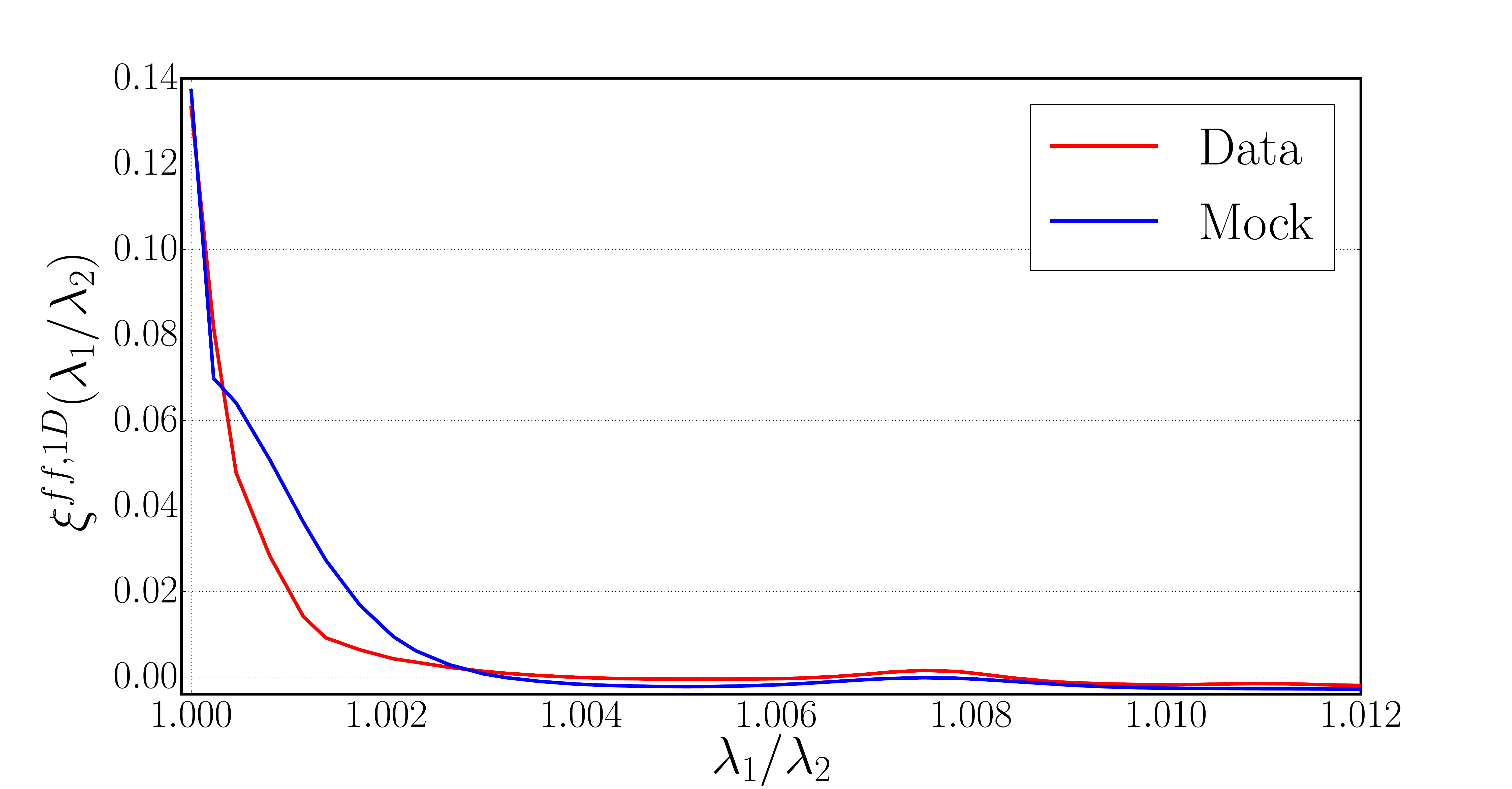}
        \includegraphics[width=\columnwidth]{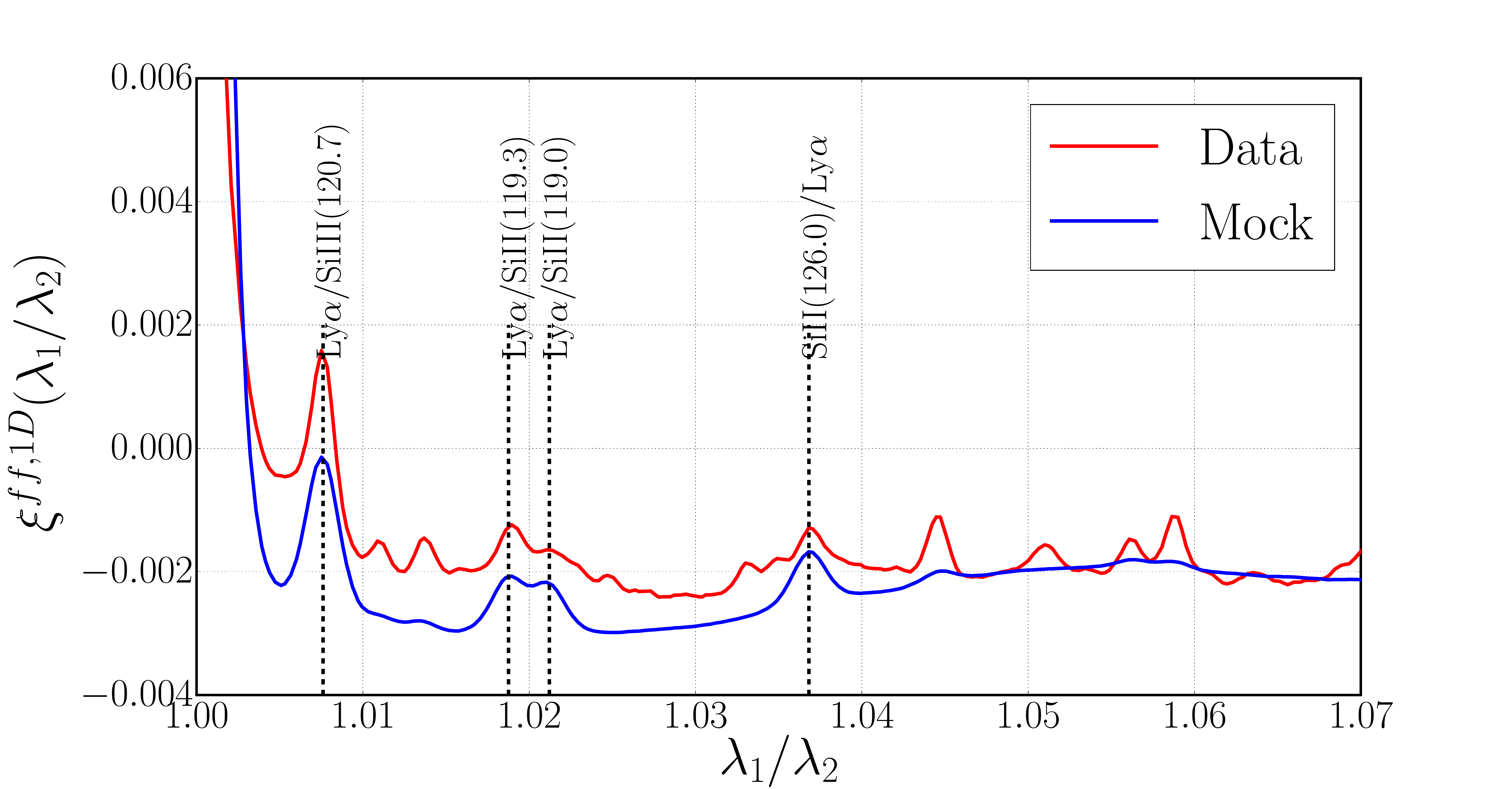}
        \caption{
                Correlation function for two pixels from the same forest,
                $\xiOneD$, as a function of wavelength
                ratio for the data and for the mocks, respectively,
                in red and blue.
                Prominent peaks due to \Lya-metal correlations are indicated.
                The metal transitions relevant to
                this study are given in Table \ref{table::metals_contribution}.
                The differences in $\xiOneD$ for data and mocks lead to differences in the covariance
                matrix for data and mocks.
        }
        \label{figure::xi1D_compare_data_mock_with_lines}
\end{figure*}

\begin{table*}
        \centering      
        \caption{
          Results of the fits of the 100 mocks of each of the three sets
          of spectra:
          \Lya~absorption only; \Lya~absorption superimposed on a quasar continuum; and
          including metal absorption.
          The weighted means and mean uncertainties ($\Delta\chi^2=1$)
          are given for the BAO-peak parameters (\aperp,\apar) and
          for the \Lya- bias parameters.
          The best-fit $\chi^2$ is listed along with $DOF$ equal to
          the number of bins minus the number of parameters.
          The final column gives $N_{6.18}$ defined as the number of mock sets
          with
        $\Delta\chi^2\equiv\chi^2(\aperpMath=\aparMath=1)-\chi^2_{\rm min}>6.18$.
        The mean of  $N_{6.18}$ would be 4.5 if the errors on (\aperp,\apar) were
            Gaussian.
        In the cross-correlation fits,
        we fix $\bqso = 3.34$ and $f = 0.95916$, and in the combined fit,
        $f = 0.95916$.
        }
        \label{table::fit_mock_sets}
        
        \begin{tabular}{l l l l l l l}

        Mock set 
        &       $\overline{\aparMath} ~ (\overline{\sigma})$ 
        &       $\overline{\aperpMath} ~ (\overline{\sigma})$ 
        &       $\overline{b_{\LyaMath}\left(1+\beta_{\LyaMath}\right)} ~ (\overline{\sigma})$
        &       $\overline{\beta_{\LyaMath}} ~ (\overline{\sigma})$
        &       $\overline{\chiSquareMin}/DOF$
        & $N_{6.18}$\\

        \noalign{\smallskip} 
        \hline \hline
        \noalign{\smallskip}

        cross-correlation:\\
        % Work/Mocks/Box_00*/Simu_00*/Results_raw_from_JeanMarc/Fit_rmin10_rmax160_freeLparaLper/cross_alone.save.pars
        % Work/Mocks/Box_00*/Simu_00*/Results_no_metals/Fit_rmin10_rmax160_freeLparaLper/cross_alone.save.pars
        % Work/Mocks/Box_00*/Simu_00*/Results/Fit_rmin10_rmax160_freeLparaLper/cross_alone.save.pars
$ \rm{Ly\alpha}  $ & $ 0.994 ~(0.025)$ & $ 1.002 ~(0.028)$ & $ -0.3858 ~(0.0046)$ & $ 1.318 ~(0.064)$ & $ 2501.16 / (2504-7)$  & 10\\
$ \rm{+Continuum}    $ & $ 0.990 ~(0.038)$ & $ 0.994 ~(0.050)$ & $ -0.3725 ~(0.0067)$ & $ 1.26 ~(0.12)$ & $ 2493.72 / (2504-7)$ & 7 \\
$ \rm{+Metals} $ & $ 0.988 ~(0.039)$ & $ 1.003 ~(0.050)$ & $ -0.3726 ~(0.0068)$ & $ 1.28 ~(0.12)$ & $ 2492.88 / (2504-11)$  & 13\\
        \noalign{\smallskip} 
        \hline \hline
        \noalign{\smallskip}
            auto-correlation:\\
$ \rm{Ly\alpha}    $ & $ 0.995 ~(0.018)$ & $ 1.002 ~(0.027)$ & $ -0.3995 ~(0.0016)$ & $ 1.412 ~(0.031)$ & $ 1252.95 / (1252-\,~6)$ & 4 \\
$ \rm{+Continuum} $ & $ 0.997 ~(0.042)$ & $ 0.986 ~(0.066)$ & $ -0.3750 ~(0.0034)$ & $ 1.243 ~(0.077)$ & $ 1260.16 / (1252-\,~6)$ & 4 \\
$ \rm{+Metals} $ & $ 0.991 ~(0.040)$ & $ 0.996 ~(0.067)$ & $ -0.3713 ~(0.0035)$ & $ 1.167 ~(0.075)$ & $ 1274.03 / (1252-10)$ & 10 \\
        \noalign{\smallskip} 
        \hline \hline
        \noalign{\smallskip}
                combined fits:\\
$ \rm{Ly\alpha}     $ & $ 0.995 ~(0.014)$ & $ 1.003 ~(0.019)$ & $ -0.3988 ~(0.0015)$ & $ 1.394 ~(0.028)$ & $ 3759.40 / (3756-10)$ & 4  \\
${\rm +Continuum}$    & $ 0.992 ~(0.030)$ & $ 1.001 ~(0.041)$ & $ -0.3751 ~(0.0032)$ & $ 1.250 ~(0.064)$ & $ 3757.43 / (3756-10)$ & 10\\
${\rm +Metals}$ & $ 0.990 ~(0.027)$ & $ 1.004 ~(0.043)$ & $ -0.3714 ~(0.0032)$ & $ 1.205 ~(0.064)$ & $ 3777.05 / (3756-14)$ & 13 \\
        \end{tabular}
\end{table*}

\begin{figure*}
        \centering
        \includegraphics[width=\columnwidth]{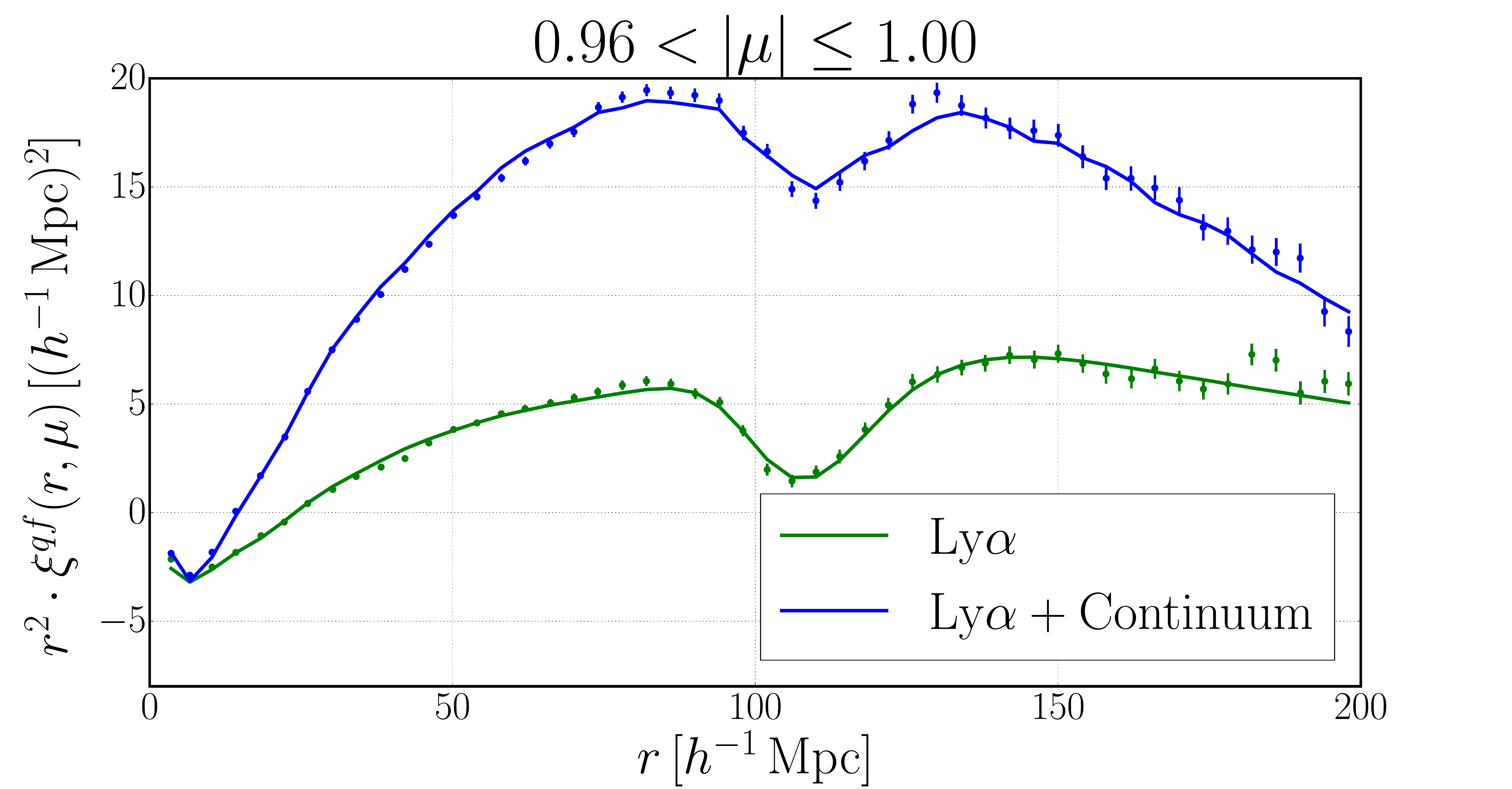}
        \includegraphics[width=\columnwidth]{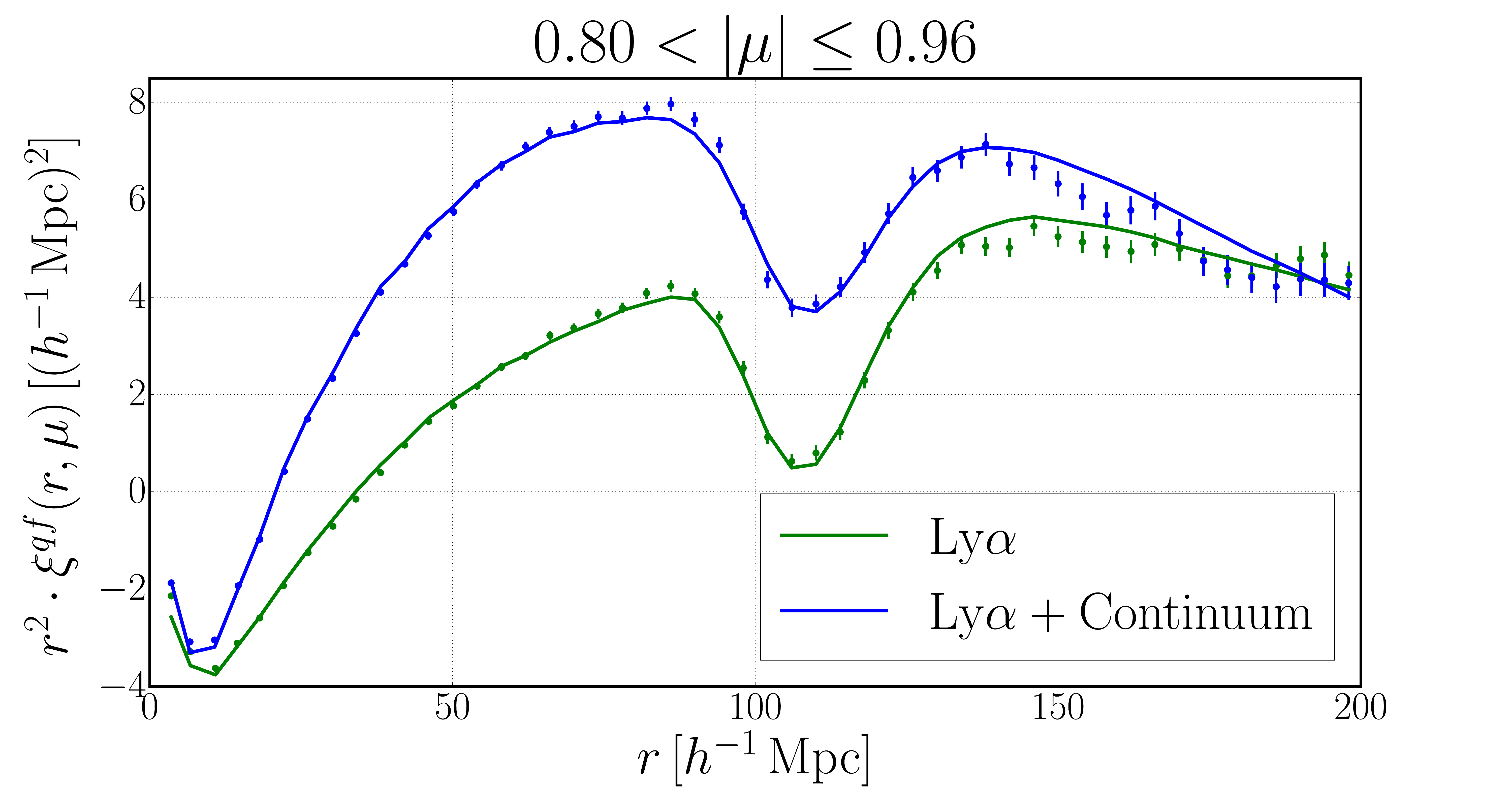}
        \includegraphics[width=\columnwidth]{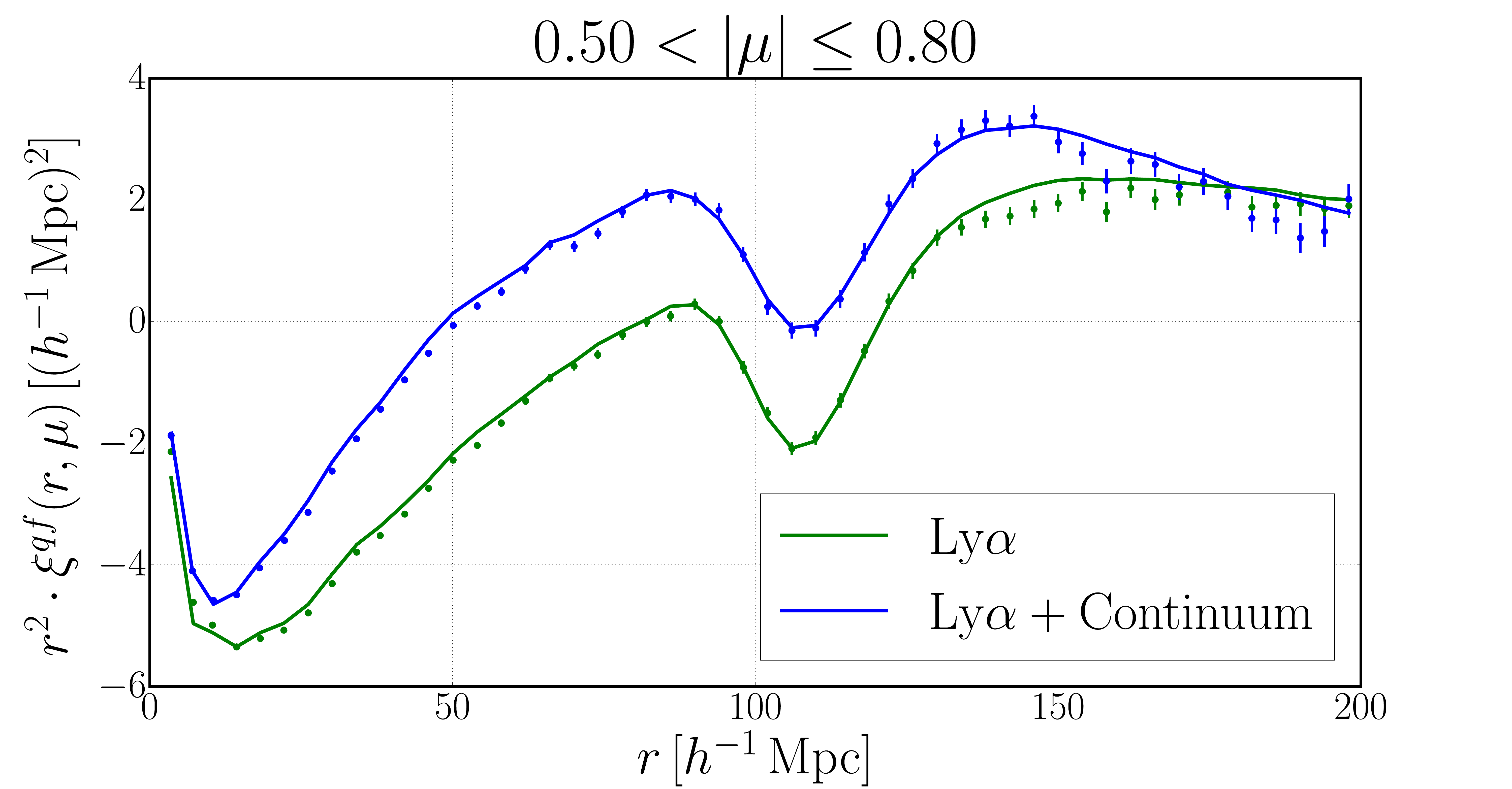}
        \includegraphics[width=\columnwidth]{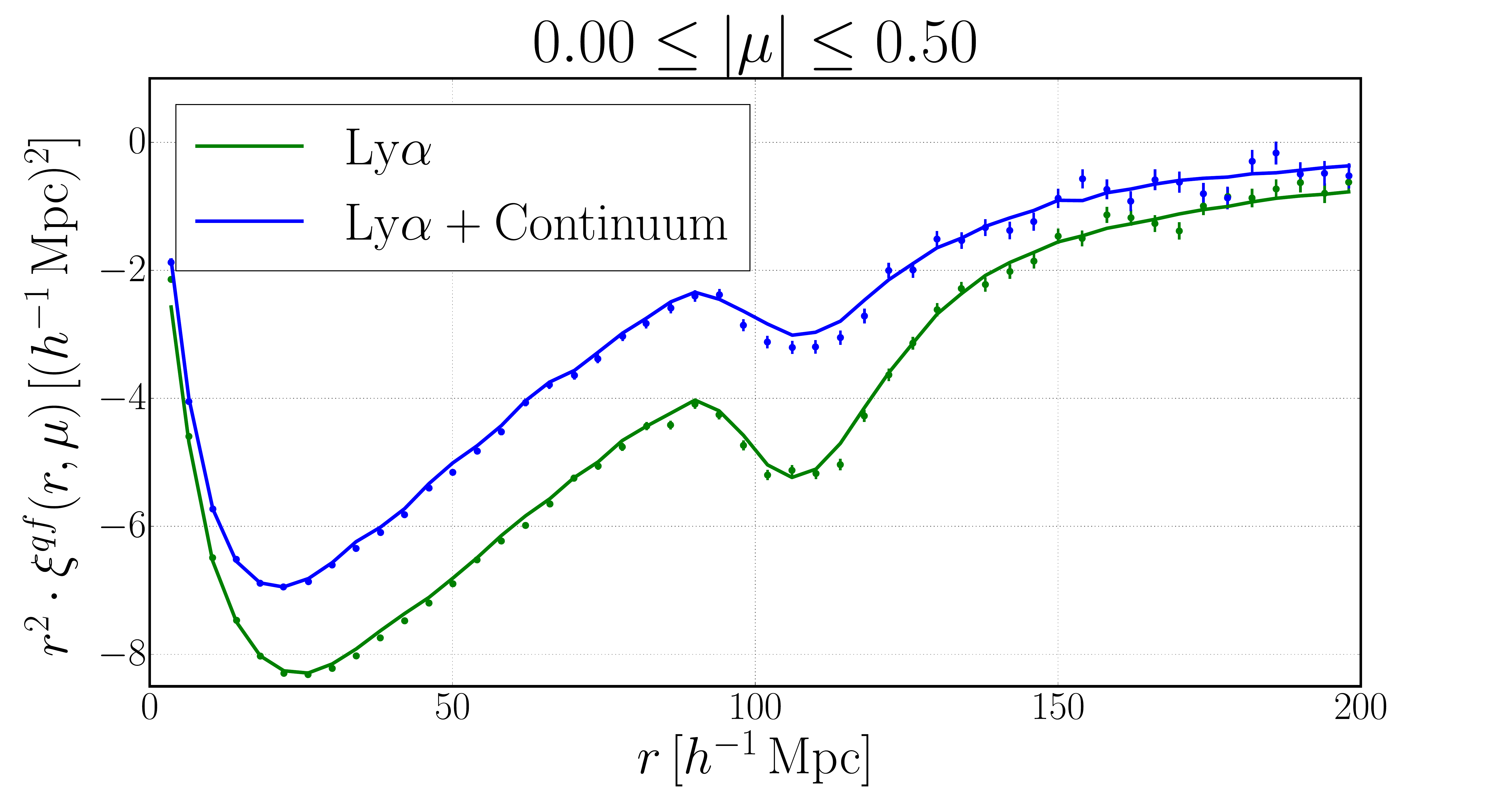}
        \caption{
        Cross-correlation of the stack of the 100 mocks in
        four  bins of $\mu=r_{\parallel}/r$, with the points representing the
        reconstructed correlation function and the lines representing the fit
        correlation function (over the range $10<r<160~\hMpc$).
        Green and blue represent
        the \Lya{} and  \Lya{}+Continuum types, respectively.
        The agreement between lines and points indicates that the distortion
        due to continuum fitting is well modeled
        by the distortion matrix $D_{AA^\prime}$ (\ref{equation::distortion_matrix_measure}).
        The cross-correlation is multiplied by a factor $r^{2}$ to show the BAO scale.
        }
        \label{figure::xi_wedge_000_rescale2_compare_mocks_raw_cooked_no_met_with_fits}
\end{figure*}

\begin{figure*}
        \centering
        \includegraphics[width=\columnwidth]{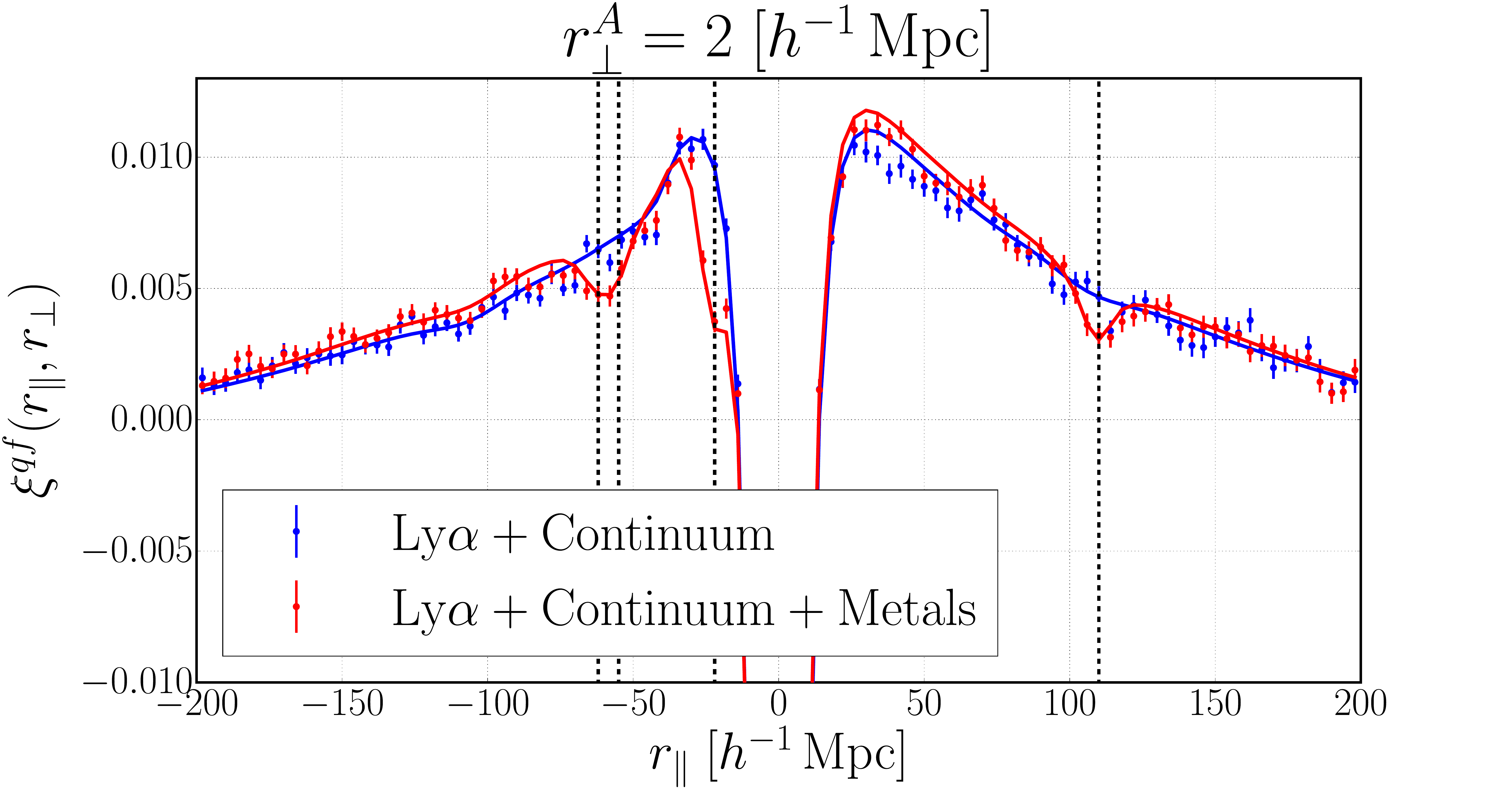}
        \includegraphics[width=\columnwidth]{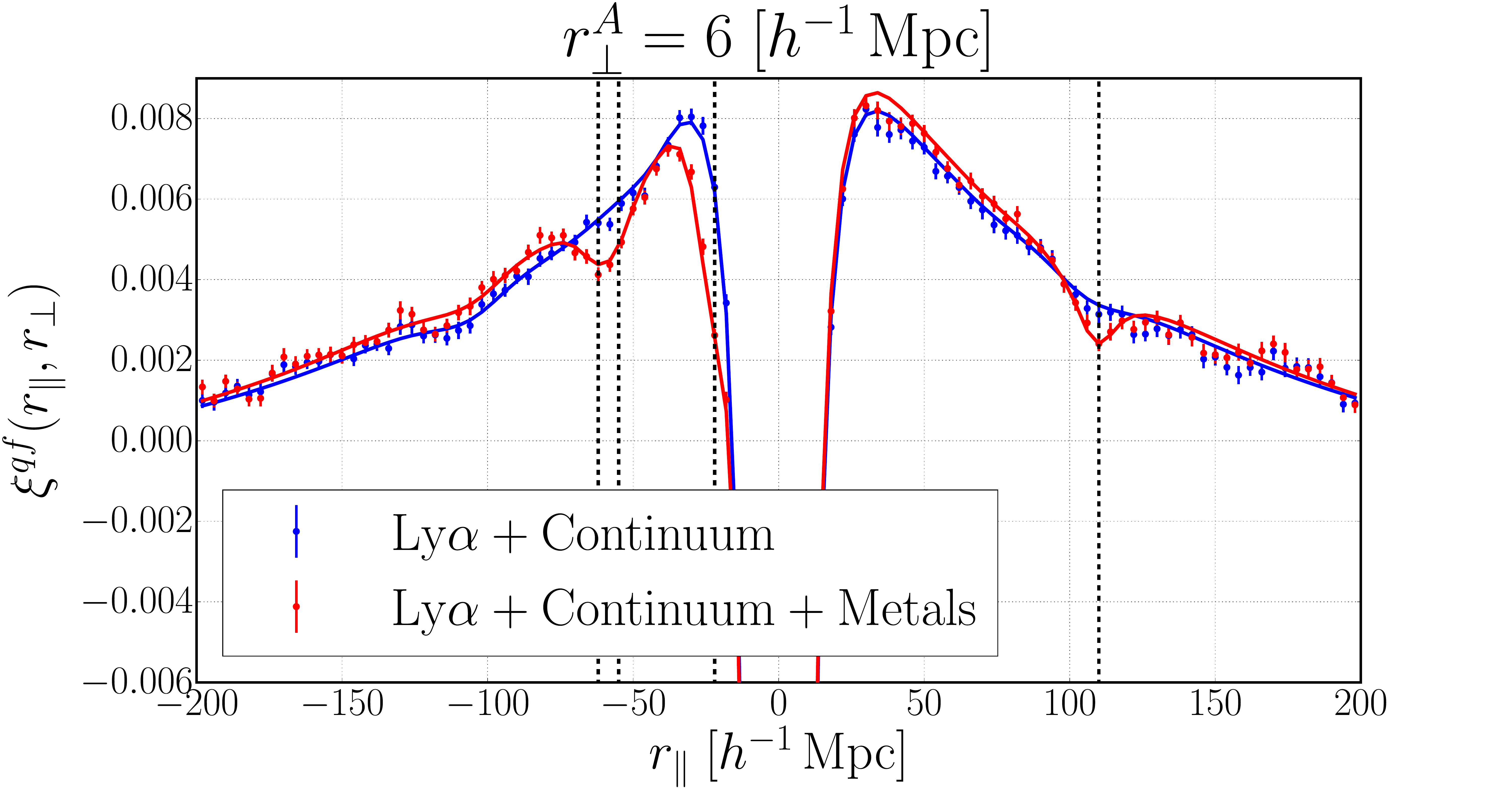}
        \caption{
        Stack of the 100 mocks cross-correlation for
        two different ranges of $\rperp$ with the points representing the
        reconstructed correlation function and the lines representing the fit
        correlation function (over the range $10<r<160~\hMpc$).
        Blue and red represent
        the \Lya{}+Continuum type and  \Lya{}+Continuum+Metals types, respectively.
        The four black dashed lines indicate the positions of the four peaks of
        the metal-quasar correlations.
        The trough at $\rpar \approx -60~\hMpc$ is
        due to the SiII(119.3)- and SiII(119.0)-quasar cross-correlations,
        at $\rpar \approx -21~\hMpc$ to the
        SiIII(120.7)-quasar cross-correlation,
        and at $\rpar \approx +103~\hMpc$ to the
        SiIII(126.0)-quasar cross-correlation.
        }
        \label{figure::xi_slice_000_compare_mocks_raw_cooked_with_fits}
\end{figure*}

\begin{figure*}
        % Codes/codes.py :: plot_compare_correlation_new_version_mocks()
        \centering
        \includegraphics[width=\columnwidth]{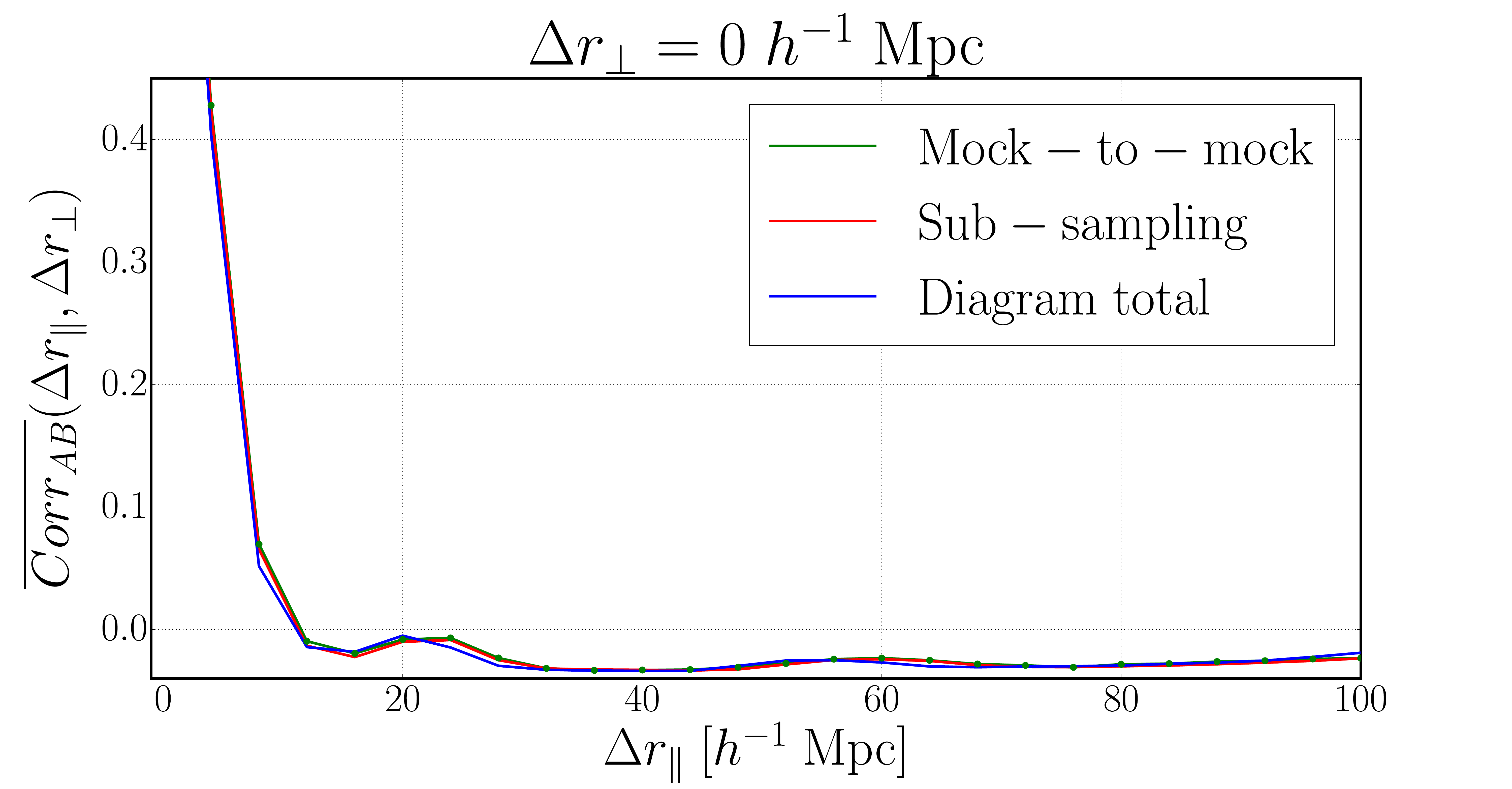}
        \includegraphics[width=\columnwidth]{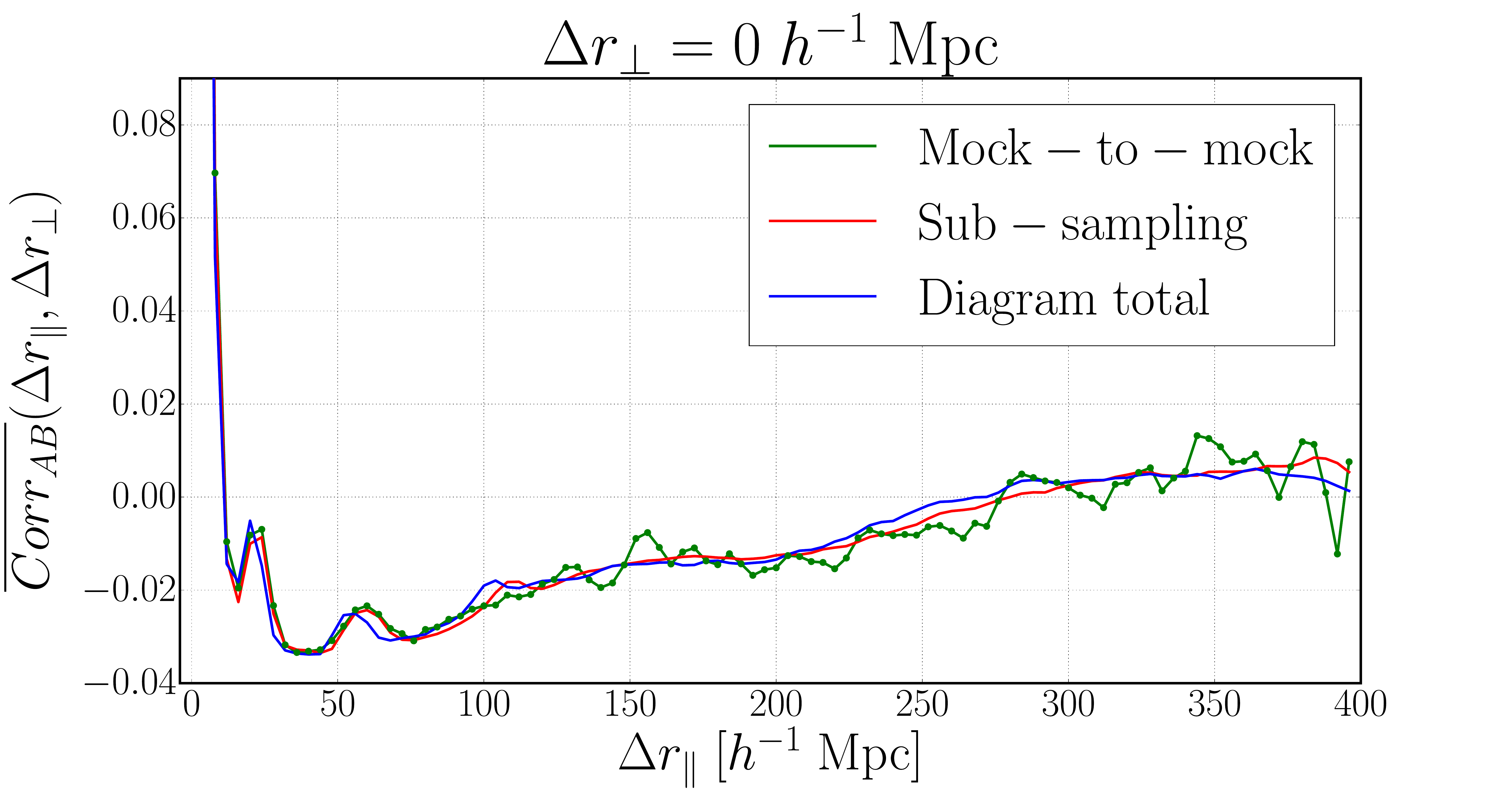}
        \includegraphics[width=\columnwidth]{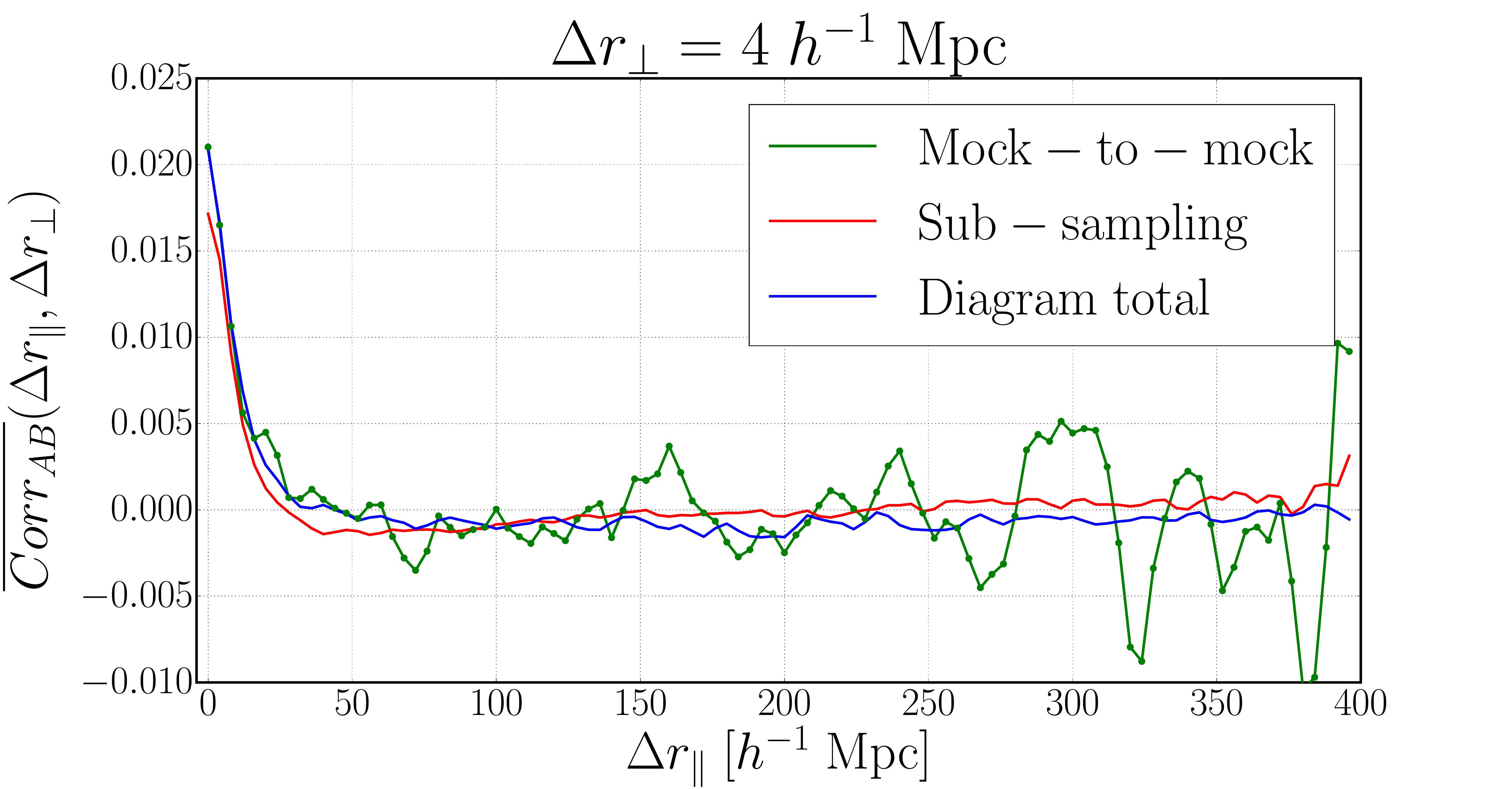}
        \includegraphics[width=\columnwidth]{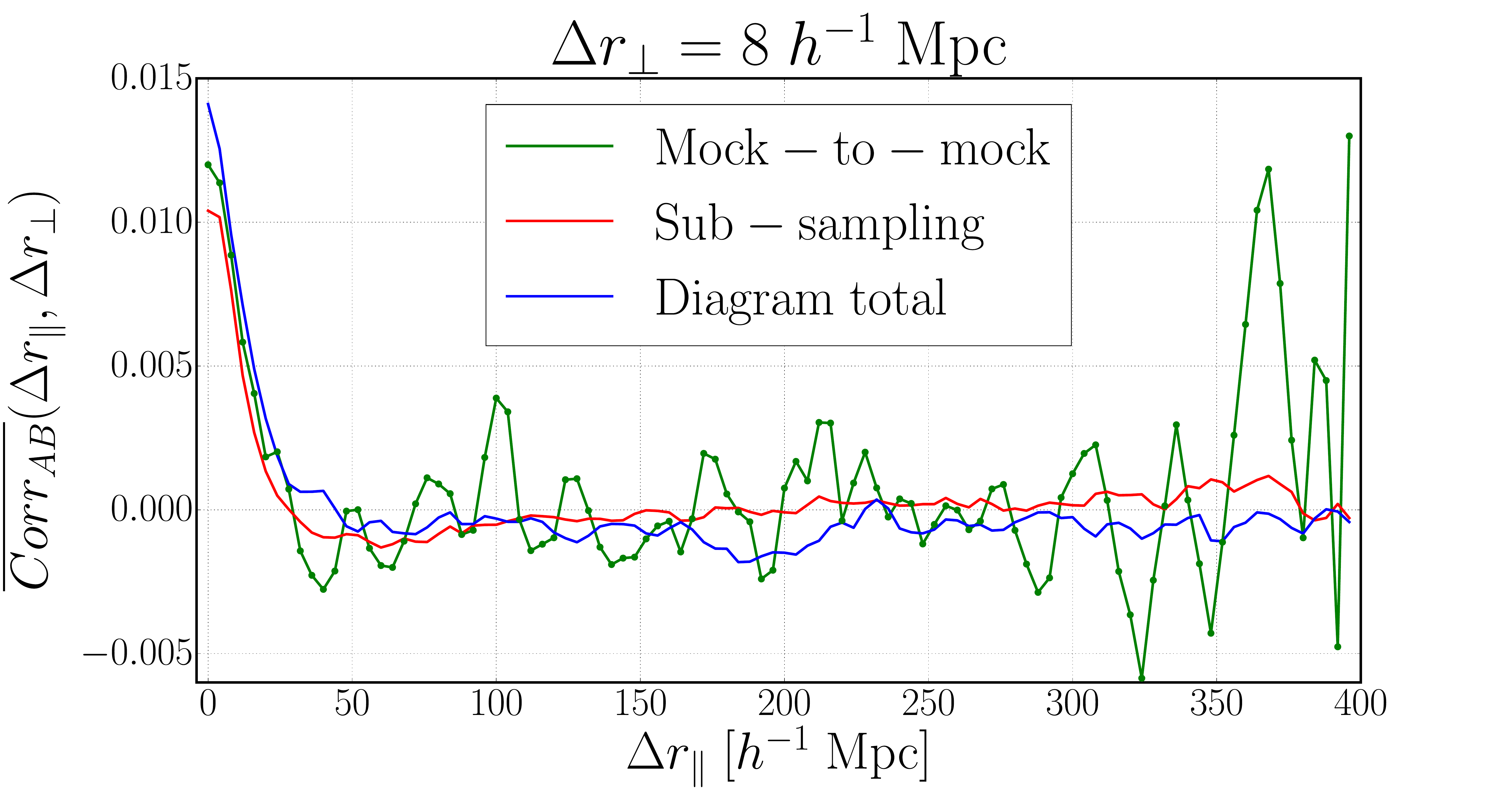}
        \caption{
          Mean normalized covariance matrix of the mocks,
          $Corr_{AB}\equiv C_{AB}/\sqrt{C_{AA}C_{BB}}$,
          as a function of
                $\Delta \rpar = |\rpar^{A}-\rpar^{B}|$ for the smallest values of
                $\Delta \rperp = |\rperp^{A}-\rperp^{B}|$.
                The top panels are for $\Delta \rperp=0$, with the right panel
                showing only the points with
                $Corr_{AB}<0.1$.
                The bottom two panels are for $\Delta \rperp=4~\hMpc$ (left)
                and $\Delta \rperp=8~\hMpc$ (right).
                Shown are the correlations given by the mock-to-mock, by the 
                mean of sub-sampling on each mock set, and, for one set,
                by the sum of all the diagrams.
        }
        \label{figure::correlation_matrix_mock}
\end{figure*}

\section{Validation of the analysis with mocks}
\label{section::mocks}

At the time of 
the cross-correlation analysis of \citet{2014JCAP...05..027F},
the only mock data sets that
were available \citep{2012JCAP...01..001F,2015JCAP...05..060B}
contained  \Lya~forests where the underlying density field
was traced by the transmission field
but not by the
associated quasars.
The forest-quasar cross-correlation therefore vanishes for these
mock data sets.
Because of this property, essential features
of the cross-correlation analysis could not be checked with the
analysis of the mock data.

For the analysis presented in this paper, we have produced a new
set of mock spectra where the \Lya{}  forests are properly correlated with
the quasars.  This correlation
is done using the technique of \citet{2011A&A...534A.135L}
where quasars are  placed at peaks
of a Gaussian-random field.
The transmission field generated with the same Gaussian field is
thus correctly correlated with the quasars.
These simulations are the first to include the four measured correlations:
the 1D correlation of two pixels in the same forest,
$\xiOneD$, the 3D auto-correlation of  pixels in different forests, $\xiff$,
the 3D auto-correlation of two quasars, $\xiqq$, and, most importantly
for this study, the 3D
cross-correlation of pixel-quasar on different forests, $\xiqf$.
A disadvantage of this approach is that the quasars
and forests are not at the same positions and redshifts
as those in the real data.  However, this technique should still allow
us to 
test the estimation of the covariance matrix and of the precision of
the fit parameters, and to search for systematic errors.

\subsection{Description of the mock data sets}

The production of the mock data sets proceeded as follows.
A Gaussian random field of density fluctuations is generated in a 
cubical volume of $79~(\hGpc)^{3}$
with the linear matter power-spectrum of CAMB \citep{2000ApJ...538..473L}.
We adopted a flat  \LCDM{} fiducial cosmology with
parameters given in Table \ref{table::cosmology_parameters}.
The corresponding box of line-of-sight velocities is generated at the same time.
This density field,  $\delta_{LR}$, has a low resolution since the cubic cells 
of the box have a side of $3.15~\hMpc$. We set the center of the box
at $z=2.5$, compute the resulting redshift in each cell, and multiply
$\delta_{LR}$ by the corresponding linear growth factor.
The velocities are also evolved to the redshift of the cell.
The size of the box along the line of sight corresponds to the redshift
range $1.71 < z < 3.66$ and the box covers $9078~\mathrm{deg^{2}}$ at $z=2.5$.

We draw quasar positions randomly within cells where the field is above
a threshold. This threshold is set such as to get a bias of $3.6$ relative
to matter distribution at $z=2.5$, which is consistent with the results of
\citet{2005MNRAS.356..415C}.
We do not vary the threshold with the redshift,
resulting in a real-space quasar correlation function that does not evolve with redshift
within the box.  This approach is a significantly better approximation 
than a constant bias.
A random selection of these cells is rejected in order to reproduce the
variation of the quasar number density with redshift.
Finally, the quasar redshifts are shifted according to the line-of-sight velocity of their cell.
The simulations up to the generation of the transmission
are then essentially as in \citet{2011A&A...534A.135L}.
One limitation is that the generated lines of sight are parallel.
We analyze the mock spectra accordingly, so in a slightly different way than the real data.

Our simulations average over a scale of $3.15~\hMpc$ and
therefore miss a significant amount of transverse
small-scale power in true forests, which are smoothed 
at the Jeans length, $\sim 100$~kpc.
To compensate for this lack of power,
20 high-resolution simulations with $16^3$ cells
of size $\sim0.2~\hMpc$ were performed.
The delta field from a randomly chosen  high-resolution simulation was added to
the delta of each large cell to provide the missing small-scale power.

The transmission, $F$, was computed as
\begin{equation}
        F = \exp \left[ -a(z)  \exp \left( b g(z) \delta \right) \right],
        \label{equation::gunn_peterson}
\end{equation}
where $\delta$ is the Gaussian field,
$g(z)$ is the linear growth factor, $b=1.58$ 
\citep{1997MNRAS.292...27H},
and  $a(z)$ is set to reproduce the measured 1D power spectrum
\citep{2006ApJS..163...80M}.

The next step is to take into account the effect of velocity field.
The transmission in each pixel of a spectrum is  transformed to the optical
depth, $\tau = - \ln{F}$, the pixel is moved in wavelength according to
the value of the velocity, and the value of the optical depth is modified
according to the gradient of the velocity. Finally, the optical depth
is transformed back to the transmission.
The resulting field  follows the Kaiser formula
(Eq. \ref{equation::bias_beta_definition}) in the range
$k<0.2~(\hMpc)^{-1}$ relevant for BAO, with a value of $\beta \approx 1.2$.

The mock expander, described in \citet{2015JCAP...05..060B},
transforms the transmission, $F$, to a  flux, $f$.
This process takes into account the
resolution of the SDSS-III BOSS spectrograph, the continuum and magnitude
properties of the BOSS quasars, and the level of noise of the data.

We also add absorption due to metal transitions near the \Lya{} transition:
SiII(126.0), SiIII(120.7), SiII(119.3), and SiII(119.0)
(Table \ref{table::metals_contribution}).
Absorption due to transitions far from the \Lya{} transition,
such as CIV(154.9),
are due to matter at low redshift, and are nearly uncorrelated
with the quasars in this study.
We use the ``procedure 1'' of \citet{2017A&A...603A..12B}
to generate absorption by metals.
Parameters of the metal transmission field
are set in order to reproduce their presence in the observed 
$\xiOneD$, the correlation between pixels of the same forest,
shown
in Figure \ref{figure::xi1D_compare_data_mock_with_lines}.
The peaks in the figure are due to correlations in absorption by
two different transitions at the same physical position.
The peaks present in the data but not in the mocks are due to 
$\mathrm{metal}_{1}$ -  $\mathrm{metal}_{2}$ correlations that
are not correctly modeled in the procedure.
These correlations have no effect on the quasar-forest cross-correlation.

Ten Gaussian-random-field boxes of $79~(\hGpc)^{3}$ volume were produced.
For each of them we use ten different random seeds to define the quasar
positions, which provides ten mock quasar catalogs. This approach is reasonable
since the quasars occupy only 1.1\% of the total number of cells above
threshold. When producing the \Lya{} spectra corresponding to the
resulting 100 mock quasar catalogs, different random seeds were used
for each quasar catalog, both for the noise and for the quasar continua.
Since our quasar and \LyaForest{} samples are strongly shot-noise limited,
the 100 sets of mock catalog and spectra are essentially uncorrelated.

For each of the 100 mock data sets, three types of spectra
were produced and analyzed.
This procedure allows us 
to understand the impact of the different
physical aspects and physical parameters
introduced along the mock production.
The results of the fits on the three types are summarized
in Table \ref{table::fit_mock_sets}.
The three types, in order of increasing realism,  are:
\begin{enumerate}

\item \Lya{}:
  The forest pixel values are the transmission field of the \Lya{}
  in the IGM. The quasar continuum,
  the metals absorption of the IGM, and the BOSS
  spectrograph resolution and noise
  are not introduced.
  When analyzing this type, the
        distortion matrix is set to the unit matrix.

      \item \Lya{}+Continuum:
        The quasar continuum, and the BOSS spectrograph
        resolution and noise
        are added to the \Lya{} mocks.
        This type allows us to understand and test our ability to
        model the distortion
        introduced by the lack of knowledge of the true continuum of the quasar.
        
      \item \Lya{}+Continuum+Metals: This type adds metals
        of Table \ref{table::metals_contribution} to the
        \Lya{}+Continuum mocks.
\end{enumerate}

Figures 
\ref{figure::xi_wedge_000_rescale2_compare_mocks_raw_cooked_no_met_with_fits}
and  \ref{figure::xi_slice_000_compare_mocks_raw_cooked_with_fits}
show stacks of the 100 mock sets for the three types.
Figure 
\ref{figure::xi_wedge_000_rescale2_compare_mocks_raw_cooked_no_met_with_fits}
illustrates how the  distortion matrix
$D_{AA^\prime}$  accounts for the change in the correlation
function due to continuum fitting.
Figure \ref{figure::xi_slice_000_compare_mocks_raw_cooked_with_fits}
shows the presence of the metals in the low $\rperp$ bins.

\subsection{Fits of individual mock sets}
\label{subsection::mockfits}

\label{subsection::Validation_of_the_covariance_matrix}

Individual mocks sets were analyzed with the aim of validating
the techniques used to analyze the data.
In particular, we wished to verify the accuracy of the
covariance matrix and search for biases in the determination
of the BAO peak position.

The covariance matrix for the data was calculated using the two
methods described in
Sect. \ref{subsection::The_Lya_forest_quasar_cross_correlation::The_covariance_matrix}.
One of the goals of the analysis of the mock spectra was to confirm
the validity of these methods by observing directly
the mock-to-mock variation of the correlation function.
The comparison of the  covariance determined by this direct
method with the two methods used for the data is shown
in Fig. \ref{figure::correlation_matrix_mock}.

The procedure for fitting the mock correlation function was
the same as that for the data with the following exceptions.
Because
only the linear power spectrum was used to generate
the mock spectra, we have $\FNL(\vec{k})=1$ and
$\VNL(k_{\parallel})=1$ for the mocks.
As stated previously, because of the size of
the cells of the mocks, we let free the two parameters
$\vec{R}=(R_{\parallel},R_{\perp})$.

The results of the fits of the 
100 mocks  are summarized in Table
\ref{table::fit_mock_sets}, which shows
the weighted mean of the best-fit values of  $\aparMath, \aperpMath,
\blya (1+\betalya)$ and $\betalya$.
Most importantly, the mean values of \aperp{} and \apar{}
are within 1\% of the expected value of unity, indicating
no significant bias in the determination of the BAO peak position.
The table gives the 
mean of the one-sigma errors of the four parameters.
These means are not far from those observed for fits of the data
(Table \ref{table::bestfit_best_model_fit_parameters_cross_auto_combined}).
The mean $\chi^2$ for the mock fits are near unity per degree of freedom,
confirming that the covariance matrix of $\xi(\rperp,\rpar)$ is well estimated.
The last column of the table lists the number of mocks sets with values
$\Delta\chi^2\equiv\chi^2(\aperpMath=\aparMath=1)-\chi^2_{\rm min}$ that
exceed 6.18.
This number is generally greater than 4.5, the number expected for Gaussian errors on (\aperp,\apar).
This result, confirmed by the Monte Carlo simulations of Appendix \ref{append::fastMC},
is unsurprising because the model is not a linear function
of these variables.

\begin{figure}
        % Codes/codes.py :: plot_compare_correlation_new_version_mocks()
        \centering
        \includegraphics[width=\columnwidth]{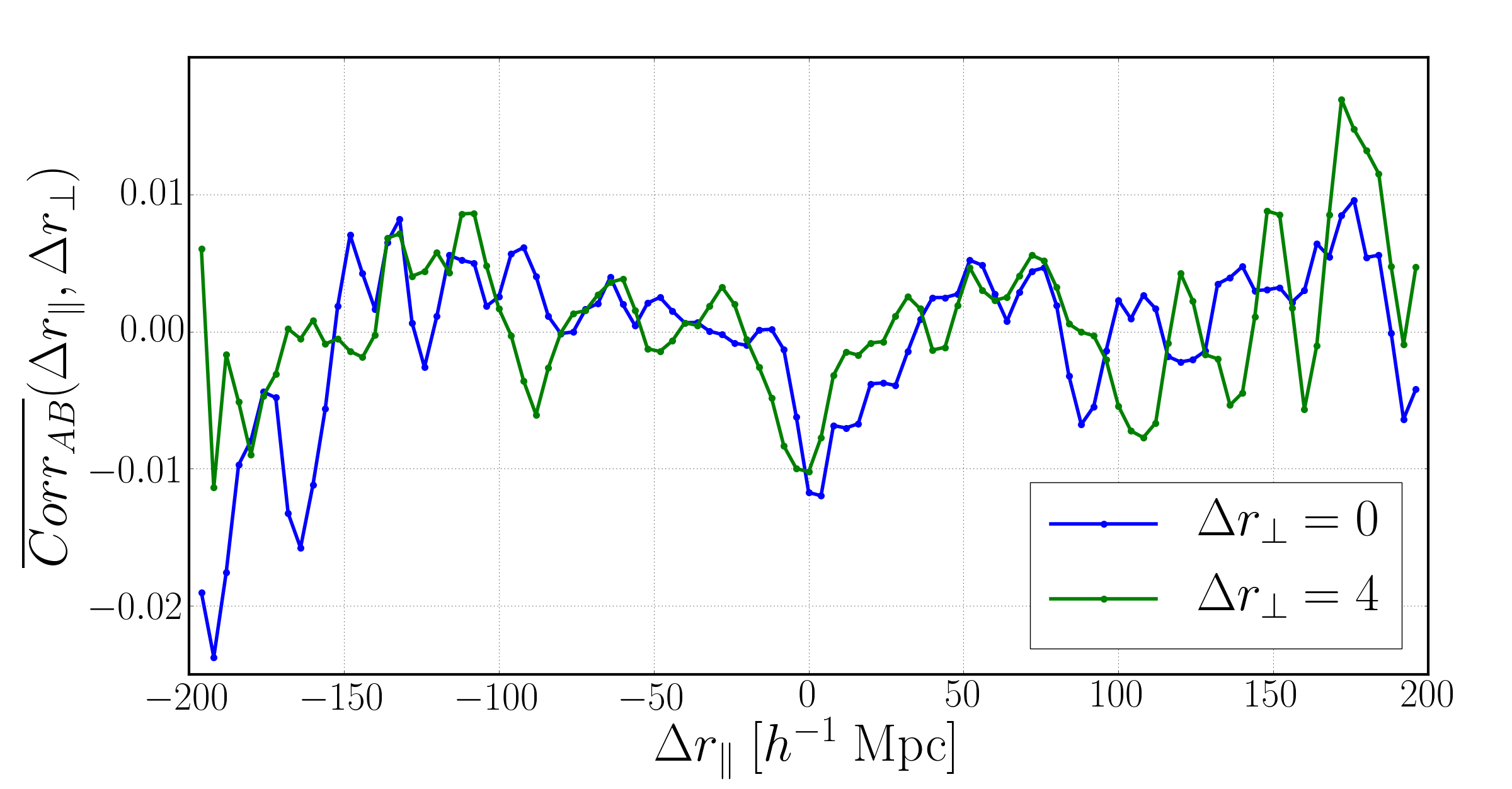}
        \caption{
          Mean normalized covariance,
$Corr_{AB}=C_{AB}/\sqrt{C_{AA}C_{BB}}$ of the
           auto- and cross-correlation functions in the
          two lowest $\rperp$ bins, as derived from mock-to-mock
          variations of the correlation function. 
        }
        \label{figure::correlation_matrix_mock_auto_cross}
\end{figure}

\subsection{Combined fits of the cross- and auto-correlation}
\label{subsection::Validation_of_the_cross_auto_covariance_matrix}

\begin{figure*}[ht]
  \centering
  \includegraphics[width=\columnwidth]{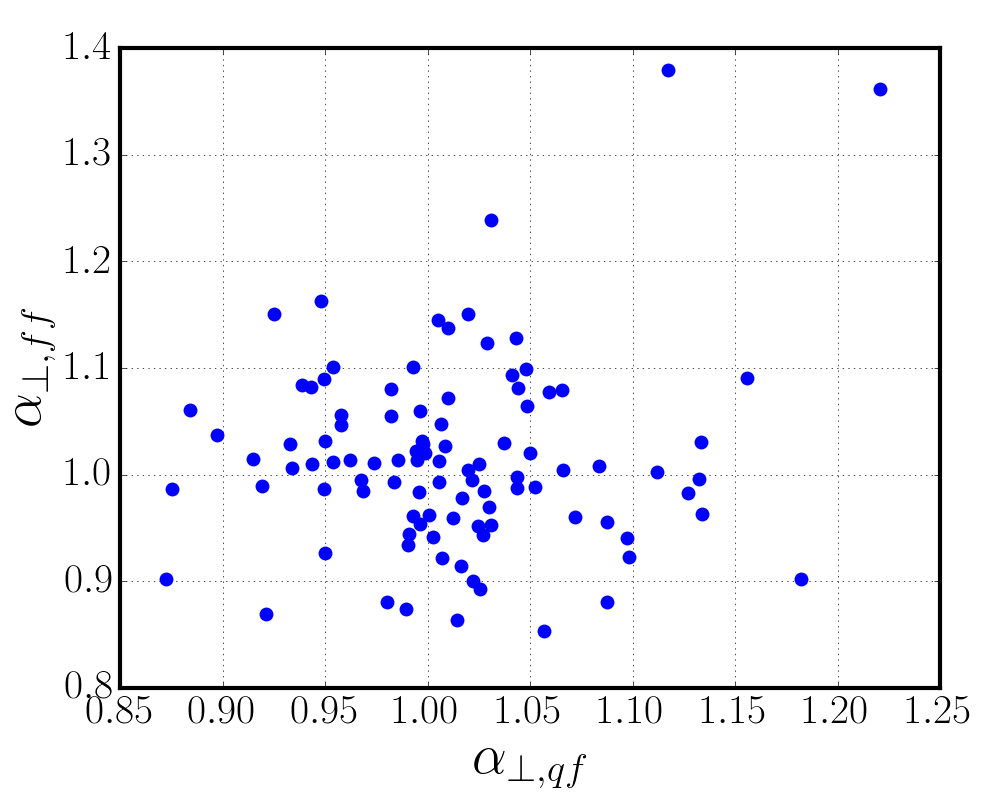}
  \includegraphics[width=\columnwidth]{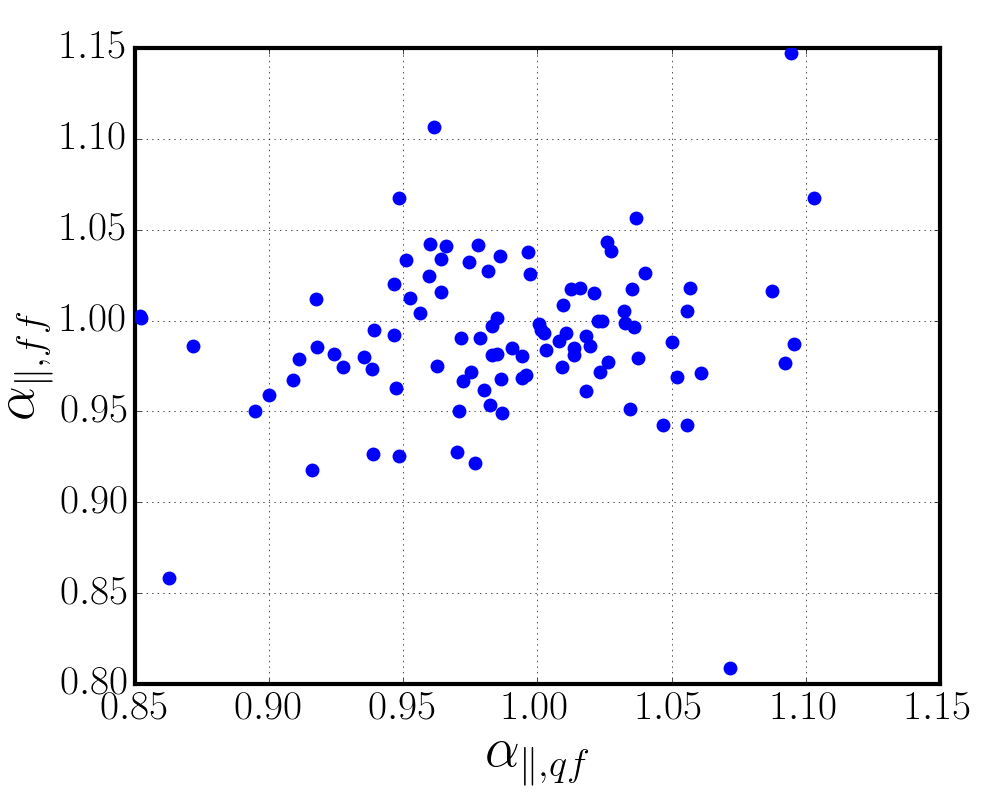}
  \caption{Scatter plot of the BAO peak position parameters
    measured with the cross-correlation versus those measured with the  auto-correlation
    for the 100 \Lya+Continuum+Metal mocks.
  }
  \label{figure::alpha_cross_auto}
\end{figure*}

As with the data, the cross- and auto-correlation functions of
the mocks can be combined either by performing a joint
fit of the two functions, or by combining the values
of $(\aperpMath,\aparMath)$ measured separately with the two functions.
The former requires the covariance matrix between the cross-
and auto-correlations while the latter requires 
the covariance of the two measurements
of $(\aperpMath,\aparMath)$.
The mock-to-mock variations of the auto- and cross-correlations shown
in Fig. \ref{figure::correlation_matrix_mock_auto_cross}
indicate that the covariance of the
two correlation functions is negligible.
The correlation of the auto and cross measurements
of $(\aperpMath,\aparMath)$ are presented in 
Fig.~\ref{figure::alpha_cross_auto} and Table  \ref{table::cross_auto_cor}.
As expected, they 
are consistent with zero. 

\begin{table}
        \centering
        
        \caption{
          Correlations between the $(\aperpMath,\aparMath)$ measured by
          the cross- and auto-correlation functions derived from
          the mock-to-mock variations of best-fit values for the 100 mocks.
        }
        \label{table::cross_auto_cor}
        
        \begin{tabular}{l l}
        \noalign{\smallskip} 
        \hline \hline
        \noalign{\smallskip}
$\rho(\aperpMath^{\rm cross},\aparMath^{\rm cross})$  & $-0.325  \pm 0.087$\\
$\rho(\aperpMath^{\rm auto},\aparMath^{\rm auto})$  &  $ -0.428  \pm 0.089$\\
        \noalign{\smallskip} 
        \hline \hline
        \noalign{\smallskip}
$\rho(\aperpMath^{\rm cross},\aperpMath^{\rm auto})$ & $0.004\pm0.096$\\ 
$\rho(\aparMath^{\rm cross},\aparMath^{\rm auto})$ & $0.11\pm0.11 $\\
$\rho(\aperpMath^{\rm cross},\aparMath^{\rm auto})$ & $-0.13\pm0.12$\\ 
$\rho(\aparMath^{\rm cross},\aperpMath^{\rm auto})$ & $0.093\pm0.09 $\\

        \end{tabular}
\end{table}

\section{Cosmological interpretation}
\label{section::cosmo}

\begin{figure}
        \centering
        \includegraphics[width=\columnwidth]{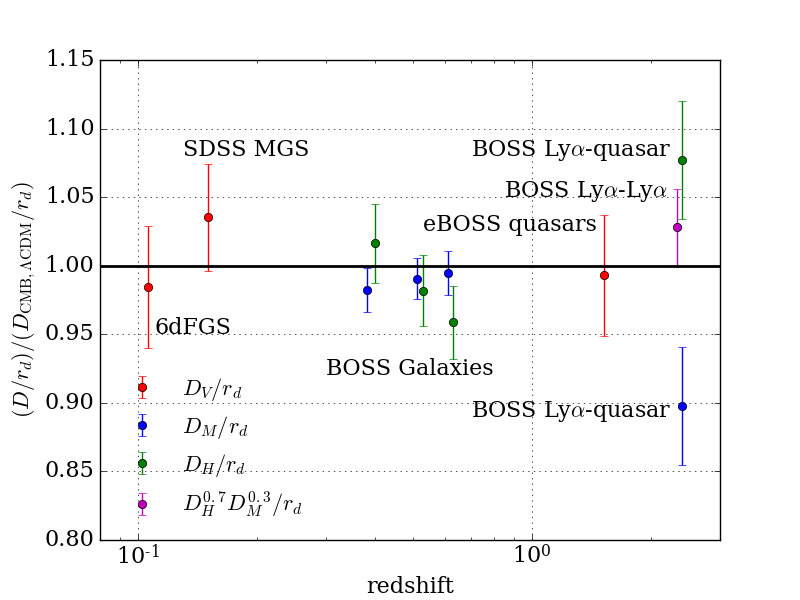}
% Work/Cosmology/plotbao.py
        \caption{BAO measurement of
          $\DMm(z)/r_d$ and $\DHh(z)/r_d$ and combinations thereof,
          compared
          to the prediction the  flat-\lcdm~model favored by CMB-anisotropy
          measurements \citep{2016A&A...594A..13P}.
          The BAO measurements come from the 6dFGS \citep{2011MNRAS.416.3017B},
          SDSS-MGS \citep{2015MNRAS.449..835R}, BOSS Galaxies
          (\citet{2017MNRAS.470.2617A}, ``BAO-only''),
          eBOSS quasars \citep{2017arXiv170506373A},
          the \Lya~forest flux auto-correlation
          \citep{2017A&A...603A..12B}, and the \Lya-quasar cross-correlation
          (this work).
        }
\label{figure::baosummary}
\end{figure}

\begin{figure}[ht]
  \centering
  \includegraphics[width=\columnwidth]{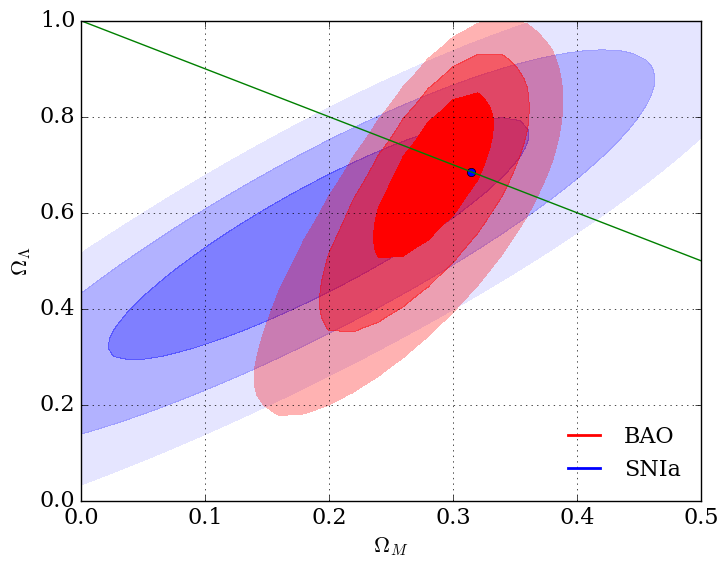}
% Work/Cosmology/omolcontours.png
  \caption{Constraints on $(\Omega_M,\Omega_\Lambda)$.
    The red  contours give the one, two, and three standard deviation constraints
    from the $D_V/r_d$ measurements of
    \citet{2011MNRAS.416.3017B} and \citet{2015MNRAS.449..835R}
    and the $(\DMm/r_d,\DHh/r_d)$ measurements of 
    \citet{2017MNRAS.470.2617A}, \citet{2017A&A...603A..12B}, and this work.
    The blue contours give the SNIa constraints of \citet{2014A&A...568A..22B}.
    The black point indicates the Planck flat-\lcdm~value
    of $(\Omega_M,\Omega_\Lambda)$.  This
    point has $\chi^2=14.8$ for $DOF=12$.
   }
  \label{figure::omolcontours}
\end{figure}

The measurements of the BAO peak with the \Lya~auto-correlation and
the quasar-\Lya~cross-correlation yield the constraints on
$\DMm(z\sim2.4)/r_d$ and $\DHh(z\sim2.4)/r_d$ that are presented in   
Fig. \ref{figure::chi2_scan_data_alpha_paral_alpha_perp_combined}.
The auto-correlation measurement of \citet{2017A&A...603A..12B}
produced a value of $\DHh^{0.7}\DMm^{0.3}/r_d$ about one standard deviation
from the flat-\LCDM~model that yields the CMB anisotropy spectrum
measured by \citet{2016A&A...594A..13P}.
The cross-correlation measurement presented here is 1.8 standard
deviations from the CMB prediction, and the combined measurement
differs by 2.3 standard deviations from this prediction.

While the results presented here represent ``tension'' with
CMB-inspired flat~\LCDM~model, the complete set of BAO
measurements presented in Fig. \ref{figure::baosummary} are in
good agreement with this model.
The CMB model has $\chi^2=14.8$ for 12 data points.
The contributions to this $\chi^2$ 
from the two low-redshift $D_V/r_d$ measurements are
$\Delta\chi^2=0.12$ \citep{2011MNRAS.416.3017B} and
$\Delta\chi^2=0.82$ \citep{2015MNRAS.449..835R}.
The measurements of $(\DMm/r_d,\DHh/r_d)$ at $0.2<z<0.8$ contribute
$5.40/6$~points  (\citet{2017MNRAS.470.2617A}, ``BAO-only'')
while the \Lya~auto-correlation at $z=2.33$ contributes
$2.18/2$~points 
\citep{2017A&A...603A..12B}.
The cross-correlation measurement presented here contributes
$6.27/2$~points, corresponding to a $1.8\sigma$ deviation
from the \LCDM~values.
This tension has no
simple, well-motivated solution \citep{2015PhRvD..92l3516A}, which suggests
that it results from a statistical fluctuation.

The BAO measurements by themselves yield the constraints
on the \LCDM~parameters $(\om,\ol)$ shown in Fig. \ref{figure::omolcontours}.
The flat-\LCDM~CMB-inspired model is about one standard deviation
from the best fit,
which has $\chi^2=12.5$ for $(12-3)$ degrees of freedom and the best-fit parameters:
\begin{multline}
  \om = 0.288\pm0.033 \hspace*{5mm}
  \ol = 0.695\pm0.115 \hspace*{5mm}
  \Omega_k = 0.02\pm0.14 \\ 
  H_0 \frac{r_d}{147.33~{\rm Mpc}}=(68.5\pm1.5)~{\rm km\,s^{-1}Mpc^{-1}} .
\label{baoomegas}
\end{multline}
Imposing $\Omega_k=0$ results in $\om=0.292\pm0.019$,
in good agreement with the CMB value
$\om=0.315\pm0.017$ \citep{2016A&A...594A..13P}.

The BAO best-fit values (\ref{baoomegas}) use the
primary \Lya~auto-correlation
result without a broadband added to the correlation function.
Inclusion of a broadband for the auto-correlation changes
the best-fit values by $\sim0.3\sigma$:
$\om=0.275\pm0.034$, $\ol=0.657\pm0.125$, and $\Omega_k=0.07\pm0.15$.

While the result (\ref{baoomegas})
strongly disfavors matter-only models (i.e., $\ol=0$),
it does not strongly imply that the expansion
is accelerating at the present epoch.
This is because
we have used data at $z>1$ where the expansion
was decelerating, so any statement about present-day acceleration
is model-dependent.
A recent report \citep{2016NatSR...635596N}
that low redshift measurements require
acceleration only at $<3\sigma$
significance
stimulated a re-examination of the evidence.
The general conclusion is that, in the absence of unidentified
luminosity evolution, the SNIa data \citep{2014A&A...568A..22B}
do support acceleration at $>4\sigma$ significance
\citep{2016ApJ...833L..30R,2017A&A...600L...1H,2017A&A...602A..73T}.
The BAO data do not provide such
precision because at low redshift
the number of available galaxies to measure the correlation
function is small.
If one uses
the four BAO data points
in Fig. \ref{figure::baosummary} with $z<0.4$,
one finds that the best non-accelerating model ($q_0=\om/2-\ol>0$)
has $(\om,\ol)=(0,0)$ with $\chi^2=8.1$.
This can be compared with $\chi^2=0.7$ for the best-fit model
and $\chi^2=2.1$ for the Planck-2016 model.
Acceleration is preferred at the $2.5\sigma$ level.

\section{Summary and conclusions}
\label{section::conclusion}

Using the entirety of the BOSS data set,
this paper has presented a measurement of the
cross-correlation of quasars and the \Lya~flux transmission
at redshift 2.4.
Apart from the improved statistical precision over our previous
measurement, we have benefited from an improved pipeline and better
modeling of the effects of continuum fitting.
The availability of mock data sets with quasar-forest correlations
was essential for verifying the reliability of the analysis.

  The modeling of continuum-fitting distortions done here
  opens up the possibility of constraining cosmology with
  the full correlation function, in addition to the BAO peak.
  However, this would require further studies to determine the sensitivity
  of such constraints to poorly constrained astrophysics:
  DLA absorption, UV fluctuations, and the transverse proximity
  effect.  These studies would probably require an 
  analysis of  the data  with multiple redshift bins.
  Relativistic effects \citep{2014PhRvD..89h3535B,2016JCAP...02..051I}
  should also be included in the model.
Further improvement on these results will be forthcoming from the
ongoing eBOSS project \citep{2016AJ....151...44D} and
the upcoming DESI
\citep{2016arXiv161100036D},
HETDEX
\citep{2008ASPC..399..115H}
and WEAVE
\citep{2016sf2a.conf..259P} projects.

The position of the BAO peak is $1.8\sigma$ from
the flat-\lcdm~model favored by CMB anisotropy measurements \citep{2016A&A...594A..13P}.
Combined with the \Lya-flux-transmission auto-correlation
measurement of \citet{2017A&A...603A..12B},
the BAO peak at $z=2.4$ is $2.3\sigma$ from the expected
value.
Despite  this tension, the ensemble of BAO measurements is in good agreement
with the CMB-inspired flat-\lcdm~model.
The measured auto- and cross-correlation, 
the best-fit results, and $\chi^{2}$ scan
are publicly available.\footnote{https://github.com/igmhub/picca/data/duMasdesBourbouxetal2017/.}

\begin{acknowledgements}

We thank Christophe Magneville for help in the production of the mock
data sets.

Funding for SDSS-III has been provided by the Alfred~P. Sloan Foundation,
the Participating Institutions, the National Science Foundation, and
the U.S. Department of Energy Office of Science. The SDSS-III web
site is http://www.sdss3.org/.

SDSS-III is managed by the Astrophysical Research Consortium for the
Participating Institutions of the SDSS-III Collaboration including the
University of Arizona, the Brazilian Participation Group,
Brookhaven National Laboratory, Carnegie Mellon University,
University of Florida, the French Participation Group,
the German Participation Group, Harvard University,
the Instituto de Astrofisica de Canarias,
the Michigan State/Notre Dame/JINA Participation Group,
Johns Hopkins University, Lawrence Berkeley National Laboratory,
Max Planck Institute for Astrophysics,
Max Planck Institute for Extraterrestrial Physics,
New Mexico State University, New York University,
Ohio State University, Pennsylvania State University,
University of Portsmouth, Princeton University,
the Spanish Participation Group, University of Tokyo,
University of Utah, Vanderbilt University,
University of Virginia, University of Washington, and Yale University.

The French Participation Group of SDSS-III was supported by the
Agence Nationale de la Recherche under contracts
ANR-08-BLAN-0222 and ANR-12-BS05-0015-01.
M.B., M.M.P., and I.P. were supported by the A*MIDEX project (ANR- 11-IDEX-0001-02) funded by the “Investissements d’Avenir” French Government program, managed by the French National Research Agency (ANR), and by ANR under contract ANR-14-ACHN-0021”.
A.F.R. and N.P.R. acknowledge support from the STFC and the Ernest Rutherford Fellowship scheme.

\end{acknowledgements}

\bibliographystyle{aa}
\bibliography{crossDR12}

\appendix

\section{Covariance matrix}
\label{appendix::covariance_matrix}

\begin{figure}
        \centering
        \includegraphics[width=\columnwidth]{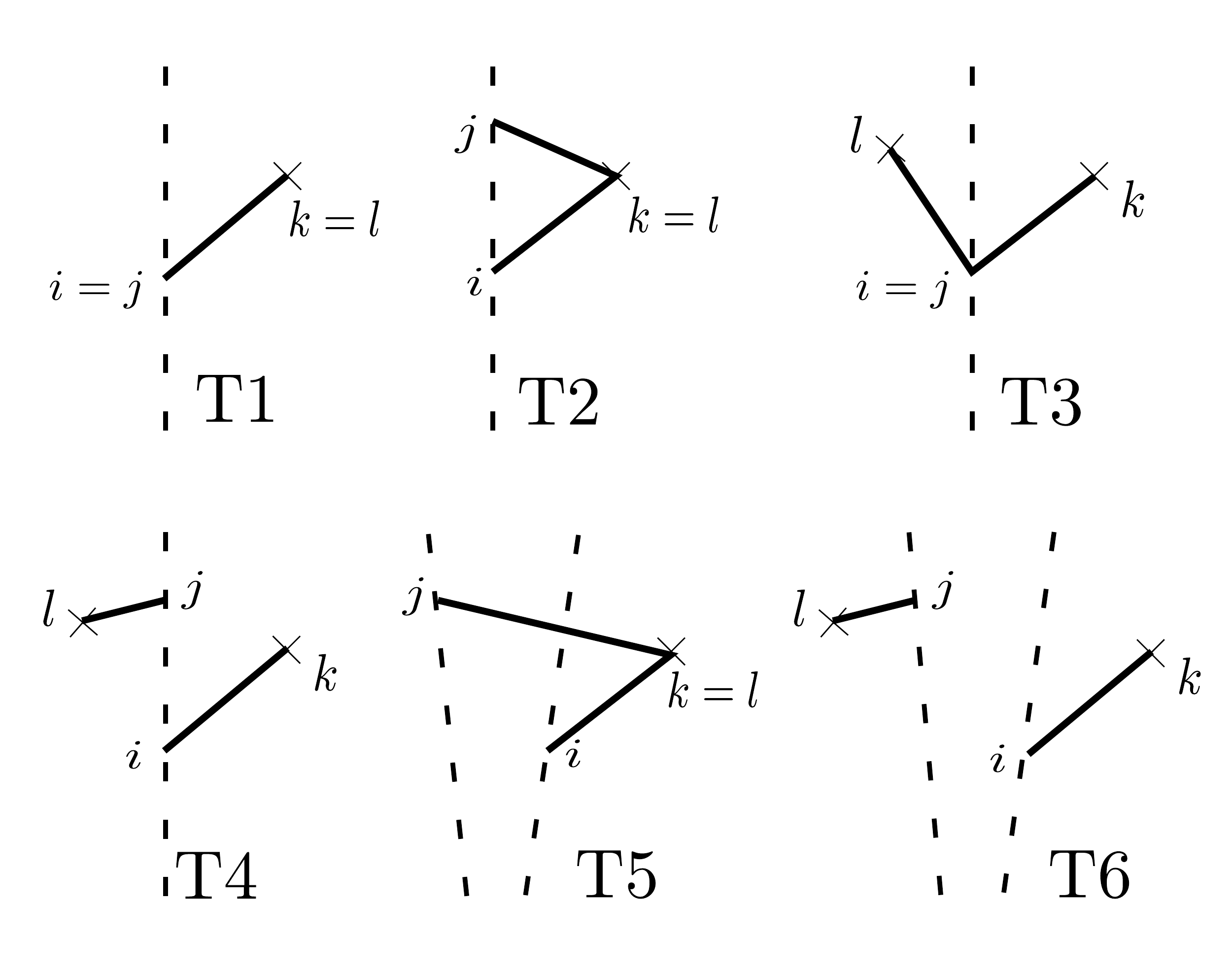}
        \caption{
                Six different diagrams of pairs of pixel-quasar pairs.
                The dashed lines refer to the forests, the crosses refer to the
                quasar position. The variance is dominated by T1 and T2.
                The off-diagonal terms of the covariance are dominated by T2.
                The diagrams T3 and T4 cancel out at large scale, the diagram T5 has a small
                contribution at small scales and T6 is negligible.      
        }
        \label{figure::appendix::covariance_matrix::covariance_diagram}
\end{figure}

\begin{figure*}
        %
        % Codes/codes.py :: plot_compare_correlation_new_version()
        %
        \centering
        \includegraphics[width=\columnwidth]{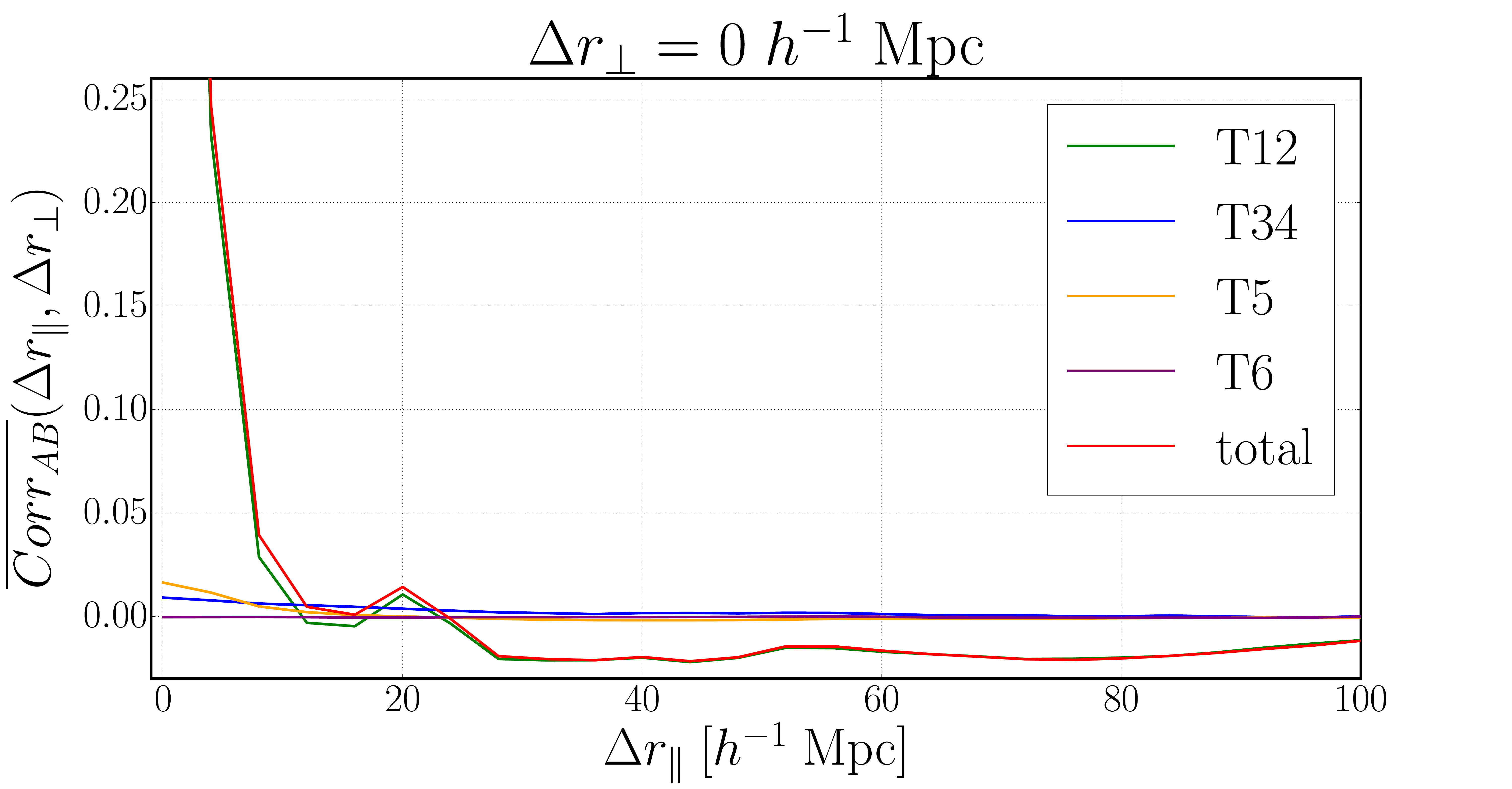}
        \includegraphics[width=\columnwidth]{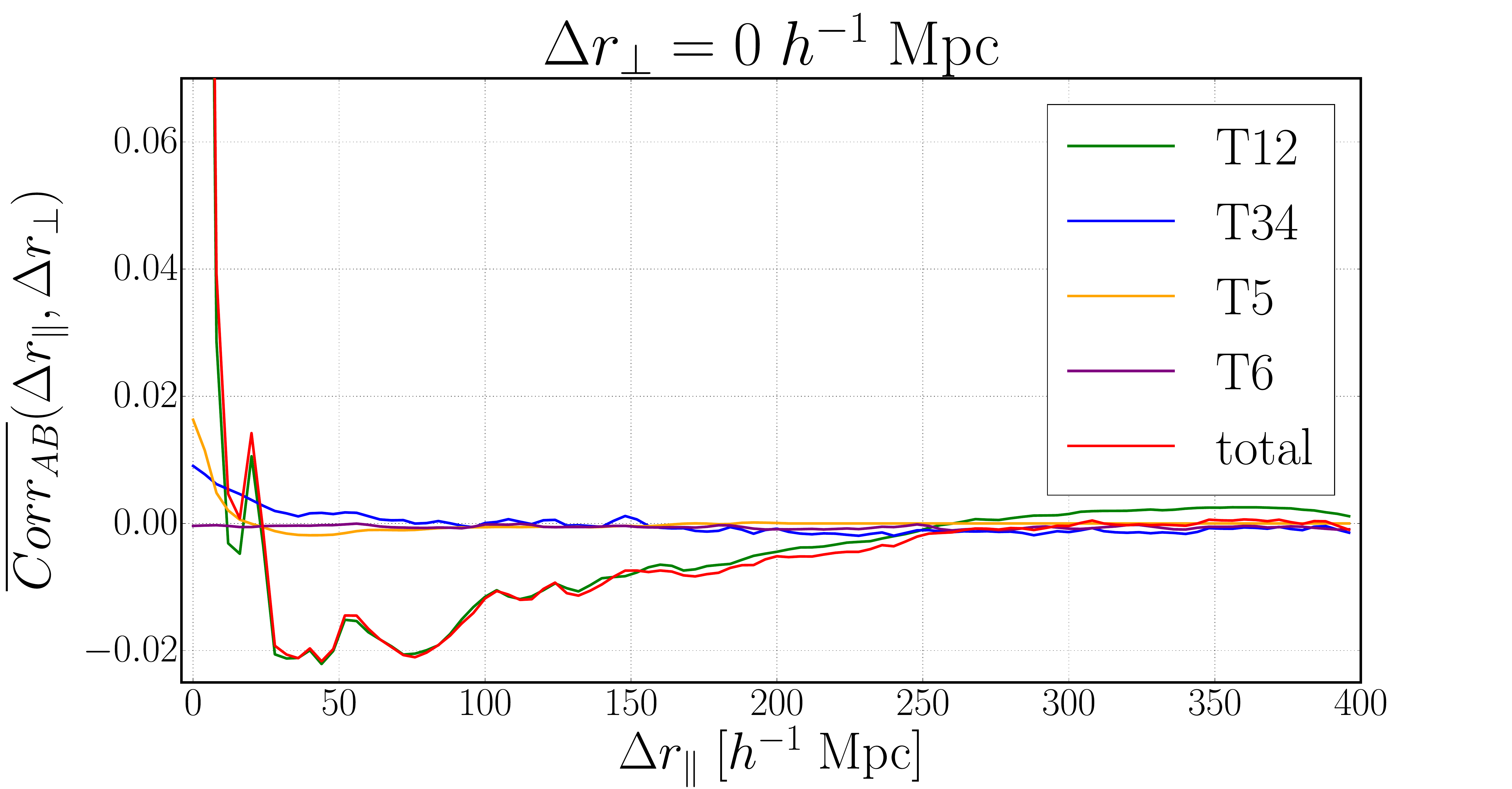}
        \includegraphics[width=\columnwidth]{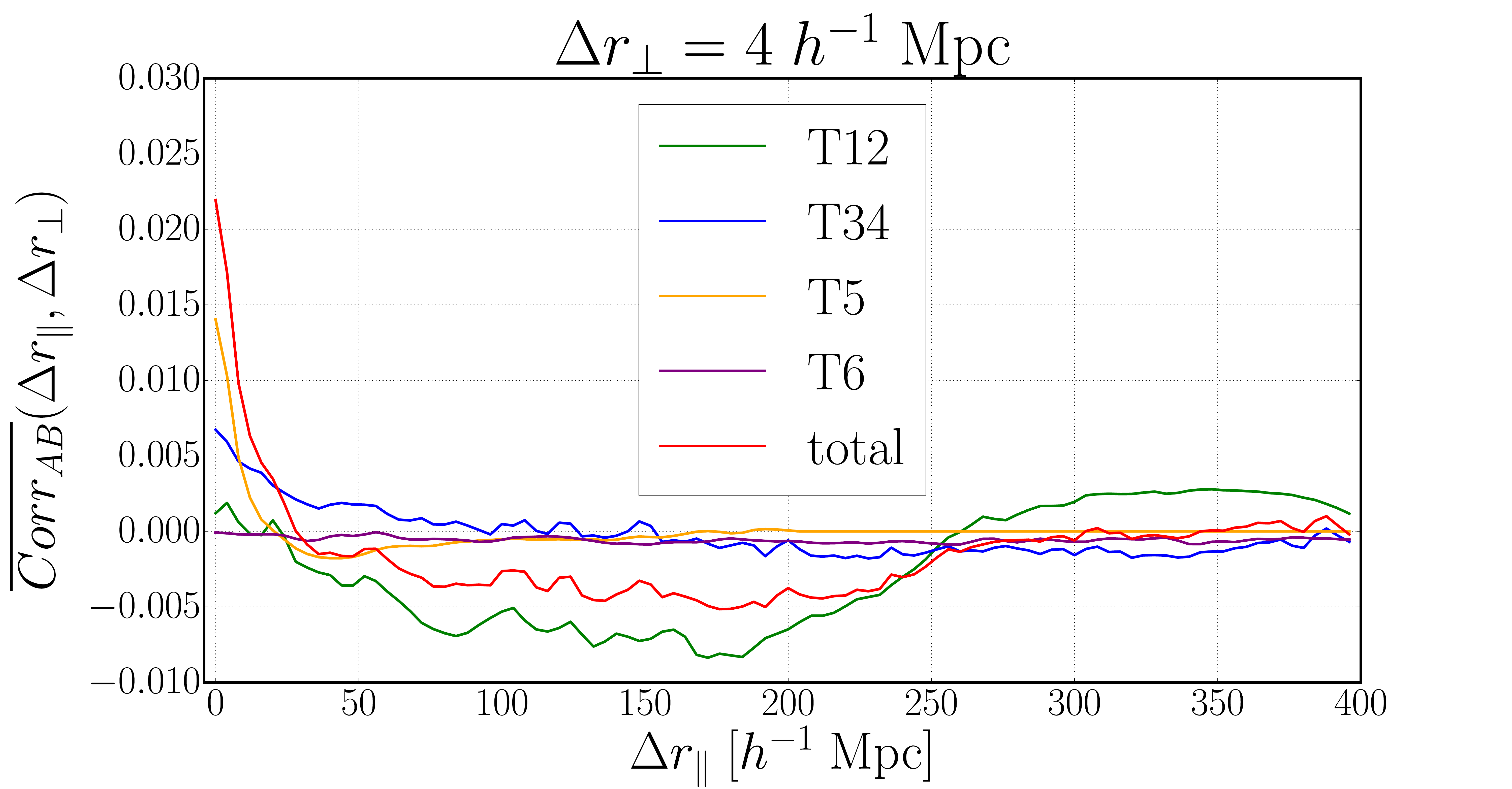}
        \includegraphics[width=\columnwidth]{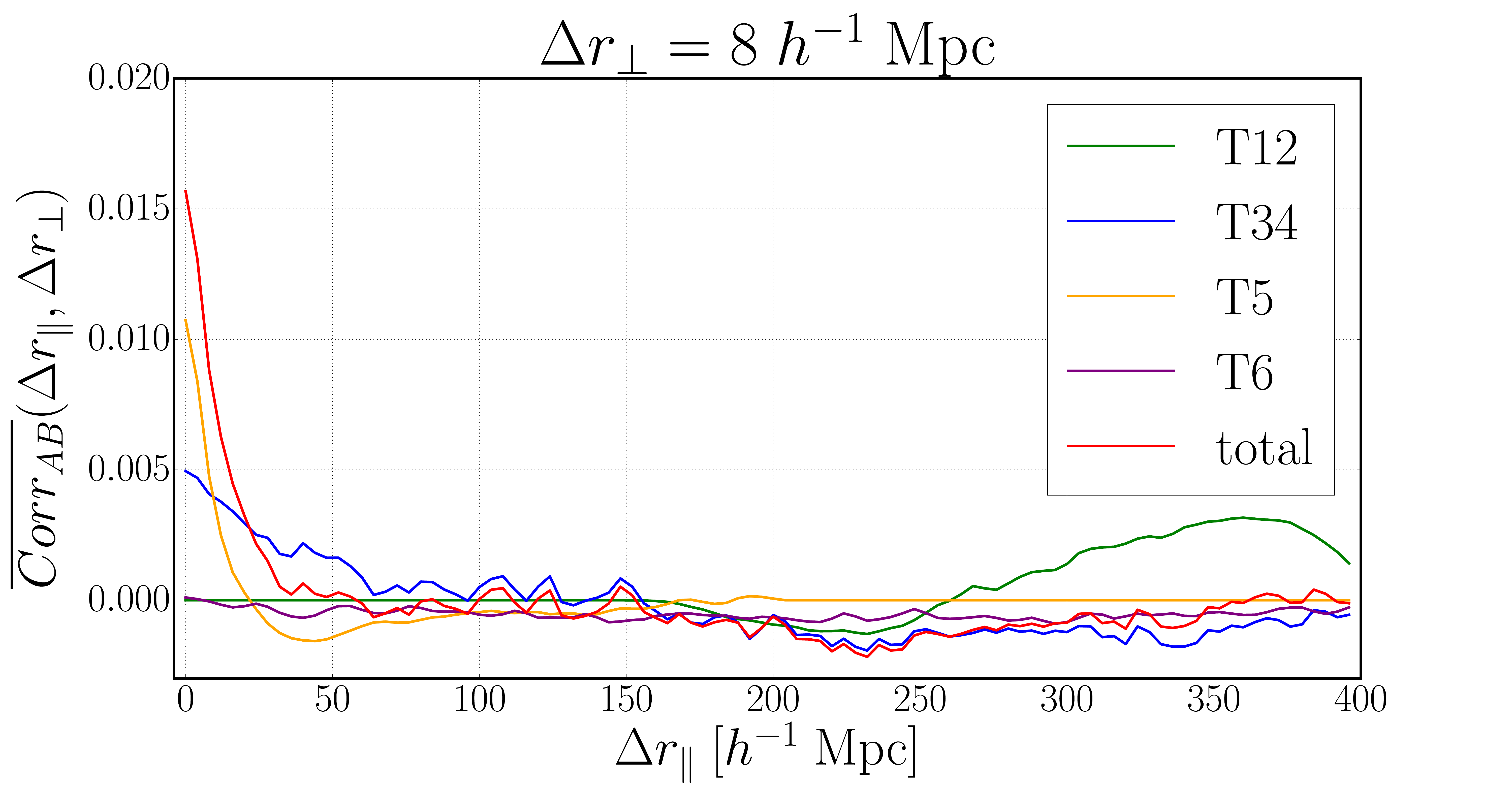}
        \caption{
          Mean normalized covariance matrix of the data,
          $Corr_{AB}=C_{AB}/\sqrt{C_{AA}C_{BB}}$ as a function of
                $\Delta \rpar = |\rpar^{A}-\rpar^{B}|$ for the smallest values of $\Delta \rperp = |\rperp^{A}-\rperp^{B}|$.
                The top panels are for $\Delta \rperp=0$ with the right panel
                displaying only the points with
                $Corr_{AB}<0.1$.
                The bottom panels are for $\Delta \rperp=4~\hMpc$ (left)
                and $\Delta \rperp=8~\hMpc$ (right).
                Shown are the correlations given by each diagram: T12 (green), T34 (blue), T5 (yellow), and T6 (purple),
                along with the sum of all the diagrams (red).
        }
        \label{figure::appendix::correlation_matrix_in_listDiagrams}
\end{figure*}

The calculation of $C_{A B}$
via Equation \ref{equation::appendix::xi_cross_covar_estimator_2}
can be decomposed into
six different diagrams, presented in Figure
\ref{figure::appendix::covariance_matrix::covariance_diagram}.
In the six diagrams of this figure, the dashed lines indicate the \Lya{}-forests
and the crosses indicate  the quasars.
In Diagrams T1 and T3, the two pixel-quasar pairs share the same pixel,
$i=j$. In these cases the pixel-pixel correlation is given by the variance
of pixels at its observed wavelength:
$\langle\delta_{i} \delta_{i}\rangle = \xiOneD(\lambda_{i},\lambda_i)$.
In Diagrams T2 and T4, the two pixels from the two pairs belong to the
same forest. Here  the pixel-pixel correlation is given by the
1D correlation:
$\langle\delta_{i} \delta_{j}\rangle = \xiOneD(\lambda_{i},\lambda_{j}/\lambda_{i})$.
This correlation is presented for the simulations and for the data in
Figure \ref{figure::xi1D_compare_data_mock_with_lines}.
In Diagrams T5 and T6, the two pixels belong to different forests,
and the pixel-pixel correlation is given by the 3D \Lya{}-forest
auto-correlation:
$\langle\delta_{i} \delta_{j}\rangle = \xiff(\vec{r}_{ij})$.
This correlation is studied in
\citet{2017A&A...603A..12B}.

Figure \ref{figure::appendix::correlation_matrix_in_listDiagrams} presents
the contribution to the correlation
matrix (Eq. \ref{equation::correlation_matrix}) of the
six diagrams and their sum.
The elements of $\overline{Corr_{AB}}$ are given
as a function of $\Delta \rpar = |\rpar^{\,A}-\rpar^{\,B}|$
for the smallest values of $\Delta \rperp = |\rperp^{\,A}-\rperp^{\,B}|$.
The top left panel
shows the correlation matrix for $\Delta \rperp=0$,
and the top right panel displays an expanded image. These two panels
are the reflection of the $\xiOneD$ presented in Figure
\ref{figure::xi1D_compare_data_mock_with_lines}. Some of the  \Lya-metal
lines are visible.
The bottom left and right panels display the correlation matrix for
$\Delta \rperp= 4~\hMpc$ and for $\Delta \rperp= 8~\hMpc$.
As expected, these correlations
are very small.

Diagrams
T1 and T2 dominate the variance $C_{AA}$ and T2 the covariance $C_{AB}$ when
the bins $A$ and $B$ have similar transverse separation $\rperp$.
T2 vanishes for very different $\rperp$.
Due to the projection of the $\delta$ (Sect.
\ref{section::Measurement_of_the_transmission_field}),
Diagrams T3 and T4 have only a maximum sub-percent contribution to the correlation matrix
and cancel out at large scale. T5 has a small contribution at small scales
and T6 is negligible.

\begin{figure*}
        % Work/Data/Cross_alone/Fits_pyLya/Fit_rmin10_rmax160_metals_laurentzian_HCD_UV_QSORadiation
        % Work/Data/Cross_alone/Fits_pyLya/Fit_rmin10_rmax160_metals
        % Work/Data/Cross_alone/Fits_pyLya/Fit_rmin10_rmax160_metals_laurentzian_HCD_UV_QSORadiation_0_2_0_6_distort/

        \centering
        \includegraphics[width=\columnwidth]{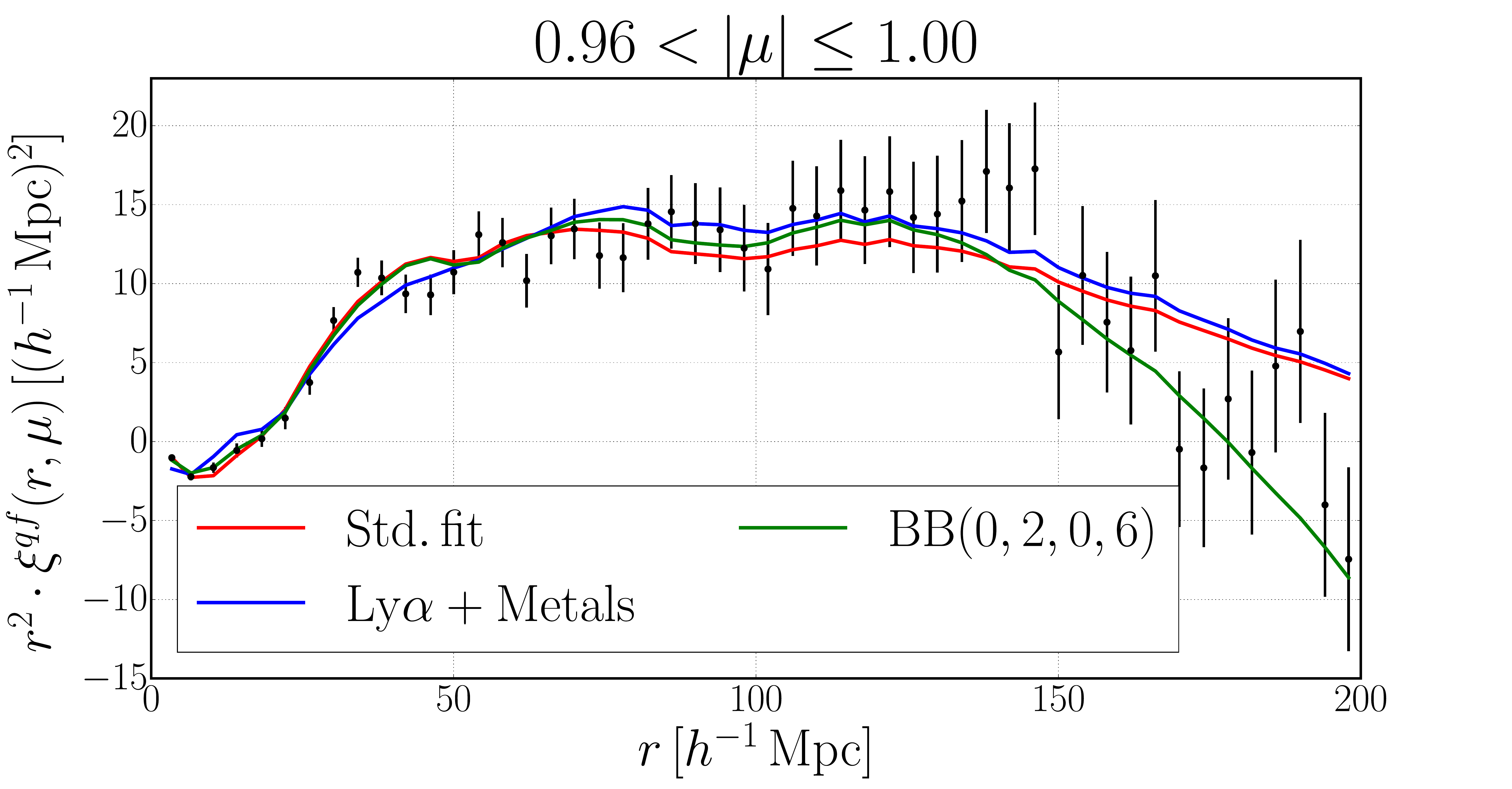}
        \includegraphics[width=\columnwidth]{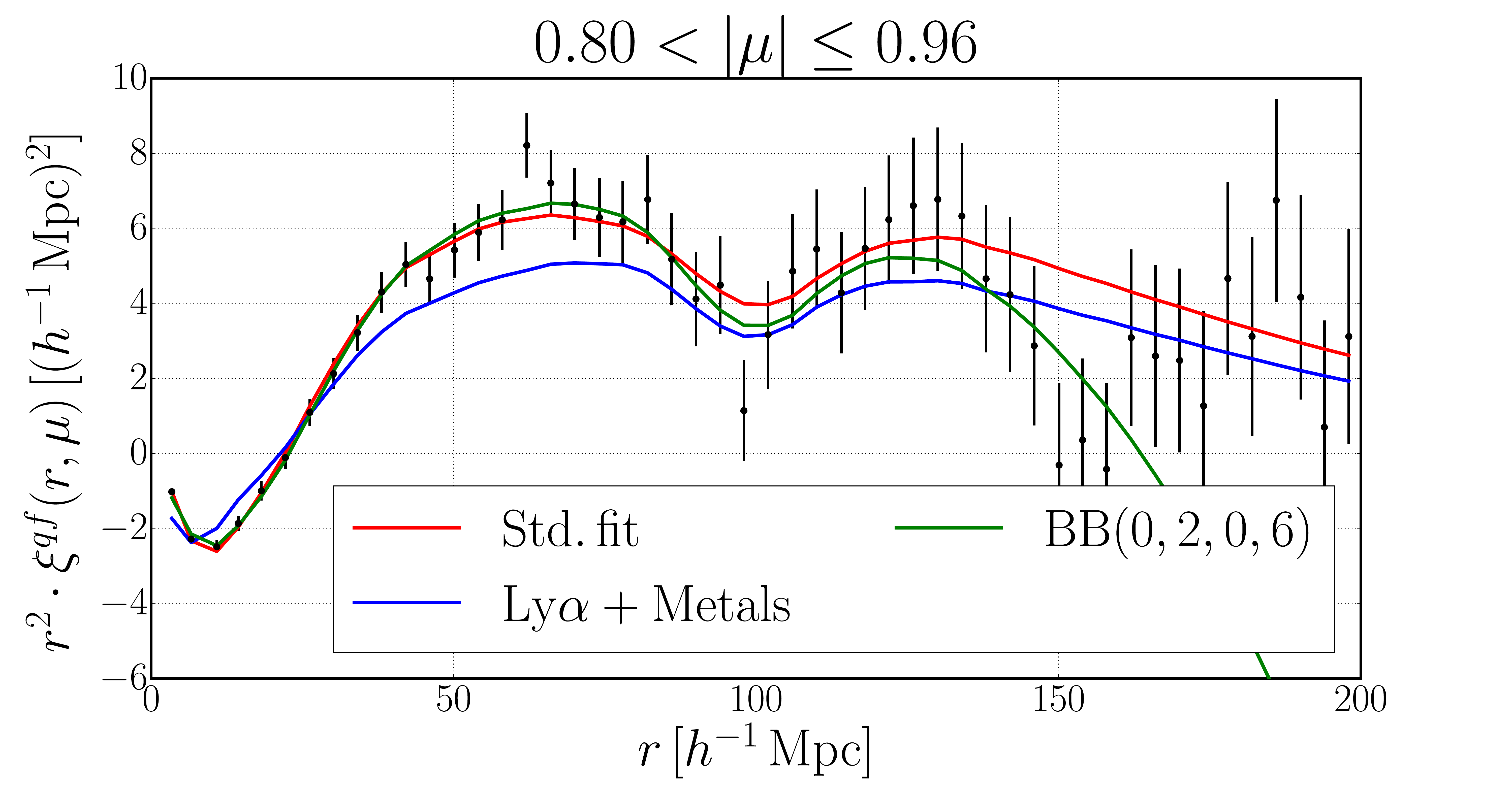}
        \includegraphics[width=\columnwidth]{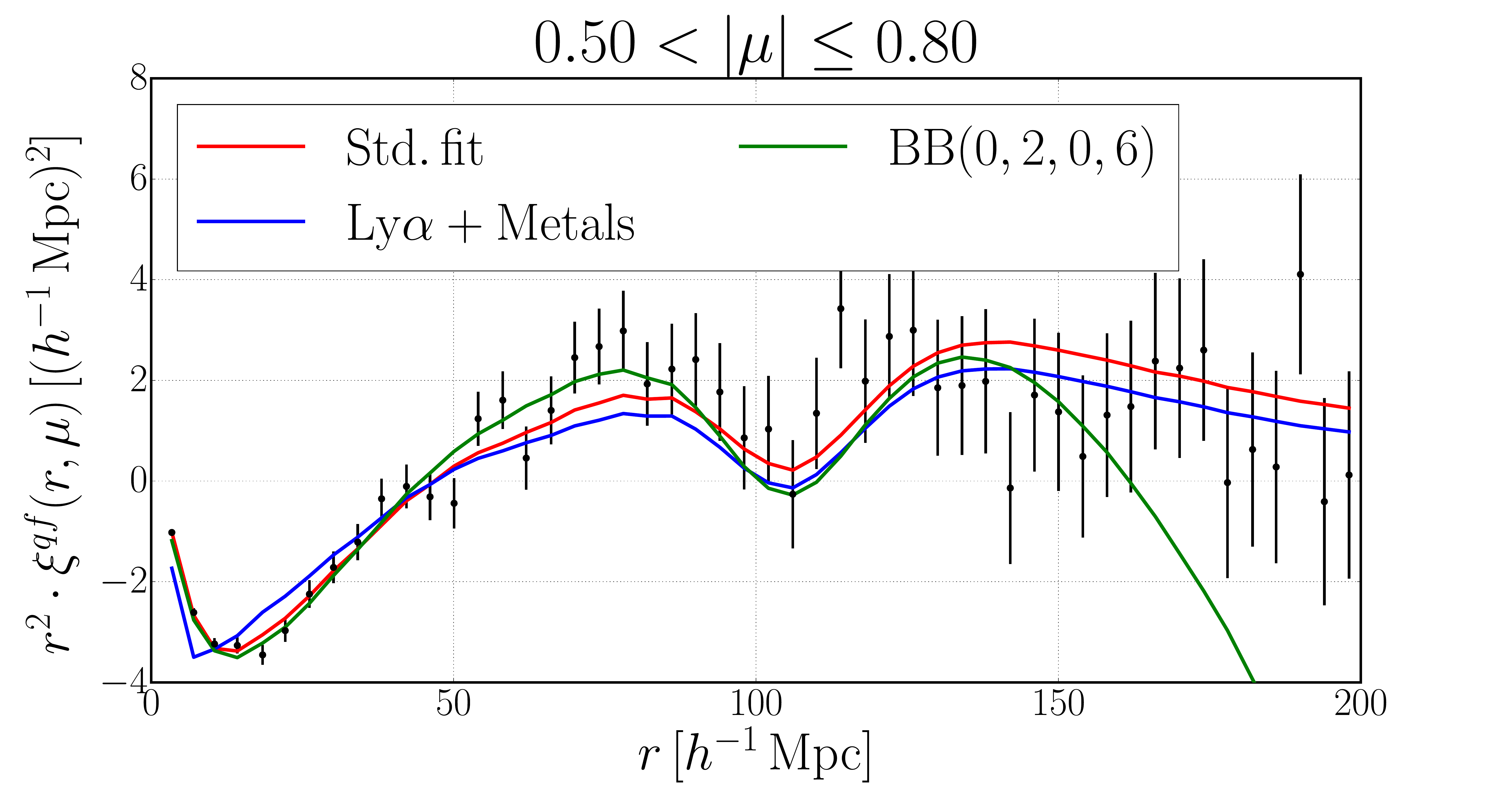}
        \includegraphics[width=\columnwidth]{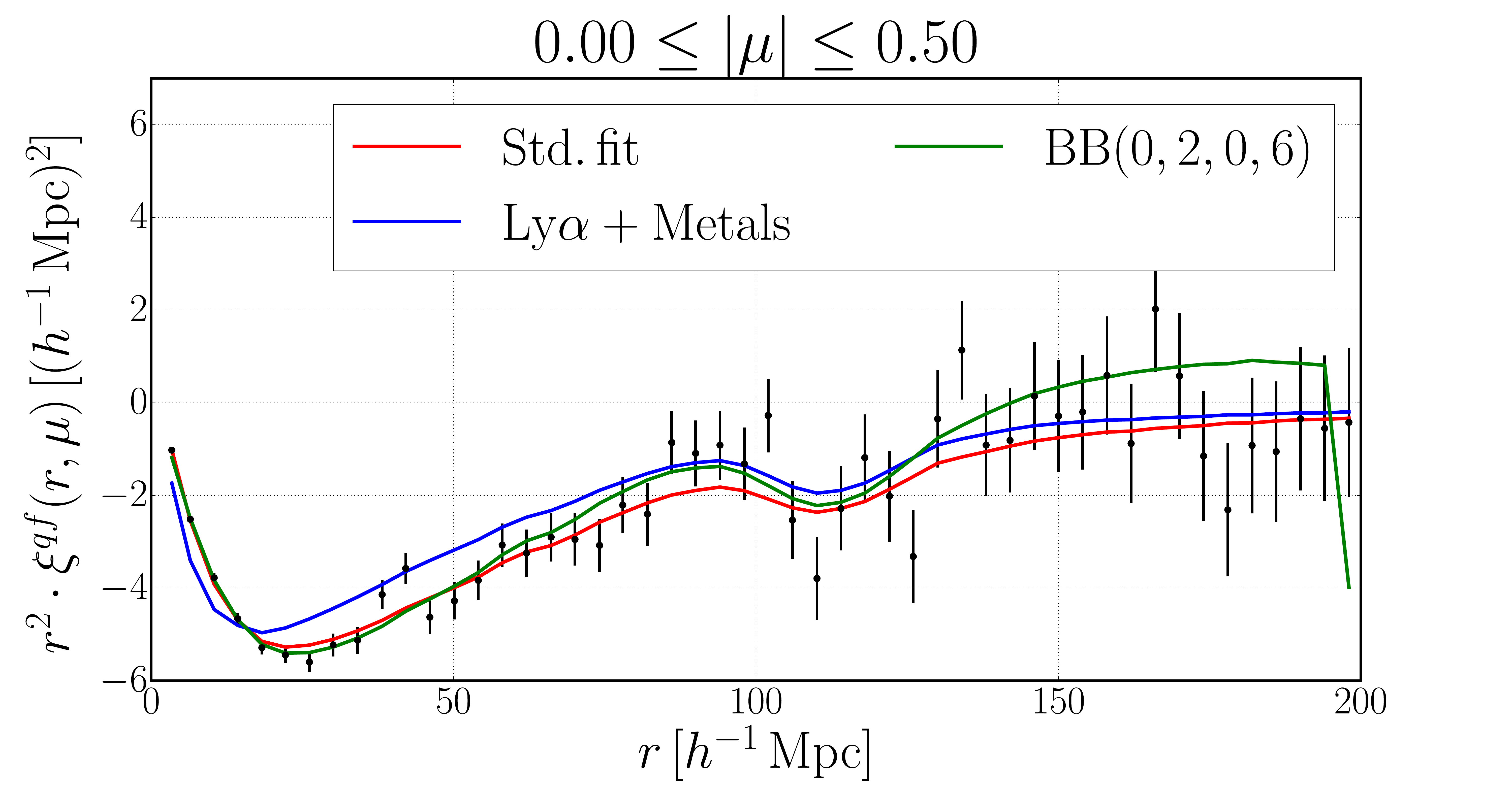}
        \caption{
          As in Fig. \ref{figure::xi_data_best_fit_4_weges_mu} but showing
          three models fit to the data.
          Red lines indicate the
                standard fit used to measure the BAO parameters
                (Sect. \ref{subsection::Fits_for_the_peak_position::Results}),
                blue lines the standard fit if the velocity distribution
                of quasars was null, and green lines  the standard fit
                with the addition of a broadband defined by
                (\ref{equation::broadband_1}) with
                $(\imin,\imax,\jmin,\jmax)=(0,2,0,6)$. 
                The cross-correlation is multiplied by a factor $r^{2}$ to show the BAO scale.
        }
        \label{figure::xi_wedge_000_rescale2_data}
\end{figure*}

\section{Non-standard fits of the correlation function}
\label{append::nonstandardfits}

In order to test the robustness of our measurement of (\aperp,\apar), we performed
fits in which the standard model was modified.
All fits yield compatible
values and precision of the two BAO parameters, and provide
confidence in the validity of our measurement.

For all of these models, Table \ref{table::fit_data_model} lists
the best-fit values of the four main parameters:  \apar, \aperp,
$b_{\mathrm{Ly}\alpha} (1+\beta_{\mathrm{Ly}\alpha}),$ and $\beta_{\LyaMath}$,
as well as
the $\chiSquareMin$ of the fit.
The first group of fits gives the results for increasingly
complicated physical models, starting with a model with
only \Lya{}~absorption and including successively metals, HCDs, UV
fluctuations, and the transverse proximity effect.
The last of this group,
\Lya{}+Metals+$z_q$+HCD+UV+TP, corresponds to the standard fit
of Sect. \ref{subsection::Fits_for_the_peak_position::Results}.
The first two fits
("\Lya{}" and "$\mathrm{Ly\alpha+Metals}$"),
which do not take into account the velocity
dispersion of quasars, have a
high $\chi^{2}_{\min}$.
Adding this effect ($z_q$ in Table \ref{table::fit_data_model})
reduces significantly $\chi^{2}_{\min}$,
but the best-fit values and precision of the two
BAO parameters do not change
significantly with successive models.
However, due to correlations and degeneracies, the values and the precision
of the bias and RSD parameters of the \Lya{} field change significantly
when adding the velocity dispersion of the quasars.

The second section of Table \ref{table::fit_data_model} presents the results
with different fitting ranges.
The third section gives results for
fits where normally unfit parameters are fit.
Finally, the fourth section includes fits with additional
constraints:  the absence of a BAO peak $(\Apeak=0)$, an
isotropic BAO peak $(\aperpMath=\aparMath),$ or
imposing the fiducial cosmology $(\aperpMath=\aparMath=1)$

Figure \ref{figure::xi_wedge_000_rescale2_data} shows the measured
cross-correlation for the data in four bins of $\mu$.
Also shown are
three of the  fits listed in  Table \ref{table::fit_data_model}:
the
standard fit
(Sect. \ref{subsection::Fits_for_the_peak_position::Results}),
the standard fit if the velocity distribution
of quasars was null, and finally   the fit
with the broadband function (\ref{equation::broadband_1}).

In our earlier studies of the cross-correlation
\citep{2014JCAP...05..027F} and of the auto-correlation
\citep{2015A&A...574A..59D} we did not attempt to measure the distortion
of the correlations by the fit of the quasar continuum.
This distortion was modeled by the broadband function (\ref{equation::broadband_1}).
We now take this effect into account with the distortion matrix
(\ref{equation::distortion_matrix_measure}), so we use these broadband
functions to test for any systematic errors in the determination
of the BAO parameters' values and precision. These potential errors
could be induced by any correlations between the sidebands and the BAO peak
position (smooth and peak components in Eq. \ref{equation::pk_p_nl}).
In this study, we tested a large number of broadband functions modeled
by Equation \ref{equation::broadband_1}, keeping $(i,j)$ within
reasonable values:
$(i_{\min},i_{\max}) \in [-4,3]$ and $(j_{\min},j_{\max}) \in [0,6]$.
In a similar way, for the broadband functions
modeled by Equation \ref{equation::broadband_2}, we tested reasonable values
of $(i,j)$:
$(i_{\min}, i_{\max}, j_{\min}, j_{\max}) \in [-4,3]$.
These choices allowed an investigation of a wide range of broadbands
without introducing an excessive number of
parameters and  unrealistic features
in the cross-correlation.
All of these different broadband functions do not change 
the values of the two BAO parameters by more than $0.5\sigma$.
The precision of the two BAO parameters is also not significantly
degraded by the presence of broadbands.
This behavior is in contrast to the auto-correlation function
\citep{2017A&A...603A..12B} where the broadband terms
over the range $40<r<180~\hMpc$
degraded significantly the precision on \aperp.

In addition to the changes in the fitting procedure described
in the previous section, we also tested the robustness of
the BAO peak position determination by dividing the data into
roughly equal subsamples that would be expected to yield compatible
peak positions.
The results of these ``data splits'' are listed in Table \ref{table::fit_data_split}.
The splits divide the data according to
the relative distances of the quasar and forest pixel
($\rpar<0,\,\geq0$),
the pair redshift ($z_{\rm pair}$), the Galactic hemisphere (NGC, SGC), the quasar
position on the observing plates (fiber Id$<500,\,\geq500$),  quasar target
sample (CORE, notCORE) \citep{2012ApJS..199....3R},
and quasar emission-line strength (Amp. CIV).
The last three data splits in the table
use indicators of the quality of the quasar spectrum:
the quasar magnitude ($i_q$) and the signal-to-noise ratio
in the forest (${\rm SNR_{Ly\alpha}}$)
and redward of \Lya~emission (${\rm SNR_{1700}}$).
None of these data splits indicate an unexpected shift in the
BAO peak.

\begin{table*}[tb]
        \centering
        
        \caption{Results of non-standard fits.
          The first group includes fits of increasingly complete physical models,
          the first assuming only \Lya~absorption and then adding successively metals,
          a Lorentzian smearing of $z_q$, high column density systems, ionizing flux (UV)
          fluctuations, and the transverse proximity effect,
          the last corresponding to the standard fit.
          The next group includes two standard fits but with non-standard fitting ranges.
          The next group includes fits with non-standard treatment of certain parameters.
          The fit ``HCD fixed'' fixes the three HCD parameters to their values found in fits
          of the auto-correlation.
          The other fits in this group leave free parameters
          that are normally fixed in the fit with the fit value given in parentheses.
          (The fit ``Gaussian'' adds an additional Gaussian
          to describe quasar redshift smearing.)
          The final group includes three fits with non-standard treatments of the BAO peak.
          In all fits we fix $\bqso = 3.87$ and $f = 0.97076$.
          Errors correspond to $\Delta\chi^2=1$.
        }
        \label{table::fit_data_model}

        \begin{tabular}{l l l l l l}

        Analysis  
    & \apar{}
    & \aperp{} 
    & $b_{\LyaMath}\left(1+\beta_{\LyaMath}\right)$
    & $\beta_{\LyaMath}$
    & $\chiSquareMin/DOF, \,\, \mathrm{probability}$  \\

        \noalign{\smallskip} 
        \hline \hline
        \noalign{\smallskip} 
        
        % In the good order
        %
        %Work/Data/Cross_alone/Fits_pyLya/Fit_rmin10_rmax160/cross_alone.save.pars
        %Work/Data/Cross_alone/Fits_pyLya/Fit_rmin10_rmax160_metals/cross_alone.save.pars
        %Work/Data/Cross_alone/Fits_pyLya/Fit_rmin10_rmax160_metals_laurentzian/cross_alone.save.pars
        %Work/Data/Cross_alone/Fits_pyLya/Fit_rmin10_rmax160_metals_laurentzian_HCD/cross_alone.save.pars
        %Work/Data/Cross_alone/Fits_pyLya/Fit_rmin10_rmax160_metals_laurentzian_HCD_UV/cross_alone.save.pars
        %Work/Data/Cross_alone/Fits_pyLya/Fit_rmin10_rmax160_metals_laurentzian_HCD_UV_QSORadiation/cross_alone.save.pars

        $ \rm{Ly}\alpha                       $ & $ 1.086\pm0.038 $ & $ 0.896\pm0.036 $ & $ -0.2796\pm0.0042 $ & $ 0.986\pm0.074 $ & $ 2830.76 / (2504-~~5),  p = 3~10^{-6} $ \\
        $ \rm{+Metals}                        $ & $ 1.084\pm0.037 $ & $ 0.896\pm0.037 $ & $ -0.2838\pm0.0043 $ & $ 1.111\pm0.085 $ & $ 2790.46 / (2504-~~9),  p = 3~10^{-5} $ \\ 
        $ +z_q                                $ & $ 1.077\pm0.039 $ & $ 0.894\pm0.040 $ & $ -0.3649\pm0.0092 $ & $ 2.69~\,\pm0.25 $ & $ 2586.43 / (2504-10),  p = 0.096 $ \\
        $ \rm{+HCD}                           $ & $ 1.080\pm0.039 $ & $ 0.896\pm0.040 $ & $ -0.370~\,\pm0.010 $ & $ 2.52~\,\pm0.24 $ & $ 2579.82 / (2504-13),  p = 0.11 $ \\
        $ \rm{+UV}                            $ & $ 1.077\pm0.040 $ & $ 0.897\pm0.041 $ & $ -0.354~\,\pm0.021 $ & $ 2.45~\,\pm0.25 $ & $ 2579.12 / (2504-14),  p = 0.10 $ \\ 
        $ \rm{+TP}                            $ & $ 1.077\pm0.038 $ & $ 0.898\pm0.038 $ & $ -0.350~\,\pm0.019 $ & $ 1.90~\,\pm0.34 $ & $ 2576.31 / (2504-15),  p = 0.11 $ \\

        \noalign{\smallskip} 
        \hline \hline
        \noalign{\smallskip} 

        % In the good order
        %
        %Work/Data/Cross_alone/Fits_pyLya/Fit_rmin10_rmax160_metals_laurentzian_HCD_UV_QSORadiation_freeLparaLper/cross_alone.save.pars
        %Work/Data/Cross_alone/Fits_pyLya/Fit_rmin10_rmax160_metals_laurentzian_HCD_UV_QSORadiation_fitForSigmanlperSigmanlpar/cross_alone.save.pars
        %Work/Data/Cross_alone/Fits_pyLya/Fit_rmin10_rmax160_metals_laurentzian_HCD_UV_QSORadiation_freeBAOAmplitude/cross_alone.save.pars
        %Work/Data/Cross_alone/Fits_pyLya/Fit_rmin10_rmax160_metals_laurentzian_HCD_UV_QSORadiation_fixBAOAmplitude/cross_alone.save.pars
        %Work/Data/Cross_alone/Fits_pyLya/Fit_rmin10_rmax160_metals_laurentzian_HCD_UV_QSORadiation_fitAiso/cross_alone.save.pars
        %Work/Data/Cross_alone/Fits_pyLya/Fit_rmin10_rmax160_metals_laurentzian_HCD_UV_QSORadiation_fixBA0ToFiducial/cross_alone.save.pars
        %Work/Data/Cross_alone/Fits_pyLya/Fit_rmin10_rmax160_metals_laurentzian_HCD_UV_QSORadiation_0_2_0_6_distort/cross_alone.save.pars
        %Work/Data/Cross_alone/Fits_pyLya/Fit_rmin40_rmax160_metals_laurentzian_HCD_UV_QSORadiation/cross_alone.save.pars
        %Work/Data/Cross_alone/Fits_pyLya/Fit_rmin10_rmax180_metals_laurentzian_HCD_UV_QSORadiation/cross_alone.save.pars
        %Work/Data/Cross_alone/Fits_baofit/BOSSDR12QSOLyaF_k_save.pars
        $ r_{\min}=40                         $ & $ 1.074\pm0.033 $ & $ 0.902\pm0.033 $ & $ -0.361\pm0.026 $ & $ 1.26\pm0.29 $ & $ 2406.25 / (2354-15),  p = 0.16 $ \\ 
        $ r_{\max}=180                        $ & $ 1.078\pm0.038 $ & $ 0.896\pm0.039 $ & $ -0.353\pm0.019 $ & $ 1.95\pm0.34 $ & $ 3352.32 / (3180-15),  p = 0.010 $ \\

        \noalign{\smallskip} 
        \hline \hline
        \noalign{\smallskip} 
  HCD fixed  &  $1.082 \pm 0.038$  &  $0.893 \pm 0.038$  &  $-0.342 \pm 0.015$  &  $2.88 \pm 0.57$  &  $2590.70 / (2504-12),  p = 0.082$  \\
        $ (R_{\parallel}, \, R_{\perp})       $ & $ 1.076\pm0.038 $ & $ 0.898\pm0.037 $ & $ -0.344\pm0.018 $ & $ 1.81\pm0.37 $ & $ 2574.88 / (2504-17),  p = 0.11 $ \\
$\; (5\pm2\;,2.5\pm2) $ \\
        $ (\Sigma_{\parallel},\Sigma_{\perp}) $ & $ 1.068\pm0.030 $ & $ 0.901\pm0.034 $ & $ -0.349\pm0.019 $ & $ 1.91\pm0.34 $ & $ 2573.54 / (2504-17),  p = 0.11 $ \\
$\;\;(<4,\;<4)$ \\
  $ \lambdauv^{-1}   $ & $ 1.076\pm0.038 $ & $ 0.898\pm0.038 $ & $-0.347\pm0.019 $ & $ 1.89\pm0.34 $ & $ 2575.97 / (2504-16),  p = 0.11 $ \\
$\;\;(6\pm8)\times10^{-6}$\\
  $(\auv,\tuv^{-1})$  &  $1.080 \pm 0.039$  &  $0.895 \pm 0.040 $  & $ -0.368 \pm 0.023$  &  $2.27 \pm 0.26 $  &  $2571.33 /
  (2504-17),  p = 0.12$  \\
  $\;\;(1.27\pm0.56,$\\
  $\;\;0.43\pm 0.34)$ \\
        $ +z_q\, \rm{Gaussian} $ & $ 1.076\pm0.038 $ & $ 0.898\pm0.038 $ & $ -0.345\pm0.020 $ & $ 1.92\pm0.36 $ & $ 2575.90 / (2504-16),  p = 0.11 $ \\
        $ \Apeak$  & $ 1.077\pm0.040 $ & $ 0.896\pm0.040 $ & $ -0.348\pm0.019 $ & $ 1.87\pm0.35 $ & $ 2576.20 / (2504-16),  p = 0.11 $ \\
  $\;\;(0.93\pm0.26)$      \\
        $ \rm{BB}(0,2,0,6)    $ & $ 1.076\pm 0.035 $ & $ 0.891\pm 0.035 $ & $-$ & $-$ & $ 2375.23/( 2354 -36), p= 0.20 $ \\
%       $ \rm{BB}(0,2,0,6)                    $ & $ 1.088\pm0.033 $ & $ 0.895\pm0.033 $ & $ -0.344\pm0.039 $ & $ 3.1~\,\pm1.5 $ & $ 2535.53 / (2504-36),  p = 0.17 $ \\

        \noalign{\smallskip} 
        \hline \hline
        \noalign{\smallskip} 

        $ \Apeak=0                          $ & $  - $ & $  - $ & $ -0.338\pm0.018 $ & $ 1.78\pm0.33 $ & $ 2590.10 / (2504-13),  p = 0.081 $ \\
        $ \aparMath=\aperpMath                       $ &  $1.003 \pm 0.028 $ & $ 1.003 \pm 0.028 $ & $ -0.347\pm0.019 $ & $ 1.95\pm0.35 $ & $ 2582.53 / (2504-14),  p = 0.096 $ \\ 
        $ \aparMath=\aperpMath=1                    $ & $  1 $ & $  1 $ & $ -0.347\pm0.019 $ & $ 1.95\pm0.35 $ & $ 2582.58 / (2504-13),  p = 0.098 $ \\ 
        
        \end{tabular}
\end{table*}

\begin{table*}

        %
        % Work/Data/Cross_alone/Fits_pyLya/Fit_rmin10_rmax160_metals_laurentzian_HCD_UV_QSORadiation/
        % Work/Data_splits/*/Fit_rmin10_rmax160_metals_laurentzian_HCD_UV_QSORadiation/
        %

        \centering
        
        \caption{
          Best-fit results for the four most important parameters 
          for different data splits as described in the text.
          The standard fit is performed on each sample
          and the errors correspond to $\Delta\chi^2=1$.
        }
        \label{table::fit_data_split}

        \begin{tabular}{l l l l l l}
        
        Test
    &   \apar{}
    &   \aperp{}
    &   $b_{\LyaMath}\left(1+\beta_{\LyaMath}\right)$
    &   $\beta_{\LyaMath}$
    &   $\chiSquareMin/DOF, \,\, \mathrm{probability}$  \\

        \noalign{\smallskip} 
        \hline \hline
        \noalign{\smallskip}

        Std. fit, full sample               & $ 1.077\pm0.038 $ & $ 0.898\pm0.038 $ & $ -0.350\pm0.019 $ & $ 1.90\pm0.34 $ & $ 2576.31 / (2504-15),  p = 0.11 $ \\
        \noalign{\smallskip}

        $ r_{\parallel}<0                 $ & $ 1.058\pm0.059 $ & $ 0.928\pm0.064 $ & $ -0.334\pm0.025 $ & $ 2.37\pm0.75 $ & $ 1214.07 / (1252-15),  p = 0.67 $  \\
        $ r_{\parallel} \geq 0            $ & $ 1.090\pm0.052 $ & $ 0.871\pm0.052 $ & $ -0.342\pm0.038 $ & $ 1.42\pm0.43 $ & $ 1337.16 / (1252-15),  p = 0.024 $  \\
        \noalign{\smallskip}

        $ z_{\rm{pairs}} < 2.3962         $ & $ 1.079\pm0.058 $ & $ 0.904\pm0.051 $ & $ -0.334\pm0.023 $ & $ 1.53\pm0.39 $ & $ 2534.08 / (2504-15),  p = 0.26 $  \\
        $ z_{\rm{pairs}} \geq 2.3962      $ & $ 1.071\pm0.052 $ & $ 0.907\pm0.057 $ & $ -0.40\,~\pm0.027 $ & $ 2.86\pm0.67 $ & $ 2607.86 / (2504-15),  p = 0.048 $  \\
        \noalign{\smallskip}

        $ \rm{NGC}                        $ & $ 1.071\pm0.042 $ & $ 0.916\pm0.046 $ & $ -0.337\pm0.021 $ & $ 1.94\pm0.40 $ & $ 2616.95 / (2504-15),  p = 0.037 $  \\
        $ \rm{SGC}                        $ & $ 1.113\pm0.091 $ & $ 0.868\pm0.065 $ & $ -0.367\pm0.054 $ & $ 1.95\pm0.68 $ & $ 2525.13 / (2504-15),  p = 0.30 $  \\
        \noalign{\smallskip}

        $ \rm{Fiber \, Id} < 500          $ & $ 1.062\pm0.059 $ & $ 0.906\pm0.050 $ & $ -0.368\pm0.028 $ & $ 2.41\pm0.54 $ & $ 2448.64 / (2504-15),  p = 0.71 $  \\
        $ \rm{Fiber \, Id} \geq 500       $ & $ 1.084\pm0.052 $ & $ 0.894\pm0.062 $ & $ -0.336\pm0.022 $ & $ 1.35\pm0.35 $ & $ 2634.90 / (2504-15),  p = 0.021 $  \\
        \noalign{\smallskip}

        $ \rm{CORE \, QSO}                $ & $ 1.090\pm0.042 $ & $ 0.873\pm0.043 $ & $ -0.36\,~\pm0.031 $ & $ 2.61\pm0.67 $ & $ 2590.52 / (2504-15),  p = 0.076 $  \\
        $ \rm{not\, CORE \, QSO}          $ & $ 1.048\pm0.051 $ & $ 1.010\pm0.097 $ & $ -0.35\,~\pm0.023 $ & $ 1.55\pm0.37 $ & $ 2613.22 / (2504-15),  p = 0.041 $  \\
        \noalign{\smallskip}

        $ \rm{Amp. \, CIV} < 7.36         $ & $ 1.079\pm0.040 $ & $ 0.856\pm0.048 $ & $ -0.367\pm0.025 $ & $ 2.36\pm0.55 $ & $ 2542.33 / (2504-15),  p = 0.22 $  \\
        $ \rm{Amp. \, CIV} \geq 7.36      $ & $ 1.117\pm0.086 $ & $ 0.902\pm0.048 $ & $ -0.34\,~\pm0.027 $ & $ 1.43\pm0.40 $ & $ 2550.04 / (2504-15),  p = 0.19 $  \\
        \noalign{\smallskip}

        $ \rm{SNR_{\LyaMath}} < 3.2919    $ & $ 1.016\pm0.053 $ & $ 0.932\pm0.049 $ & $ -0.35\,~\pm0.032 $ & $ 2.46\pm0.63 $ & $ 2680.68 / (2504-15),  p = 0.0039 $  \\
        $ \rm{SNR_{\LyaMath}} \geq 3.2919 $ & $ 1.117\pm0.049 $ & $ 0.863\pm0.048 $ & $ -0.349\pm0.022 $ & $ 1.72\pm0.39 $ & $ 2631.62 / (2504-15),  p = 0.023 $  \\
        \noalign{\smallskip}

        $ \rm{SNR\_}1700 < 5.16           $ & $ 1.068\pm0.045 $ & $ 0.902\pm0.045 $ & $ -0.373\pm0.028 $ & $ 2.43\pm0.57 $ & $ 2618.61 / (2504-15),  p = 0.035 $  \\
        $ \rm{SNR\_}1700 \geq 5.16        $ & $ 1.077\pm0.061 $ & $ 0.902\pm0.062 $ & $ -0.344\pm0.025 $ & $ 1.58\pm0.39 $ & $ 2697.78 / (2504-15),  p = 0.0019 $  \\
        \noalign{\smallskip}

        $ \rm{Mag_{i}} < -25.4            $ & $ 1.089\pm0.046 $ & $ 0.880\pm0.051 $ & $ -0.348\pm0.021 $ & $ 1.61\pm0.36 $ & $ 2596.31 / (2504-15),  p = 0.066 $  \\
        $ \rm{Mag_{i}} \geq -25.4         $ & $ 1.040\pm0.056 $ & $ 0.921\pm0.048 $ & $ -0.379\pm0.039 $ & $ 2.92\pm0.79 $ & $ 2554.25 / (2504-15),  p = 0.18 $  \\
        \noalign{\smallskip}

        \noalign{\smallskip}

        \noalign{\smallskip}

        \end{tabular}
\end{table*}%

\section{Monte-Carlo determination of the statistical errors of fit parameters}
\label{append::fastMC}

To make a precise estimate of the relation between
$\Delta\chi^2$ and confidence level, we
generated a large number of  simulated correlation functions using the fiducial
cosmological model and the best-fit values of non-BAO parameters, 
randomized using the covariance matrix measured with the data.
Two types of simulated correlation functions were produced: one with only \Lya-absorption
and one ``complete'' simulation that included metals, UV fluctuations, and
quasar (QSO)
radiation.
Each simulated correlation function was then fit for the model parameters
and the $\chi^2$ for the best-fit parameters compared with the best $\chi^2$ with
one or more parameters set to the known input values.
The $\Delta\chi^2$ corresponding to a given fraction of simulated correlation
functions could then be determined.

The results are summarized 
in Table \ref{table::fastmc} for fits
of the cross- and auto-correlation functions and for combined fits.
For the cross-correlation, 
the parameters $\blya(1+\betalya)$ and $\betalya$ have an associated $\Delta\chi^2$
for $CL=(68.27,95.45,99.7)\%$ that is consistent with the expected
values of $(1,4,9)$.
The values for the BAO peak parameters (\aperp,\apar) are somewhat
higher: $\Delta\chi^2\sim(1.18,4.8,11.)$ suggesting that the nominal ``$1\sigma$'' errors
should be increased by a factor $\sqrt{1.18}=1.09$.
This is true for both the \Lya-only and complete simulations.
For the pair (\aperp,\apar), the results 
indicate $\Delta\chi^2\sim(2.62,7.25)$ corresponds to
confidence levels of $(68.27,95.45\%)$.
We have adopted these values of $\Delta\chi^2$ for the errors
reported in Equations (\ref{equation::measure_alpha_perp})
through (\ref{equation::measure_Dh_over_rd}) and the
contours in Fig. \ref{figure::chi2_scan_data_alpha_paral_alpha_perp}.

The values for $\Delta\chi^2$ for the auto-correlation are similar to
those for the cross-correlation.  An exception is the value for the parameter
$\betalya$ , which has $\Delta\chi^2=1.09\pm0.02$, significantly higher than
the expected value of unity.
For the pair (\aperp,\apar), we have adopted the values
$\Delta\chi^2=(2.6,7.1)$ for the auto-correlation contours in
Fig. \ref{figure::chi2_scan_data_alpha_paral_alpha_perp_combined}.

For the combined fits, the $\Delta\chi^2$ are closer to the nominal values.
This is to be expected because the peak position is better determined, so
the model is closer to being a linear function of (\aperp,\apar) in the limited
range around (1,1).
For the combined-fit contours in
Fig. \ref{figure::chi2_scan_data_alpha_paral_alpha_perp_combined}.
we have adopted the values
$\Delta\chi^2=2.45,6.4,14.)$ for $CL=(68.27,95.45,99.7)\%$.

The ``frequentist'' intervals reported in this paper are renormalized
using the $\Delta\chi^2$ found with the simulation presented in this
section.  Bayesian ``credible intervals'' require no such renormalization
since they use directly the measured $\chi^2$ as a function of model
parameters.
The Bayesian analogs of our results (\ref{equation::measure_alpha_perp})
and (\ref{equation::measure_alpha_parallel}) for a uniform prior
on (\aperp,\apar) are
\begin{align}
\aperpMath = 0.906  \pm 0.0424\, (68.27\%)  \pm 0.0917\, (95.45\%) 
  \\
\aparMath = 1.077  \pm 0.0405\, (68.27\%)  \pm 0.0841\, (95.45\%) \;.
\end{align}

\begin{table*}
        \centering
        \caption{
          Values of  $\Delta\chi^2$ corresponding to  CL=(68.27, 95.45, 99.7\%)
          as derived from the Monte-Carlo simulation of correlation functions.
          For the four fit variables
          $x=(\blya(1+\betalya),\,\betalya,\,\aparMath,\,\aperpMath)$ and
          for the combination $(\aperpMath,\aparMath)$,
          the table gives the range of $\Delta\chi^2\equiv\chi^2(x=x_{\rm in})-\chi^2_{\rm min}$
          that includes a percentage CL of the generated data sets.
          Values are given for fits of the cross- and auto-correlations and for the
          combined fits.
          In each of these three cases, values are given for the $\sim10000$ simulations with 
          only \Lya~absorption, and for the $\sim1000$ ``complete'' simulations with
          metals, UV fluctuations, and QSO radiation.
          The uncertainties are statistical, reflecting the number of simulations.
          For the variables $\blya(1+\betalya)$ and $\betalya$,
          $\Delta\chi^2$
          is close to the nominal values $(1, 4, 9)$ expected for Gaussian statistics.
          For \aperp{} and \apar{} the number is consistently higher than
          the nominal values, as is also the case
          for the combination (\aperp,\apar), which is consistently
          larger than the nominal values (2.29, 6.18, 11.82).
        }
        \label{table::fastmc}

        \begin{tabular}{l| l l l| l l l }
          \hline
          & \multicolumn{3}{ c }{$\Delta\chi^2$: \Lya-only simulation} & \multicolumn{3}{ c }{$\Delta\chi^2$: Complete simulation}\\
 CL         & 68.27\% & 95.45\% & 99.7\% & 68.27\% & 95.45\% & 99.7\% \\
          \hline

\noalign{\smallskip}
Cross   & & & & & &  \\
  %  9945 6059
$\aparMath$ & $1.14 \pm 0.02$ & $4.76 \pm 0.09$ & $10.05 \pm 0.53$ &
$1.16 \pm 0.03$ & $4.77 \pm 0.10$ & $10.5 \pm 0.79$ \\
  %  9930 6063
$\aperpMath$ & $1.18 \pm 0.02$ & $4.86 \pm 0.08$ & $10.59 \pm 0.43$ &
$1.23 \pm 0.03$ & $4.93 \pm 0.11$ & $10.16 \pm 0.30$ \\
  %  9974 1485
$\bbetalya$ & $1.04 \pm 0.02$ & $4.13 \pm 0.09$ & $9.12 \pm 0.25$ &
$1.04 \pm 0.06$ & $4.07 \pm 0.22$ & $9.52 \pm 1.55$ \\
  %  9965 2521
$\betalya$ & $1.02 \pm 0.01$ & $4.07 \pm 0.09$ & $9.17 \pm 0.27$ & $1.00
\pm 0.03$ & $4.01 \pm 0.23$ & $8.91 \pm 1.40$ \\
  %  9992 6281
$(\alpha_{\parallel},\alpha_{\perp})$ & $2.62 \pm 0.03$ & $7.25 \pm
0.05$ & $12.93 \pm 0.32$ & $2.65 \pm 0.04$ & $7.24 \pm 0.09$ & $13.23
\pm 0.43$ \\
\noalign{\smallskip}
\hline
\noalign{\smallskip}
Auto   & & & & & &  \\
  %  9952 6163
$\aparMath$ & $1.14 \pm 0.02$ & $4.52 \pm 0.10$ & $10.68 \pm 0.43$ &
$1.16 \pm 0.02$ & $4.66 \pm 0.11$ & $9.94 \pm 0.48$ \\
  %  9959 6171
$\aperpMath$ & $1.20 \pm 0.01$ & $4.85 \pm 0.08$ & $10.84 \pm 0.59$ &
$1.20 \pm 0.03$ & $4.87 \pm 0.09$ & $10.36 \pm 0.34$ \\
  %  9966 6169
$\bbetalya$ & $0.98 \pm 0.02$ & $4.09 \pm 0.09$ & $9.25 \pm 0.40$ &
$0.96 \pm 0.02$ & $3.82 \pm 0.09$ & $8.65 \pm 0.43$ \\
  %  9955 6175
$\betalya$ & $0.99 \pm 0.01$ & $4.07 \pm 0.07$ & $9.48 \pm 0.49$ & $1.09
\pm 0.02$ & $4.42 \pm 0.07$ & $9.67 \pm 0.56$ \\
  %  9999 6193
$(\alpha_{\parallel},\alpha_{\perp})$ & $2.63 \pm 0.03$ & $7.13 \pm
0.11$ & $14.22 \pm 0.74$ & $2.65 \pm 0.05$ & $7.07 \pm 0.12$ & $13.63
\pm 0.77$ \\
\noalign{\smallskip}
\hline
\noalign{\smallskip}
Combined   & & & & & &  \\
  %  6930 2945
$\aparMath$ & $1.08 \pm 0.02$ & $4.19 \pm 0.05$ & $9.73 \pm 0.41$ &
$1.03 \pm 0.03$ & $4.20 \pm 0.09$ & $9.89 \pm 0.46$ \\
  %  9962 910
$\aperpMath$ & $1.08 \pm 0.02$ & $4.37 \pm 0.08$ & $10.22 \pm 0.53$ &
$1.11 \pm 0.07$ & $4.19 \pm 0.32$ &  ---  \\
  %  4641 2947
$\bbetalya$ & $1.03 \pm 0.02$ & $4.06 \pm 0.09$ & $9.41 \pm 0.90$ &
$0.93 \pm 0.04$ & $3.77 \pm 0.10$ & $9.31 \pm 1.01$ \\
  %  19311 2944
$\betalya$ & $1.00 \pm 0.01$ & $4.18 \pm 0.06$ & $9.18 \pm 0.15$ & $1.24
\pm 0.05$ & $4.79 \pm 0.14$ & $11.24 \pm 0.64$ \\
  %  9997 2955
$(\alpha_{\parallel},\alpha_{\perp})$ & $2.45 \pm 0.02$ & $6.58 \pm
0.10$ & $12.95 \pm 0.44$ & $2.47 \pm 0.05$ & $6.45 \pm 0.11$ & $13.5 \pm
1.52$ \\
\noalign{\smallskip}
\hline

        \end{tabular}
\end{table*}%

\end{document}